\documentclass[]{aa}

\usepackage{graphicx}

\usepackage{amsmath}
\usepackage{amssymb}
\usepackage{amsfonts}
\usepackage{latexsym}
\usepackage{txfonts}

\usepackage{natbib}

\usepackage{fixltx2e}

\usepackage{afterpage}
\usepackage{multirow} 
\usepackage{color}

\usepackage{float}
\usepackage{caption}

\usepackage[hyperindex]{hyperref}
\hypersetup{
breaklinks = {true},
colorlinks = {false},
linkcolor={black},
pdfpagemode = {None}, 
pdfborder = {0 0 1},
pdftitle = {Probing the dark-matter halos of cluster galaxies with weak lensing},
pdfsubject = {},
pdfauthor = {E. Pastor Mira, S. Hilbert,  J. Hartlap, and P. Schneider},
pdfkeywords = {none}
}

\bibpunct{(}{)}{;}{a}{}{,}

\newcommand{\kms}{\ensuremath{\mathrm{km\,s}^{-1}}}

\newcommand{\pc}{\ensuremath{\mathrm{pc}}}
\newcommand{\kpc}{\ensuremath{\mathrm{kpc}}}
\newcommand{\Mpc}{\ensuremath{\mathrm{Mpc}}}

\newcommand{\Msolar}{\ensuremath{\mathrm{M}_\odot}}

\newcommand{\degt}{\ensuremath{\mathrm{deg}}}

\newcommand{\vect}[1]{\boldsymbol{#1}}

\newcommand{\ii}{\mathrm{i}}

\newcommand{\omm}{\ensuremath{\Omega_{\rm m}}}
\newcommand{\omv}{\ensuremath{\Omega_\Lambda}}
\newcommand{\seight}{\ensuremath{\sigma_8}}

\newcommand{\Sigmas}{\Sigma_\mathrm{s}}
\newcommand{\rs}{r_\mathrm{s}}
\newcommand{\rt}{r_\mathrm{tr}}
\newcommand{\Msubf}{M_\mathrm{SUBF}}

\newcommand{\gtang}{\gamma_{\rm t}}
\newcommand{\gammat}{\gamma_{\mathrm{t}}}
\newcommand{\epsilont}{\epsilon_{\mathrm{t}}}

\newcommand{\vtheta}{\vect{\theta}}
\newcommand{\vbeta}{\vect{\beta}}

\newcommand{\ev}[1]{\left\langle{#1}\right\rangle}
\newcommand{\abs}[1]{\left\lvert{#1}\right\rvert}

\newcommand{\GGLC}{GGLC}

\begin{document}

\title{Probing the dark-matter halos of cluster galaxies with weak lensing}

\author{E. Pastor Mira\inst{1}, S. Hilbert\inst{1,2},  J. Hartlap\inst{1}, P. Schneider\inst{1}}

\institute{
 Argelander-Institut f{\"u}r Astronomie, Universit{\"a}t Bonn, Auf dem H{\"u}gel 71, D-53121 Bonn, Germany \and 
 Max-Planck-Institut f{\"u}r Astrophysik, Karl-Schwarzschild-Stra{\ss}e 1, 85741 Garching, Germany
}
\date{Received  / Accepted }
\authorrunning{Pastor Mira et al.}
\titlerunning{Probing the dark-matter halos of cluster galaxies with weak lensing}

\keywords{gravitational lensing: weak, Galaxies: clusters: general, Galaxies: evolution, dark matter}

\abstract {
  Understanding the evolution of the dark matter halos of galaxies after they
  become part of a cluster is essential for understanding the evolution of
  these satellite galaxies.}{
  We investigate the potential of galaxy-galaxy lensing to map the halo
  density profiles of galaxies in clusters.}{
  We propose a method that separates the weak-lensing signal of the
  dark-matter halos of galaxies in clusters from the weak-lensing signal of
  the cluster's main halo. Using toy cluster models as well as ray-tracing
  through N-body simulations of structure formation along with semi-analytic
  galaxy formation models, we test the method and assess its performance.
}{
  We show that with the proposed method, one can recover the density profiles
  of the cluster galaxy halos in the range 30 - 300 kpc. Using the method, we
  find that weak-lensing signal of cluster member galaxies in the Millennium
  Simulation is well described by an Navarro-Frenk-White (NFW) profile. In
  contrast, non-singular isothermal mass distribution (like PIEMD) model
  provide a poor fit.  Furthermore, we do not find evidence for a sharp
  truncation of the galaxy halos in the range probed by our method. Instead,
  there is an observed overall decrease of the halo mass profile of cluster
  member galaxies with increasing time spent in the cluster. This trend, as
  well as the presence or absence of a truncation radius, should be detectable
  in future weak-lensing surveys like the Dark Energy Survey (DES) or the
  Large Synoptic Survey Telescope (LSST) survey. Such surveys should also
  allow one to infer the mass-luminosity relation of cluster galaxies with our
  method over two decades in mass.
}{
  It is possible to recover in a non-parametric way the mass profile of
  satellite galaxies and their dark matter halos in future surveys, using our
  proposed weak lensing method.}
  
\maketitle

\section{Introduction}
Astronomical observations indicate that the majority of matter in the Universe
is of a yet unknown form that neither emits nor absorbs light
\citep[e.g.][]{2010arXiv1001.4538K,2010MNRAS.401.2148P,2009ApJ...699..539R}. A
unique way to study this so-called dark matter and its relation to luminous
matter is provided by gravitational lensing
\citep{SchneiderKochanekWambsganss_book}. Gravitational lensing has been used
to probe the matter associated with galaxies (e.g.~SLACS
survey\footnote{http://www.slacs.org}, \citealt{2004ApJ...606...67H},
\citealt{2006MNRAS.368..715M}, \citealt{2007ApJ...669...21P},
\citealt{2008JCAP...08..006M}, \citealt{2008A&A...479..655S},
\citealt{2009MNRAS.393..885T}), clusters (e.g.~\citealt{2006ApJ...648L.109C},
\citealt{2006ApJ...652..937B}, \citealt{2006MNRAS.372.1425H},
\citealt{2008A&A...481...65H}, \citealt{2009ApJ...704..672J},
\citealt{2010A&A...514A..60S}) and the large-scale structure \citep[e.g.][]
{2010A&A...516A..63S,2008A&A...479....9F,2007MNRAS.381..702B}.

According to the hierarchical structure formation paradigm, local
overdensities created in the early Universe collapse into smaller dark
matter halos, in which galaxies form. Larger halos, corresponding to
galaxy clusters, form later through accretion and mergers. As a
result, a typical galaxy cluster has a massive main halo of dark
matter with a bright central galaxy (BCG) at its center. Within this
main halo, there are many smaller subhalos hosting a satellite galaxy.
These were once isolated objects that merged with the cluster.

An important open question is how subhalos evolve after they become part of a
cluster. This is also essential to understand the evolution of the satellite
galaxies embedded in them. Simple analytic models assume that subhalos are
just stripped of their mass outside some tidal radius by gravitational tidal
forces. On the other hand numerical simulations indicate that tidal forces
also heat the subhalos causing them to expand and decrease their central
density (e.g.~\citealt{1998MNRAS.300..146G}, \citealt{2000ApJ...544..616G},
\citealt{2003ApJ...584..541H}). In this picture, both tidal stripping and
heating continually change the radial mass profiles of subhalos, eventually
destroying them.

Most observational studies of subhalos in clusters with gravitational lensing
employ parametric models for the mass distribution of the subhalos. Using
parametric models has the advantage that one can obtain useful constraints on
the model parameters even with a modest amount and quality of lensing
data. The method analyzes individual clusters and obtains the mass model
parameters as a function of the luminosity of the satellite galaxy
\citep[e.g.][]{2007A&A...461..881L,2007MNRAS.376..180N,2007ApJ...656..739H,2010arXiv1007.4815S}. However,
a major disadvantage of this approach is that it relies on strong assumptions
about the mass profiles of subhalos.

Direct measurements of the mean tangential shear profile around suitably
chosen samples of massive objects provide a more direct and less biased view
of their mean mass profiles. This approach has been successfully applied to
study the halos of field galaxies (e.g.~\citealt{2008JCAP...08..006M}), and
cluster main halos (e.g.~\citealt{2009ApJ...703.2232S},
\citealt{2010PASJ...62..811O}, \citealt{2010A&A...520A..58I}). However, this
approach has a drawback when used for cluster member galaxies and their
embedding subhalos: The resulting shear signal probes the subhalo profiles
only very close to their center, while the signal becomes dominated at larger
distances by the surrounding much more massive cluster main halo.

In this work, we propose a modification of the standard weak galaxy-galaxy
lensing approach that addresses the latter problem. The key idea is to `calibrate
out' the cluster main halo signal by subtracting from the signal measured
around the satellite galaxies the tangential shear signal around a specific
set of calibration points. As shown in this paper, the additional calibration
allows one to reliably probe the subhalo profiles to much larger radii than
with the standard galaxy-galaxy lensing approach.

We test the proposed method and assess its performance compared to the
standard approach using simulated lensing fields generated from the Millennium
Run \citep{2005Natur.435..629S}. Furthermore, we analyze the resulting
subhalos profile as predicted by our simulations and provide forecasts for the
signal-to-noise level expected for large upcoming and future surveys like the
Dark Energy Survey\footnote{\url{http://www.darkenergysurvey.org/}}(DES), or
the Large Synoptic Survey Telescope\footnote{\url{http://www.lsst.org}} (LSST)
survey. Moreover, we discuss how weak lensing can be used to study the mass
loss of subhalos during their evolution in galaxy clusters.

Our paper is organized as follows: In Sect.~\ref{sec:theory}, we briefly
discuss the theoretical background. In Sect.~\ref{sec:method} we describe our
proposed method in detail. A short overview of our simulations is given in
Sect.~\ref{sec:simulations}. In Sect.~\ref{sec:RES} we test the performance of
our method, we apply it to characterize the dark matter profiles of subhalos
in the Millennium simulation, and we forecast signals and noise for upcoming
surveys. In Sect.~\ref{sec:system} we discuss the most relevant systematic
effects for this work. The paper concludes with a summary in
Sect.~\ref{sec:sumcon}.

\section{Theory}
\label{sec:theory}
In this section, we briefly discuss the theory of gravitational lensing needed
for our work. We refer the reader to the standard literature
\citep[e.g.][]{2001PhR...340..291B,SchneiderKochanekWambsganss_book} for a
detailed discussion.

\subsection{Weak lensing basics}
\label{sec:weak_lensing}
A massive foreground structure, the lens, deflects the light emitted by a
distant galaxy in its background, the source, toward an observer. Sources at
angular position $\vbeta=(\beta_1, \beta_2)$ are seen by the observer at a
possibly different image position $\vtheta=(\theta_1, \theta_2)$. Differential
deflection causes distortions in the images of the background
galaxies. Locally, the distortion is quantified by the Jacobian
\begin{equation}
\mathcal{A}_{ij}=\partial\beta_i/\partial\theta_j
\end{equation}

\noindent of the lens mapping $\vtheta \mapsto \vbeta$. This Jacobian defines
the convergence,
\begin{equation}
 \kappa =1-\mathrm{Tr}(\mathcal{A})/2,
\end{equation}

\noindent the complex shear $\gamma = \gamma_1 + \ii\gamma_2$ where: 
\begin{subequations}
\begin{align}
 \gamma_1=&(\mathcal{A}_{22}-\mathcal{A}_{11})/2,\\
 \gamma_2=&(\mathcal{A}_{22}+\mathcal{A}_{11})/2,
\end{align}
\end{subequations}

\noindent and the reduced shear $g = \gamma/(1-\kappa)$.

Defining a complex ellipticity $\epsilon=\epsilon_{1}+\ii \epsilon_{2}$
\citep{2001PhR...340..291B} for the background sources, the differential light
deflection, and thus the mass associated with the lens, can be inferred
(statistically) from the measured shape. The relation between the intrinsic
ellipticity $\epsilon_{\mathrm{i}}$ and the observed one $\epsilon$ are
related by:

\begin{equation}
\label{eq:she_exp2}
  \epsilon=
  \begin{cases}
    \dfrac{\epsilon_{\mathrm{i}}+g}{1 + g^*\epsilon_{\mathrm{i}}}, & \text{if } \abs{g} \leq 1,\vspace{5pt} \\
    \dfrac{1+g\,\epsilon^*_{\mathrm{i}}}{\epsilon^*_{\mathrm{i}} + g^*}, & \text{if } \abs{g} > 1.
  \end{cases}
\end{equation}

In the following, we assume the validity of the weak lensing approximation
i.e.:
\begin{equation}
  \abs{\kappa} \ll 1, \; g < 1, \; g \approx \gamma \text{, and} \; \epsilon\approx\epsilon_{\mathrm{i}} +\gamma.
\end{equation}
Note, however, that we test this assumption in Sect.~\ref{sec:wap}. Under this
approximation and assuming that the $\epsilon_{\mathrm{i}}$ of the different
sources are uncorrelated:
\begin{equation}
\ev{\epsilon}=\gamma\label{eq:ave_ell}
\end{equation}

\subsection{Galaxy-galaxy lensing}
\label{sec:galaxy_galaxy_lensing}

In galaxy-galaxy lensing, one measures the tangential ellipticity of a
background galaxy image with respect to the position of a foreground lens
galaxy,
\begin{equation}
  \epsilont = -\cos(2\phi) \epsilon_1 - \sin(2\phi) \epsilon_2,
\end{equation}
where $\phi$ is the polar angle of the line connecting lens and background
image. From Eq.~(\ref{eq:ave_ell}) follows that $\ev{\epsilont} = \gammat$ in
the weak-lensing regime, where $\gammat$ denotes the tangential component of
the shear with respect to the lens position.

Under the geometrically-thin lens approximation, we can define the excess
surface mass density $\Delta\Sigma(\xi)$ at projected physical radius $\xi$
by:
\begin{equation}
  \Delta\Sigma(\xi)
  =\bar{\Sigma}(\xi )-\Sigma(\xi),
\end{equation}
where $\bar{\Sigma}(\xi)$ denotes the average surface mass density within a
circle of radius $\xi$, and $\Sigma(\xi)$ is the average surface mass density
on that circle.

The tangential shear $\gammat(\xi)$ averaged on a circle with physical
projected radius $\xi$ around a lens galaxy is related to the excess surface
mass density $\Delta\Sigma(\xi)$ by \citep{2006glsw.conf..269S}:
\begin{equation}
  \Delta\Sigma(\xi) =\gtang(\xi ) \Sigma_{\rm crit},
\end{equation}
where the critical surface mass density $\Sigma_{\rm crit}$ is a function of
the angular diameter distance to the lens $D_\mathrm{d}$, to the source
$D_\mathrm{s}$, and the distance between lens and source $D_\mathrm{ds}$:
\begin{equation}
 \Sigma_{\rm crit}=\frac{c^2}{4\pi G}
  \frac{D_\mathrm{s}}{D_\mathrm{ds}D_\mathrm{d}}.
\end{equation}

An important aspect of galaxy-galaxy lensing is to combine the different
measurements from different foreground-background galaxy pairs to increase the
signal-to-noise. Each lens-background galaxy pair $i$ with projected
separation $\xi$ at the lens, tangential ellipticity $\epsilon_{\mathrm{t},i}$
of the background galaxy image, and critical surface mass density
$\Sigma_{\mathrm{crit},i}$ provides an estimate
\begin{equation}\label{eq:expdelsig}
\widehat{\Delta\Sigma}_i(\xi) = \Sigma_{\mathrm{crit},i}  \epsilon_{\mathrm{t},i}
\end{equation}
for the excess surface mass density $\Delta\Sigma(\xi)$ of the lens. For a
sample of lens and background galaxies, one can compute the final estimate
$\widehat{\Delta\Sigma}(\xi)$ by a weighted mean:
\begin{equation}
 \widehat{\Delta\Sigma}(\xi)=\sum_i \,w_i\:  \widehat{\Delta\Sigma}_i(\xi),
\end{equation}
where the sum runs over all pairs formed by a background image and a lens
galaxy.  If one assumes that each lens galaxy has the same mass profile and
each shear estimate from a background galaxy image carries the same
uncertainty, the optimal weights are given by
\begin{equation}
  w_i=\frac{\Sigma^{-2}_{\mathrm{ crit},i}(z_{\rm s},z_{\rm l})}
  {\sum_j\Sigma^{-2}_{\mathrm{ crit},j}(z_{\rm s},z_{\rm l})}.
\end{equation}
The weights make our estimator sensitive to the redshift distribution of
lenses and sources. These weights are a simplified version of those used, e.g,
by \cite{2008JCAP...08..006M}, since we neglect redshift errors and
redshift-dependent errors in the ellipticity estimation.

In this work we also consider simulated pixelized surface mass density
maps. In those cases, we estimate directly:

\begin{equation}
\widehat{\Delta\Sigma}(\xi)=\sum_k \, w'_k \: \left(\hat{\overline{\Sigma}}_k(\xi)-\hat{\Sigma}_k(\xi)\right)
\end{equation}

\noindent around each mass concentration. The sum runs over all the lenses in
the sample. For these weights we use the number of pixels in the annulus defined
to compute $\hat{\Sigma}_k(\xi)$,

\begin{equation}
  w'_k=\frac{n_k}{\sum_jn_j}.
\end{equation}

\subsection{Analytic models for halo profiles}
\label{sec:analytic_halo_profiles}

In the analysis of the simulated excess surface mass profiles, we
consider several analytic models for the mass distribution inside
galaxy halos.

\subsubsection{Navarro-Frenk-White profiles}

On average, the three-dimensional density profiles of the virialized inner
regions of dark-matter halos in simulations are well fit by
Navarro-Frenk-White (NFW) profiles \citep[][]{1997ApJ...490..493N}:
\begin{equation}
  \rho(r)=\dfrac{\delta_{\rm c}\,\rho_{\rm m}}{(r/r_{\rm s})(1+r/r_{\rm s})^2},
\end{equation}
where $\delta_{\rm c}$ is a characteristic density, $\rho_{\rm m}$ is the mean
matter density of the universe at the halo redshift, and $r_{\rm s}$ is the
halo scale radius.

We characterize the spatial extent of NFW halos by $r_{200}$, i.e. the radius
in which the average density is 200 times the mean cosmic matter density
$\rho_{\rm m}$. A commonly used parameter for NFW profiles is the
concentration $c=r_{200}/r_{\rm s}$.

The mass $M_{200}$ within $r_{200}$ is then given by
\begin{equation}
  M_{200} = 4\pi \,\delta_{\rm c}\,\rho_{\rm m}\,r^3_{\rm s}
  \left[
    \ln(1+c)
    -\frac{1}{1+c}
  \right].
\end{equation}

Using the parameter $x=\xi/r_{\rm s}$ and the characteristic surface mass
density $\Sigma_{\rm s}=r_{\rm s}\delta_{\rm c}\rho_{\rm m}$, the projected
surface mass density $\Sigma_{\rm NFW}$ of the NFW profile at a projected
radius $\xi$ reads $\Sigma(\xi)=\Sigma_{\rm s}\cdot f(\xi/r_{rm s})$ and the
mean projected mass density $\bar{\Sigma}(\xi)=\Sigma_{\rm s}\cdot g(\xi/r_{\rm
  s})$ where $f(x)$ and $g(x)$ are defined in \cite{2000ApJ...534...34W}, (see
also \citealt{1996A&A...313..697B}).

\subsubsection{Truncated Navarro-Frenk-White profiles}

If one integrates the mass of an NFW profile up to an infinite radius, the
total mass diverges unless one introduces an additional truncation. The
truncated NFW profile we consider here was derived by
\citet{2009JCAP...01..015B}. Its density profile reads:
\begin{equation}
  \rho(r)=
  \dfrac{\delta_{\rm c}\,\rho_{\rm m}}
  {(r/r_{\rm s})(1+r/r_{\rm s})^2
    \left[ 1+(r/r_{\rm tr})^2\right]},
\end{equation}
where $r_{\rm tr}$ denotes the truncation radius. Defining $x=\xi/r_{\rm s}$, $\tau=r_{\rm tr}/r_{\rm s}$, and the function
\begin{equation}
  F(x)=
  \begin{cases}
    \dfrac{\ln\Big(1/x+\sqrt{1/x^2-1}\Big)}{\sqrt{1-x^2}} & \text{for } x<1, \\
    1 & \text{for } x=1,\vspace{5pt}\\
    \dfrac{\arccos(1/x)}{\sqrt{x^2-1}} & \text{for } x>1,
    \\ 
  \end{cases}
\end{equation}
one has
\begin{multline}
  \Sigma(x) =
  \frac{2\,\Sigma_{\rm s}\,\tau^2}{(\tau^2+1)^2}
  \Bigg[
  \frac{\tau^2+1}{x^2-1}\big(1-F(x)\big)+2\,F(x)
  \\
  -\frac{\pi}{\sqrt{\tau^2+x^2}}+\frac{\tau^2-1}{\tau\sqrt{\tau^2+x^2}}
  \log \left(\frac{x}{\sqrt{\tau^2+x^2}+\tau} \right)
  \Bigg].
\end{multline}
The mean density within a radius $x$ is:
\begin{multline}
  \bar{\Sigma}(x) = 
\frac{4\,\Sigma_{\rm s}}{x^2}
\frac{\tau^2}{(\tau^2+1)^2}
\times\\
\Bigg\{
\bigg[\tau^2+1+2(x^2-1)\bigg]F(x)+\tau\,\pi 
+(\tau^2-1)\log\tau 
\\
+\sqrt{\tau^2+x^2}
\left[
-\pi+\frac{\tau^2-1}{\tau}\log \left(\frac{x}{\sqrt{\tau^2+x^2}+\tau} \right)
\right]
\Bigg\}.
\end{multline}

\subsubsection{Truncated Non-singular Isothermal profiles}

This profile (in the following TNSI) is a truncated and spherical version of
the Pseudo-Isothermal Elliptical Mass Distribution (PIEMD) profile by
\cite{1993ApJ...417..450K}. Its three dimensional distribution reads:
\begin{equation}
   \rho(r)=
  \dfrac{\rho_{0}}{  (1+r^2/r^2_{\rm core}) ( 1+r^2/r^2_{\rm cut})},
\end{equation}
with $r_{\rm core}$ denoting the profile core radius, and $r_{\rm
  cut}$ denotes a cutoff radius. The cutoff radius $r_{\rm cut}$ plays
a role similar to the truncation radius $r_{\rm tr}$ of the truncated
NFW. The surface mass density distribution reads:
\begin{equation}
\Sigma(\xi)=\frac{\rho_{0}\,r^2_{\rm core}\,r^2_{\rm cut}\,\pi}{r^2_{\rm cut}-r^2_{\rm core}}
\left(
\frac{1}{\sqrt{r^2_{\rm core}+\xi^2}}
-\frac{1}{\sqrt{r^2_{\rm cut}+\xi^2}}
\right).
\end{equation}
By integrating, one obtains the mean surface mass density
\begin{equation}
\bar{\Sigma}(\xi)
=\frac{2\,\rho_{0}\,r^2_{\rm core}\,r^2_{\rm cut}\,\pi}{\xi^2(r_{\rm cut}+r_{\rm core})}
\left(
1-
\frac{\sqrt{r^2_{\rm cut}+\xi^2}-\sqrt{r^2_{\rm core}+\xi^2}}{(r_{\rm cut}-r_{\rm core})}
\right).
\end{equation}

\section{Galaxy-galaxy lensing for subhalos}
\label{sec:method}

\begin{figure}
 \centerline{\includegraphics[width=.3\textwidth,angle=0]{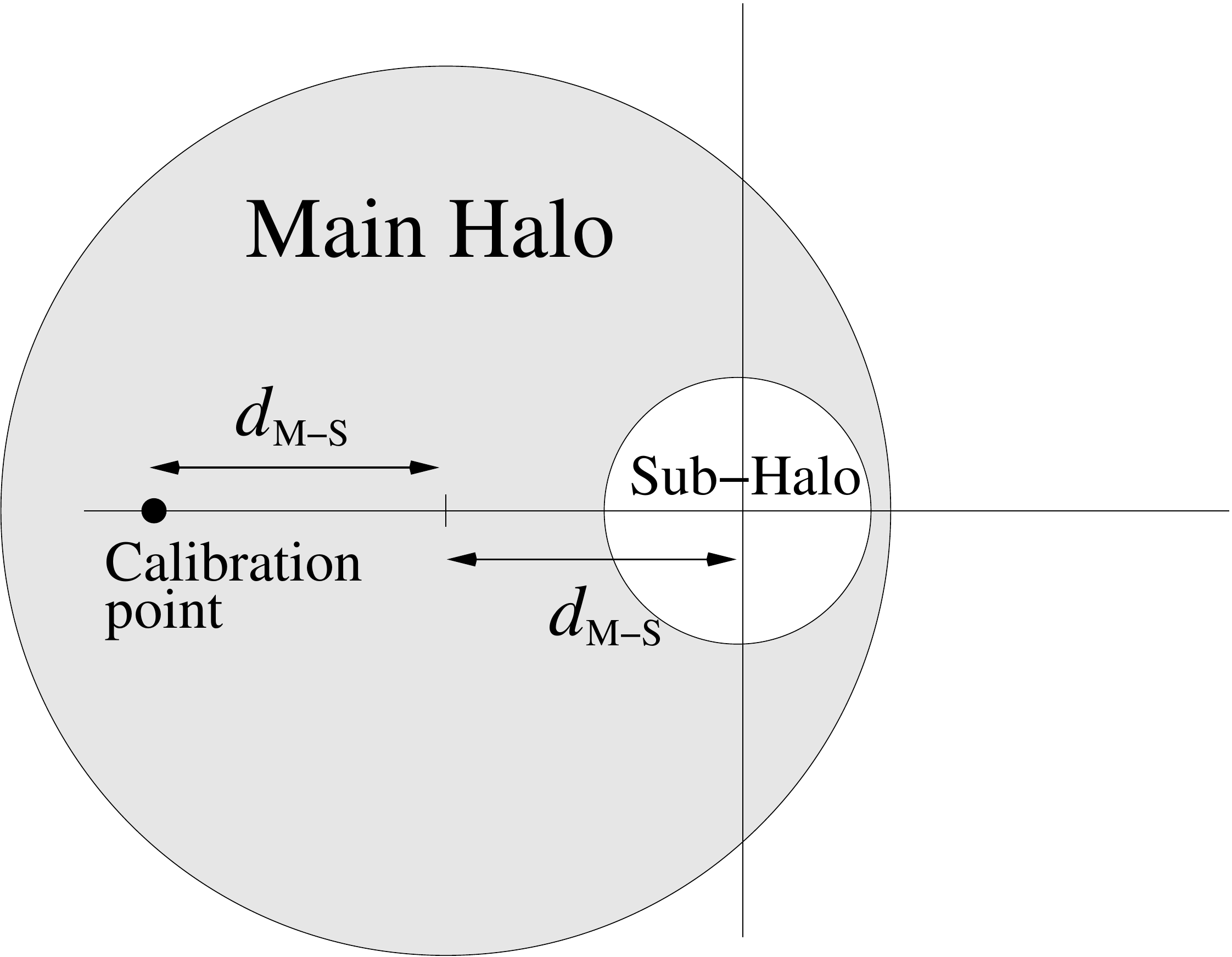}}
 \caption{ \footnotesize Sketch of a cluster with the main halo and a subhalo
   with a cluster galaxy at its center. Also shown is the calibration point
   used to estimate the main halo contribution to the lensing signal around
   the cluster galaxy and the same distance $d_{\rm M-S}$ from the cluster
   center.}\vspace{-5pt}\label{fig:mainsket}
\end{figure}

\begin{figure}
  \includegraphics[width=.5\textwidth,angle=0]{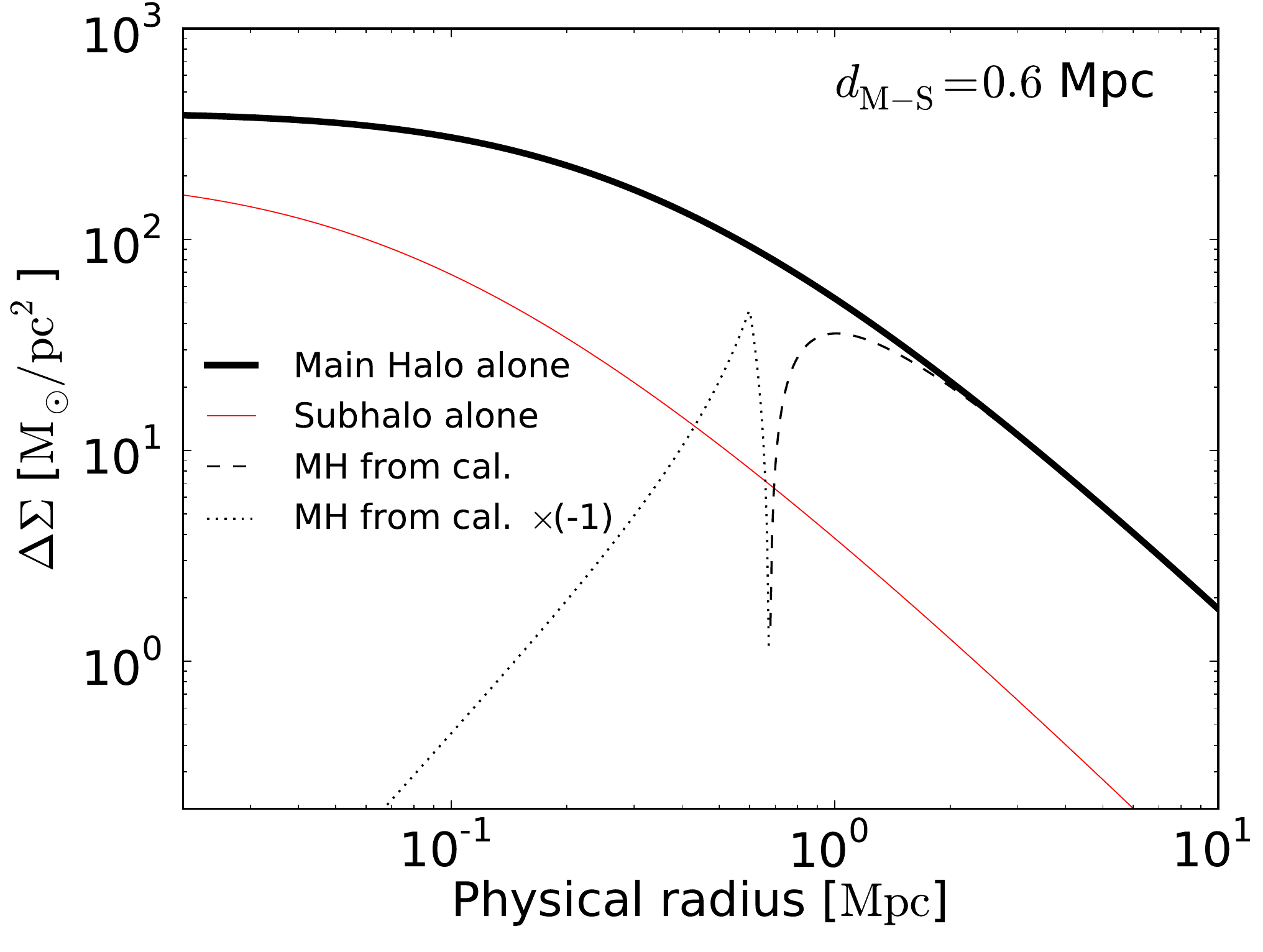}
  \vspace{-10pt}
  \includegraphics[width=.5\textwidth,angle=0]{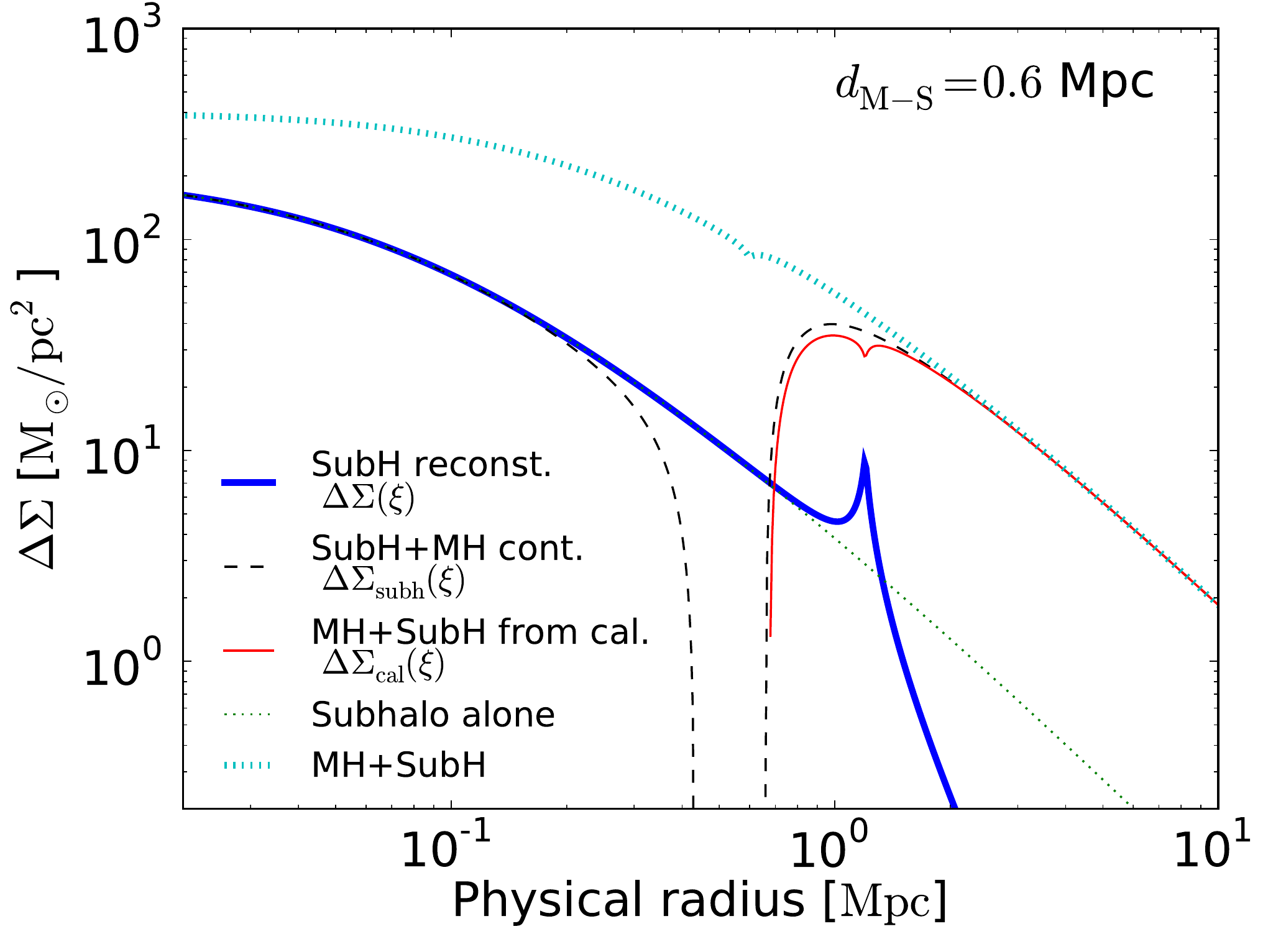}
  \caption{\footnotesize Galaxy-galaxy lensing signals of cluster main halo
    and subhalo (assuming NFW profiles). The subhalo is at 0.6 Mpc from the
    cluster center. Top panel: the subhalo around the subhalo center (thin
    solid red line), the main halo around the main halo center (thick solid
    black line), and the main halo around the calibration point (dashed/dotted
    line for positive/negative part). Bottom panel: the subhalo around the
    subhalo center (thin dotted green line), the subhalo plus main halo around
    the subhalo center (black dashed line), the subhalo plus main halo around
    the main halo center (cyan dotted line), the subhalo plus main halo around
    the calibration point (red solid line), and the reconstructed subhalo
    (blue thick solid line).} \label{fig:mainsketsig} \vspace{-10pt}
\end{figure}

Fig.~\ref{fig:mainsket} shows a schematic view of a galaxy cluster
consisting of a cluster main halo and a subhalo. When galaxy-galaxy lensing in
its standard form is applied to probe the mass distribution around satellite
galaxies (subhalos), the resulting lensing signal has a strong contribution
from the cluster main halo.


One can use an analytic model of the main halo to take its contribution into
account. However, this relies on strong assumptions about the main halo mass
profile. We propose a non-parametric scheme to estimate the main halo
contamination and subtract it from the signal. As illustrated in
Fig.~\ref{fig:mainsketsig}, a calibration point can be used to estimate the
contamination from the main halo. The calibration point is located at the same
distance $d_{\rm M-S}$ from the main halo center as the subhalo center, but on
the opposite side. If the cluster is point-symmetric around its center, we can
subtract the signal around the calibration point from the signal around the
subhalo, and remove the contamination from the main halo. In the following, we
will call this method \emph{galaxy-galaxy lensing with calibration} (GGLC).
  
Note, however, that the calibration does not work for very large projected
radii $\xi$. For $\xi>2\times d_{\rm M-S}$ the method fails since the
calibration signal includes the subhalo signal itself. Moreover, as shown in
Fig.~\ref{fig:mainsketsig}, the signals around the subhalo and the calibration
point are very steep for $\xi$ around $d_{\rm M-S}$. Thus the reconstructed
subhalo signal there will be very susceptible to errors introduced by shot
noise or the intrinsic ellipticity of the background galaxy. This makes it
impossible to reconstruct the subhalo profile for $\xi>d_{\rm M-S}$ in
practice.

Since the point-symmetry of any cluster sample is not perfect, our calibration
is not only limited from the theoretical point of view. We discuss this in
detail in Sect.~\ref{sec:restest}.

The excess surface mass density $\Delta\Sigma_{\rm subh}(\xi)$ measured around
a sample of lens galaxies can be thought to consist of (at least) two
parts. The first contribution is due the (sub)halos of the lenses we are
targeting. This part usually dominates the signal close to the lens
galaxy. There, $\Delta\Sigma_{\rm subh}(\xi)$ can be interpreted as the excess
surface mass density profile of an average lens galaxy halo. The second
contribution is due to the (sub)halos of other galaxies, and the clusters'
main halo. This contribution dominates on larger scales. There,
$\Delta\Sigma_{\rm subh}(\xi)$ cannot be interpreted directly as the average
lens galaxy halo profile.

For a realistic study we need a large sample of clusters, and we do not need
each cluster to be perfectly point symmetric. We only need the cluster {\it on
  average} to be {\it sufficiently point-symmetric}. In the following the
lensing results we present are given by

\begin{equation}
  \overline{\Delta\Sigma}(\xi)=\overline{\Delta\Sigma}_{\rm subh}-\overline{\Delta\Sigma}_{\rm cal}(\xi).
\end{equation}

\section{Simulations}
\label{sec:simulations}

\begin{figure*}
  \includegraphics[width=0.48\textwidth]{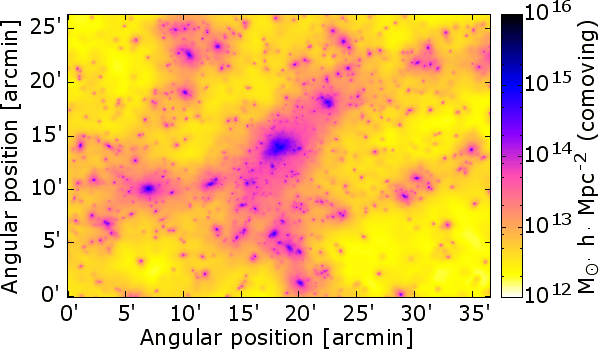}\hspace{.5cm}
  \includegraphics[width=0.48\textwidth]{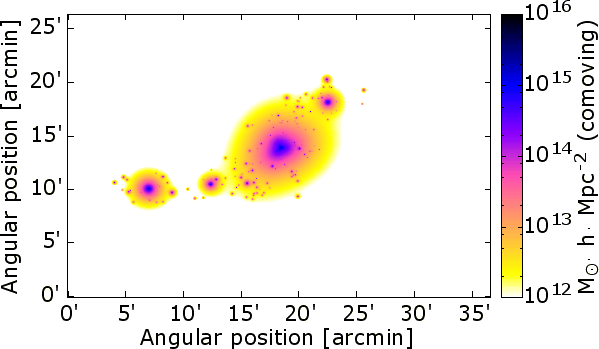}
  \caption{\footnotesize  Projected mass maps of a cluster in the MS (left panel) and the
    synthetic cluster created from it (right panel).} \label{fig:subhchara}
\end{figure*}

In order to test our proposed galaxy-galaxy lensing method and provide
forecasts for the expected signal, we use simulated lensing fields. The fields
are obtained by ray-tracing through the Millennium Simulation in combination
with mock galaxy catalogs from semi-analytic galaxy formation models.

\subsection{The Millennium Simulation}
\label{sec:MR}

The Millennium Simulation (MS) \citep{2005Natur.435..629S} is a large
$N$-body simulation of cosmic structure formation in a flat
$\Lambda$CDM universe. It uses $2160^3$ dark-matter particles of
$8.6\times 10^8h^{-1}\,\Msolar$ in a periodic simulation volume of
$500h^{-1}\,\Mpc$ side length. The assumed cosmological parameters
are: a Hubble constant $h=0.73$ (in units of $100\,\kms\Mpc^{-1}$), a
matter density $\omm=0.25$ and a cosmological constant energy density
$\omv=0.75$ (in units of the critical density), and a normalization of
the matter power spectrum $\seight=0.9$.
 
An important aspect of the simulation is how structure is defined. All
particles which are closer than a certain threshold value are
linked. Independent groups of particles are grouped with this Friends of
Friends (FOF) algorithm. These FOF groups are considered main halos in the
simulation. Substructures within these main halos are identified by the
SUBFIND algorithm \citep{2001MNRAS.328..726S}: Each FOF group is subjected to
a gravitational binding analysis, and every group of at least 20
gravitationally bound particles forms a subhalo. The algorithm also estimates
a mass $M_{\rm SUBF}$ for each subhalo that equals the total mass of the bound
particles.\footnote{Note, that there is no unique way of estimating
  the mass of a subhalo, and that even the same algorithm gives different
  values depending on the subhalo position inside the main halo
  \citep{2010arXiv1008.2903M}.}

We define a cluster as a FOF group whose main halo has a mass according to the
SUBFIND algorithm larger than $10^{14}\:\mathrm{M}_{\odot}/h$. We use this
criterion to certify that the selected FOF groups have a large fraction of
their mass gravitationally bound and to avoid non-virialized objects.

The MS itself does not take into account baryonic physics. Galaxy formation
within the evolving matter distribution of the simulation can however be
followed by semi-analytic galaxy formation models \citep[see,
e.g.,][]{2001MNRAS.328..726S,2004MNRAS.349.1101D}. In this work we use the
semi-analytic catalogues by \cite{2007MNRAS.375....2D} to infer the properties
of galaxies in the MS. The model used by them assumes that at the center of
every main halo and subhalo there is a galaxy whose evolution is determined by
the merging history of the host dark-matter halo and gas-physical
processes. These processes are approximated by a set of empirical equations
with parameters tuned to make the resulting galaxy population consistent with
many observed galaxy properties.

Furthermore, the semi-analytic model allows galaxies whose subhalos has been
dissolved during a halo merger event onto a larger halo to survive for some
time before merging onto the central galaxy of the main halo. The properties
and merging timescales of these orphan galaxies are not known very accurately
(from observations or theory), but their contribution to the lensing signal
can be significant. In our primary analysis we neglect them. In
Sect.~\ref{sec:type2} we discuss how our results are changed if these galaxies
are included in the analysis.

\subsection{Lensing simulations}
\label{sec:rt}

The lensing effects caused by the matter structures in the MS are calculated by
the multiple-lens-plane ray-tracing algorithm described in
\citet[][]{2009A&A...499...31H}. The ray-tracing is used to create 128
simulated fields of view of $2\times2\,\degt^2$. Convergence and shear in
these fields are stored on regular grids of $4096^2$ pixels in the image plane
for multiple source redshifts. In addition, the lensed image properties of all
semi-analytic galaxies in each field with stellar masses $\geq 10^9
h^{-1}\,\Msolar$ are computed. The lens galaxy population for the
galaxy-galaxy lensing simulations are selected from these lensed mock galaxy
catalogs.

The ray-tracing simulations are also used to compute pixelized maps of the
projected surface mass density $\Sigma(\boldsymbol{\beta})$ of thin redshift
slices of the simulated lensing light cones. We use these maps first to create
simplified mock clusters for performance tests of the calibration method; and
second to make a preliminary analysis of our signals.

The source galaxy populations for the galaxy-galaxy lensing simulations are
created by randomly drawing image positions within the simulated fields
(assuming a uniform spatial distribution in the image plane). The redshifts of
the source galaxies are drawn from a distribution
(\citealt{1996ApJ...466..623B})

\begin{equation}
 p(z)=z^\alpha  \, \exp\left(-(z/z_0)^{\beta}\right)\:
\frac{\beta}{(z_0)^{1+\alpha}\:\:\Gamma\left(\frac{1+\alpha}{\beta}\right)},
\end{equation}

\noindent where $\alpha=2$, $\beta=1.5$ and $z_0$ is a function of the median
redshift of the survey.  We then identify the lens plane that is closest in
redshift to each galaxy and interpolate onto it the shear and magnification
from the ray-tracing. We add an intrinsic ellipticity to the shear in order to
mimic real galaxy images (Eq.~\ref{eq:she_exp2}). The modulus of the intrinsic
ellipticity is drawn from a Gaussian distribution with zero mean and standard
deviation $\sigma_{\vert \epsilon_{\rm i} \vert}$. In Sect.~\ref{sec:simsur}
we discuss in more detail our simulated lensing catalogues.

\subsection{Mock cluster mass maps}
\label{sec:synthetic_clusters}

To test our calibration scheme, we use a sample of mock clusters. This allows
us to know the `true' mass profiles of main halo and subhalos, and thus to
quantify how well our method can retrieve these profiles. Since we want to
assess the performance of the method in situations as realistic as possible,
we use the cluster sample we have from ray-tracing through the Millennium
Simulation as a starting point. For each cluster in our cluster sample we
create mock clusters at various levels of simplification consisting of an
elliptical, or non-elliptical main halo, or no main halo at all.

The main halo of each cluster in the ray-tracing catalog is characterized
first by fitting an elliptical truncated NFW profile to the projected mass
profile of the cluster.\footnote{Projected NFW profiles have been shown to be
  a good fit for projected cluster halo profiles of the Millennium Simulation
  \citep{2010MNRAS.404..486H}.} The resulting mass distribution $\Sigma_{\rm
  ellip}(\vbeta)$ is point-symmetric. To create non-symmetric main-halos, the
original projected mass $\Sigma(\vbeta)$ of the cluster is first divided by
the best-fitting elliptical truncated NFW profile $\Sigma_{\rm ellip}(\vbeta)$. We then calculate the coefficients
\begin{equation}
\begin{split}
   a_n &= \sum_p \frac{\Sigma(\vbeta_p)}{\Sigma_{\rm ellip}(\vbeta_p)} \cos\left[ n \phi(\vbeta_p) \right],\\ 
   b_n &= \sum_p \frac{\Sigma(\vbeta_p)}{\Sigma_{\rm ellip}(\vbeta_p)} \sin\left[ n \phi(\vbeta_p) \right],
\end{split}
\end{equation}

\noindent where the index $p$ runs over all pixels in the considered
patch. The angle $\phi(\vbeta_p)$ is the polar angle of the pixel from the
center of the cluster. The synthetic non-elliptical mass distribution is then
\begin{multline}
  \Sigma_{\rm non-el}({\vbeta})=\Sigma_{\rm ellip}({\vbeta})
  \times\\
  \left( 1+\sum_{n=1}^3\left\{ \cos\left[ n \phi(\vbeta_p) \right] +\sin\left[ n \phi(\vbeta_p) \right] \right\} \right).
\end{multline}

The mock subhalo profiles are based on the projected mass profiles of subhalos
in the MS. We group subhalos in our lensing simulations into bins of subhalo
mass $\Msubf$. Using the proposed calibration method on the projected mass
maps, we calculate the bin-averaged excess surface density $\Delta\Sigma(\xi)$
as a function of radius $\xi$ for each bin. We fit NFW profiles to the
resulting profiles in the range $0.03\,\Mpc \leq \xi \leq 0.2\,\Mpc$. The
relations between the subhalo mass $\Msubf$ and the best-fitting parameters
$\Sigmas$ and $\rs$ of the NFW profiles are then fit by power laws, yielding
\begin{align} 
 \frac{\rs}{\Mpc} &= 1.1\times10^{-6} 
  \left(\frac{\Msubf}{\Msolar}\right)^{0.365} \text{ and}
  \\
  \frac{\Sigmas}{\Msolar\, \pc^{-2}} &= 1.92 \times 10^{-3}
  \left(\frac{\Msubf}{\Msolar}\right)^{0.24}.
\end{align}

\noindent The mass of the NFW profile without truncation diverges if we integrate up to
an infinite radius. Therefore, we truncate our halos to make them more
realistic. We try different truncation schemes, keeping the former
relations for $\rs$ and $\Sigmas$. We experimented with subhalos without
truncation, significantly truncated ($r_{\rm tr}=2\,\rs$) and with a
truncation above the maximum radius for which we can recover the
subhalo lensing signal ($r_{\rm tr}=6.66\,\rs$) in the least massive subhalos.

Each synthetic cluster is constructed by combining the previously described
mock main halo and the mock subhalo profiles, whereby the centers of the main
halo and the subhalos are used as centers for the mock halos. The mock
clusters are then put onto grids having the same resolution as the projected
mass maps of the original clusters. Thanks to our procedure, the mock clusters
properly take into account
\begin{itemize}
\item the masses and sizes of the main halo and the subhalos,
\item the spatial distribution of the subhalos within the cluster, and
\item the correlation between the main halo shape and the position of the halo center and the subhalos.                                    
\end{itemize}
An example cluster is shown in Fig.~\ref{fig:subhchara}.

\section{Results}\label{sec:RES}

\subsection{Performance tests}\label{sec:restest}

\begin{figure}
  \begin{flushleft}
    \includegraphics[width=.49\textwidth,angle=0]{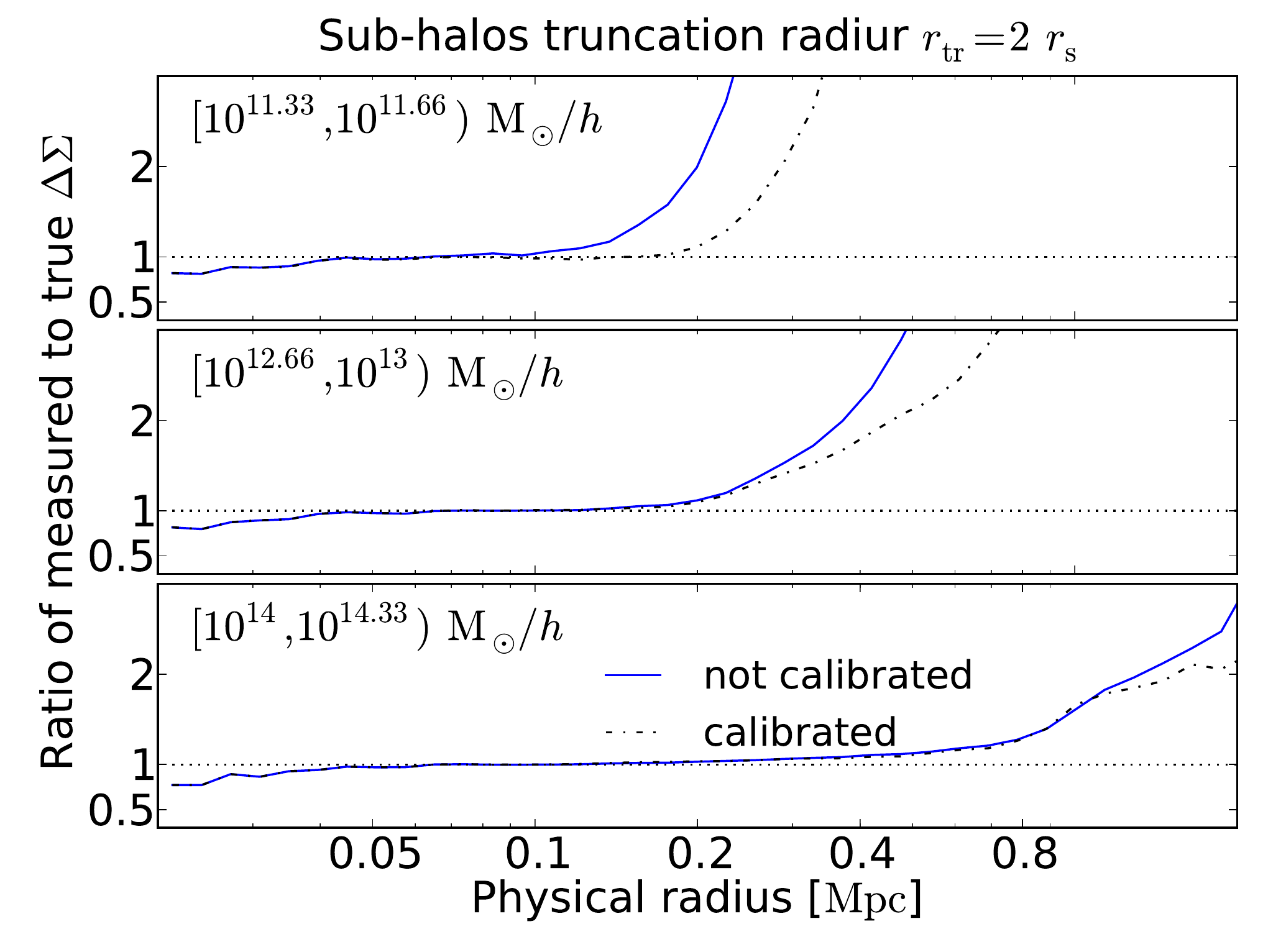}
  \end{flushleft}
  \vspace{-15pt}
  \caption{\footnotesize Ratio of the measured $\Delta\Sigma(\xi)$ to the one
    we input, using \GGLC{} and standard GGL. The dotted black line, where the
    ratio is one, is a visual reference.}\label{fig:calnocal}
  \vspace{-10pt}
\end{figure}

\begin{figure}
  \begin{flushleft}
    \includegraphics[width=.49\textwidth,angle=0]{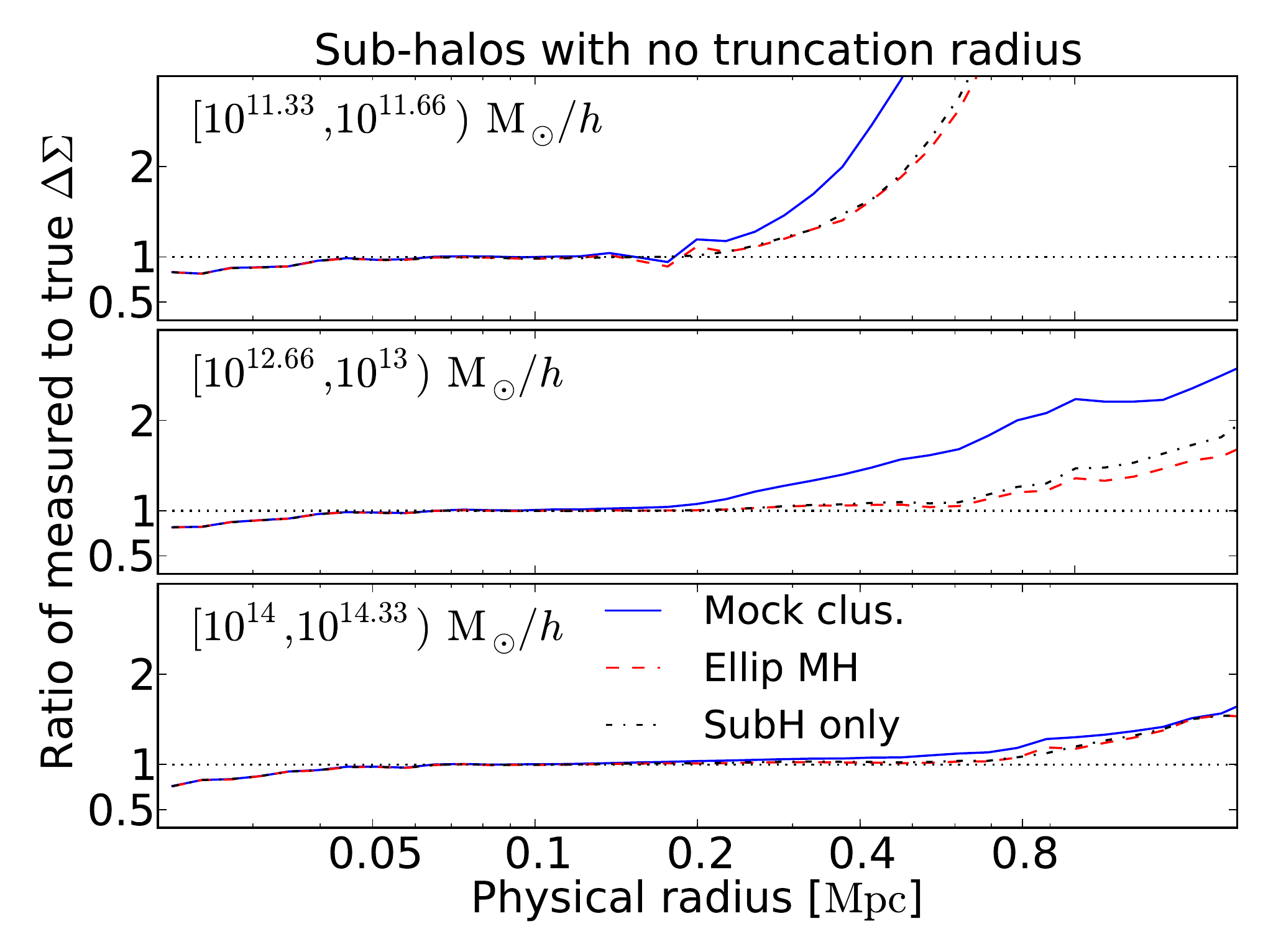}\\
    \includegraphics[width=.49\textwidth,angle=0]{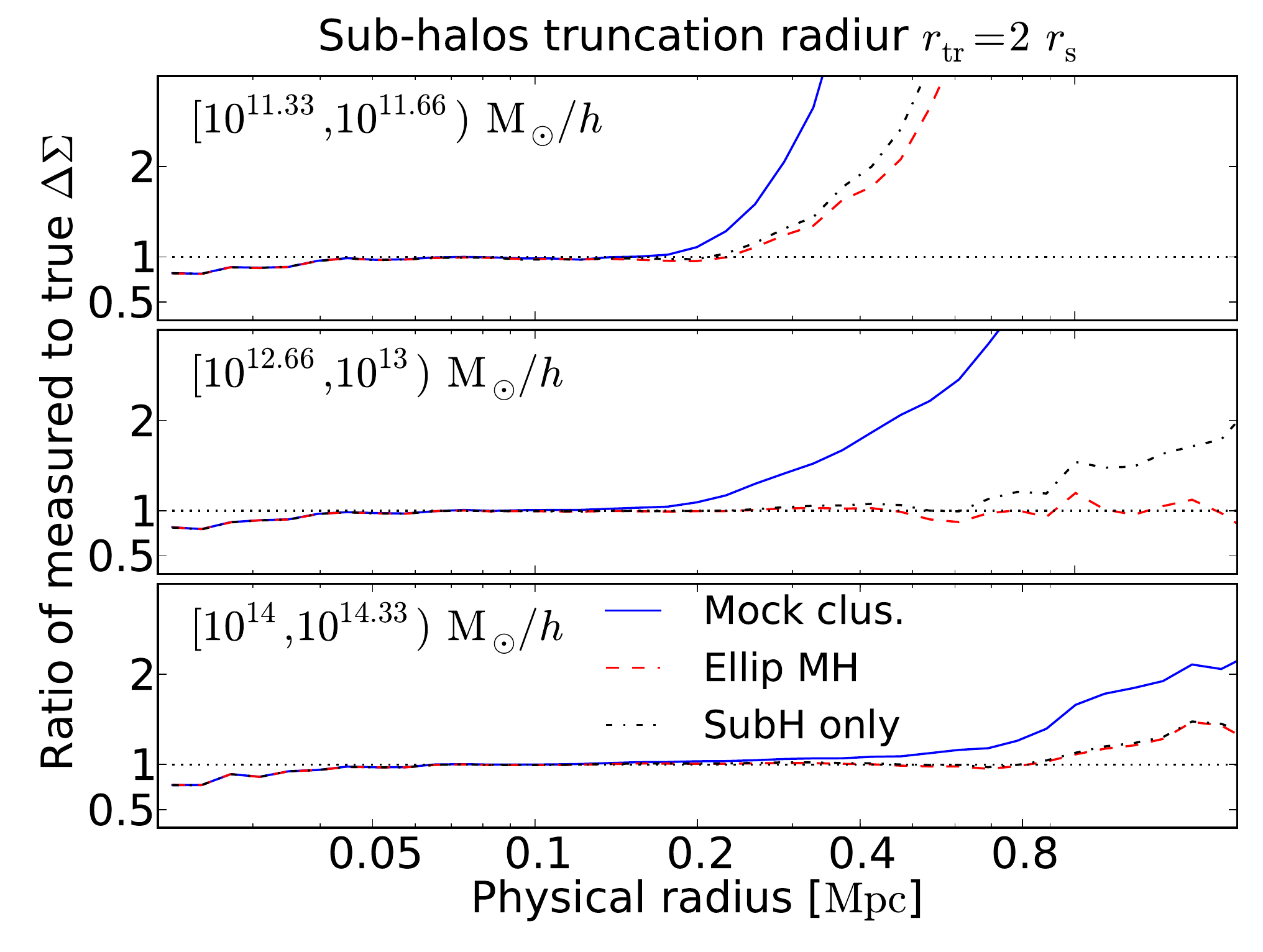}
  \end{flushleft}
  \vspace{-15pt}
  \caption{\footnotesize Results from mock clusters. Ratio of the measured
    $\Delta\Sigma(\xi)$ to the one we input using \GGLC{}, versus the
    projected physical radius $\xi$. Top figure: non-truncated NFWs. Bottom
    figure: $r_{\rm tr}=2\,r_{\rm s}$. We plot only the results for a
    selection of mass bins, the mass ranges are given. Black
    dotted-dashed lines: no main halo was added. Red dashed lines: clusters
    with an elliptical main halo. Blue solid lines: clusters with a
    non-elliptical main halo. The dotted black line, where the ratio is one,
    is a visual reference.}\label{fig:cal}
  \vspace{-10pt}
\end{figure}

The method we proposed in Sect.~\ref{sec:method} (\GGLC{}) allows
theoretically to clean the galaxy-galaxy lensing signal of a subhalo sample
from the main halo contamination. Here, we investigate the performance of the
method using the synthetic cluster sample. We quantify the range where we can
recover the signal below a certain bias and compare it to the reliably range
when using the `standard' GGL.

We divide the whole mock subhalo sample into several bins according to
$\Msubf$ as listed in Table~\ref{tab:chara_massb}. For each bin, we calculate
the stacked excess surface mass density profiles $\Delta\Sigma(\xi)$ as a
function of projected radius $\xi$ one would obtain if the subhalos were
completely isolated. We then compare this `true' profile to the signal
measured from our synthetic clusters.

In Fig.~\ref{fig:calnocal}, we present the difference between GGL and \GGLC{}
for one of the truncation schemes. The different truncation schemes offer the
same qualitative results. The bias with respect to the true value is reduced in all
cases. For the most massive subhalos, comparable in mass to the main halo, the
improvement is reduced. On the other hand, for the least massive subhalos the
range in radius $\xi$ with a small enough bias doubles. This is a significant improvement for
those subhalos which are more subjected to tidal fields and therefore
better tracers of the effects that we want to study.

Fig.~\ref{fig:cal} shows the ratio of the measured and true signal for samples
of subhalos in the clusters with truncated and non-truncated NFW subhalo
profiles and various models for the mass distribution of the cluster main
halo. For each mass bin in each panel, we present the results from different
situations. We present $\Delta\Sigma(\xi)$ for the following cases: coming
only from the subhalos, i.e.~the main halo was not added; for an elliptical
main halo plus the subhalo population; and from a realistic cluster, using a non-elliptical main halo.

The distribution of the sub-halos is not perfectly point-symmetric with
respect the cluster center. When we measure around a particular subhalo, we
``see'' the surrounding ones and the contribution to our measurement does not
average out. This also shows a correlation of the subhalo positions
independent of the mass distribution of the main halo.

A completely elliptical halo can always be calibrated properly. The observed
difference with respect to the signal with subhalos only, is due to the
resolution of our maps (which we also use for the synthetic cluster sample),
combined with the sensitivity of the calibration for radii larger than
$d_{\rm M-S}$ (Sect.~\ref{sec:method}). Internal checks were able to
verify the source of this discrepancy.

As a compromise between a small bias and a sufficiently large range in $\xi$,
we measure our profiles for those radii where the bias is less than a 10\%
from the input model. We will only be able to detect a truncation radius if
this scale is well inside the range we can measure. The difference between the
truncation models does not substantially change the region that we eventually
define as suitable to derive the mass profile. Since as we show later on, our
simulations do not present a detectable truncation of the mass density
profile, we focus in the following on subhalos \textit{without} truncation.

With GGLC, the measured subhalo profile recovers the true profile within
$10\%$ accuracy up to a radius of $\sim 200\,\kpc$ for less massive subhalos
(with $\Msubf \approx 10^{11.5}h^{-1}\,\Msolar$ and $\rs \approx 0.03\,\Mpc$),
and up to $\sim 500\,\kpc$ for more massive subhalos (with $\Msubf \approx
10^{14}h^{-1}\,\Msolar$). We can also see that on scales below $40\,\kpc$, the
measured signal is affected by the limited resolution of the synthetic mass
maps.

\begin{table}
  \caption{\footnotesize Binning according to $M_{\rm SUBF}$ of all subhalos in clusters for
    $z<0.9$: Mass ranges, number of subhalos, and mean masses of each subhalo
    bin used for assessing the performance of \GGLC{}. Note that mass ranges
    are given in $h^{-1}\Msolar$, while mean masses are given in $\Msolar$. }
  \label{tab:chara_massb}
  \centering
  \begin{tabular}{r c c c}
    \hline  
    \bf Bin&\bf Mass range [$\mathrm{M_\odot/h}$]\phantom{$\Big($} &\bf \#Subhalos &\bf ${M_{\rm SUBF}}$ [$\mathrm{M_\odot}$]\\
    \hline 
    1  & [$10^{11.33}:10^{11.66}$)               &  31831 &  $4.27\times 10^{11}$ \\
    2  & [$10^{11.66}:10^{12\phantom{.22}}$)      &  16025 &  $9.14\times 10^{11}$ \\    
    3  & [$10^{12\phantom{.22}}   :10^{12.33}$)   &  8273  &  $1.97\times 10^{12}$ \\
    4  & [$10^{12.33}:10^{12.66}$)               &  3834  &  $4.23\times 10^{12}$ \\
    5  & [$10^{12.66}:10^{13\phantom{.22}}$)      &  1816  &  $9.15\times 10^{12}$ \\
    6  & [$10^{13\phantom{.22}}   :10^{13.33}$)   &  810   &  $1.96\times 10^{13}$ \\
    7  & [$10^{13.33}:10^{13.66}$)               &  385   &  $4.27\times 10^{13}$ \\
    8  & [$10^{13.66}:10^{14\phantom{.22}}$)      &  162   &  $8.92\times 10^{13}$ \\    
    9  & [$10^{14\phantom{.22}}   :10^{14.33}$)   &  74    &  $1.88\times 10^{14}$\\
    \hline     \hline 
  \end{tabular}
\end{table}

\begin{table}
  \caption{\footnotesize Range in radius $\xi$ for which the bias is below
    10\%, for each subhalo bin according to $M_{\rm SUBF}$ in clusters for
    $z<0.9$. The subhalos have a separation from the main halos center $d_{\rm
      M-S}>0.5\:\mathrm{Mpc}$. The mass binning can be found in Tab.~\ref{tab:chara_massb}. }
  \label{tab:mass_range1}
  \centering
  \begin{tabular}{c c}
    \hline 
    \bf$M_{\rm SUBF}$ bin & \bf $\xi$ range\\
    \hline 
    1 &  [0.043:0.18]          \\
    2 &  [0.043:0.18]          \\    
    3 &  [0.043:0.18]          \\
    4 &  [0.043:0.27]          \\
    5 &  [0.043:0.27]          \\
    6 &  [0.043:0.41]          \\
    7 &  [0.043:0.41]          \\
    8 &  [0.043:0.5\phantom{0}]\\    
    9 &  [0.043:0.5\phantom{0}]\\
    \hline     \hline 
  \end{tabular}
\end{table}

The main halo's contribution to the bias in $\Delta\Sigma(\xi)$ is more
important for subhalos with a small separation from the cluster center $d_{\rm
  M-S}$. In fact, including subhalos with a too small $d_{\rm M-S}$, may
decrease the quality of the measured $\Delta\Sigma(\xi)$. Having this in mind,
we consider different minimum values for $d_{\rm M-S}$.

\begin{figure}
    \includegraphics[width=.495\textwidth,angle=0]{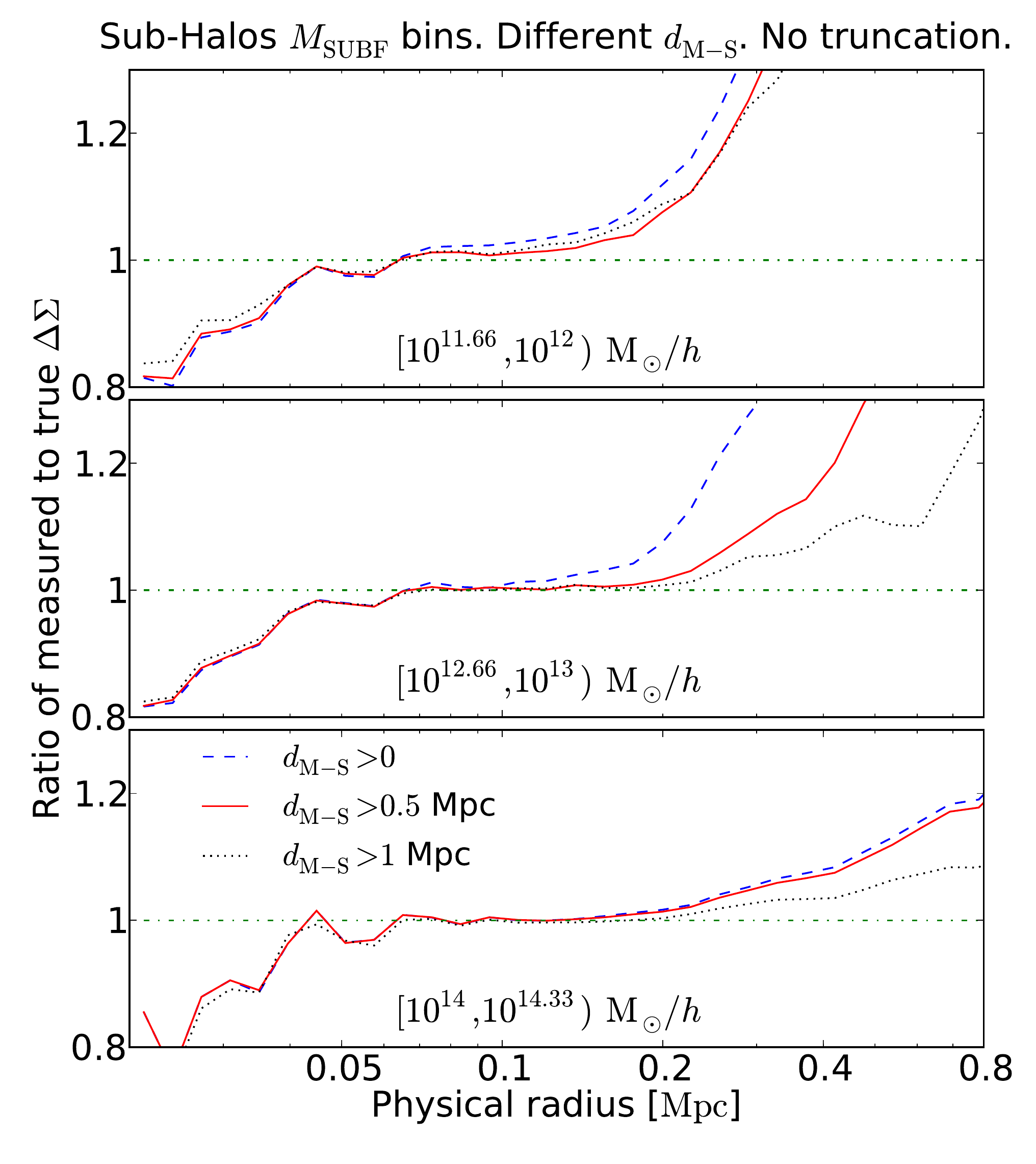}
  \vspace{-15pt}
  \caption{\footnotesize Results from mock clusters, ratio of the
    measured $\Delta\Sigma(\xi)$ to the one we input, versus the projected
    physical radius $\xi$. Only the results for a selection of mass bins, for
    the case with non-truncated NFWs are presented. The mass ranges are
    given. Blue dashed lines: whole subhalo sample. Solid red lines: subhalos
    with $d_{\rm M-S}>0.5$ Mpc. Black dotted lines: subhalos with $d_{\rm
      M-S}>1$ Mpc. The dotted black line, where the ratio is one, is a visual
    reference.}
  \label{fig:calsep} \vspace{-10pt}
\end{figure}

The results are shown in Fig.~\ref{fig:calsep}. Here we present again the
ratio of the measured $\Delta \Sigma(\xi)$ to the input value versus physical
radius $\xi$. This time we compare different selections of subhalos. The
signal improves if we discard subhalos with $d_{\rm
  M-S}<0.5\:\mathrm{Mpc}$. The most massive subhalos are more likely to be
recent mergers and lie far away from the cluster center, so the effect is
small in this case (lowest panel).

Increasing the minimal $d_{\rm M-S}$, depletes the sample and limits our
result to subhalos in the outskirts of the cluster. For $d_{\rm
  M-S}<1\:\mathrm{Mpc}$ the range where the bias is below 10\% increases, but
only at the expense of depleting our sample. For the less massive halos, the
depletion of the sample worsens the signal. For the most massive subhalos, a
larger minimal $d_{\rm M-S}$ improves the results, and seems to be a better
choice. Nevertheless, in this work we shall discard subhalos with $d_{\rm
  M-S}<0.5\:\mathrm{Mpc}$, regardless of subhalo type for simplicity. We
present the range in $\xi$ defined with our mock clusters for all $M_{\rm
  SUBF}$ mass bins in Tab.~\ref{tab:mass_range1}.

\subsection{Subhalo-galaxy lensing in the Millennium simulation}
\label{sec:charac}

\subsubsection{Subhalo mass profiles for different $M_{\rm SUBF}$}
\label{sec:massbin}

\begin{figure}
  \centering
  \includegraphics[width=0.5\textwidth]{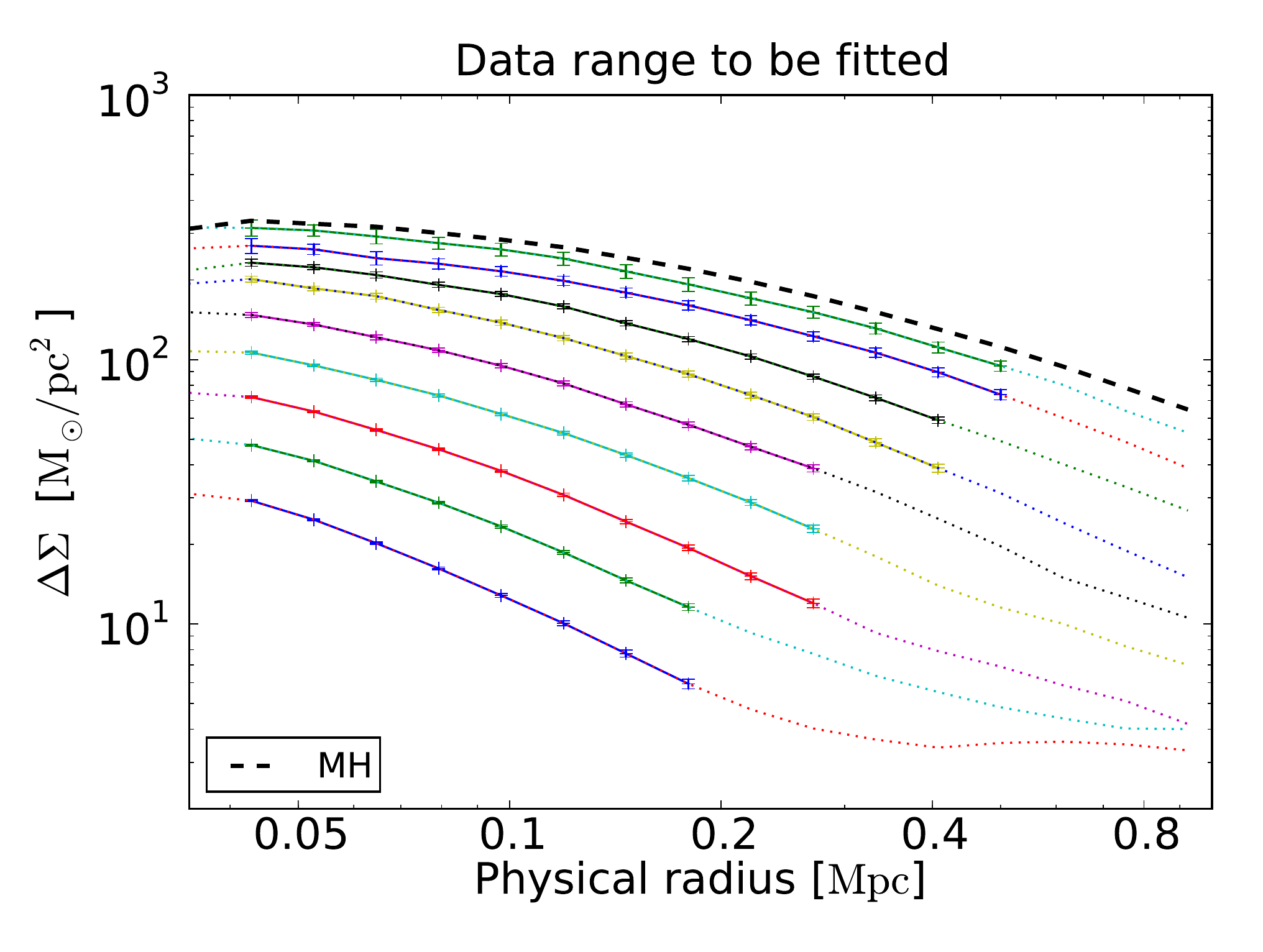}
  \vspace{-15pt}
  \caption{\footnotesize The $\Delta\Sigma(\xi)$ profile measured from the
    projected mass maps $\Sigma(\boldsymbol{\beta})$, for the different mass
    bins. Thick lines show the final range used for the analysis. The range in
    $\xi$ was defined through the test in Sect.~\ref{sec:restest}. In dashed
    black we show the average signal for the host cluster as a visual
    reference.}
  \label{fig:massb}
\end{figure}

We now analyze the mass profiles of subhalos in the MS as obtained by applying
the \GGLC{} to the projected mass maps from the simulation. The subhalo mass
ranges, numbers, and mean masses of the subhalo samples used are listed in
Table~\ref{tab:chara_massb_minsep}. For the analysis, we only consider
subhalos in clusters with redshift $z\leq 0.9$ and with $d_{\rm
  M-S}<0.5\:\mathrm{Mpc}$. The resulting average excess surface mass density
profiles are shown in Fig.~\ref{fig:massb}.

In order to characterize the subhalo mass profiles we try to find the best
fitting model. We consider an NFW profile, a truncated NFW profile and a TNSI
profile.  For the NFW model, we choose the characteristic surface mass density
$\Sigmas$ and the scale radius as free parameters. For the truncated NFW
model, we take the truncation radius $\rt$ as additional parameter. For the
TNSI model, we fix the core radius $r_{\rm core} = 10^{-4}\,\Mpc$ and only
fit the cut radius $r_{\rm cut}$ and the central density $\rho_{0}$. This
allows us to analyze the choices made by \citet{2007A&A...461..881L} and
\citet{2007MNRAS.376..180N}.

We conduct a Bayesian analysis to get the best fitting parameters for each
model in each mass bin. We assume a Gaussian likelihood in each case. The
corresponding covariance matrices are computed from the data using a bootstrap
algorithm. We construct each bootstrap realization by randomly drawing 
sub-halos with replacements. We do so for each mass bin, producing $10\,000$
realizations in each case.

\begin{table}[h]\vspace{-10pt}
  \caption{\footnotesize Binning of all subhalos in clusters for
    $z<0.9$ according to $M_{\rm SUBF}$. All subhalos with $d_{\rm M-S}>0.5$
    Mpc. Note the samples are slightly different to those in
    Tab.~\ref{tab:chara_massb}. The ranges are given in $\mathrm{M_\odot/h}$
    while the mean mass are in $\mathrm{M_\odot}$.}
  \label{tab:chara_massb_minsep}
  \centering
  \begin{tabular}{r r r c }
    \hline 
    \bf Bin&\bf Range [$\mathrm{M_\odot/h}$]\phantom{$\Big($} &\bf \# Subhalos &\bf ${M_{\rm SUBF}}$ [$\mathrm{M_\odot}$]\\
    \hline 
    1  & \phantom{$\hat{M^3}$}[$10^{11.33}:10^{11.66}$)             & 23229& $4.27\times 10^{11}$   \\
    2  & [$10^{11.66}:10^{12\phantom{.22}}$)    & 11821&  $9.14\times 10^{11}$  \\    
    3  & [$10^{12\phantom{.22}}   :10^{12.33}$) &  6151&  $1.98\times 10^{12}$  \\
    4  & [$10^{12.33}:10^{12.66}$)             &  2837&  $4.25\times 10^{12}$   \\
    5  & [$10^{12.66}:10^{13\phantom{.22}}$)    &  1476&  $9.22\times 10^{12}$  \\
    6  & [$10^{13\phantom{.22}}   :10^{13.33}$) &   728&  $1.97\times 10^{13}$   \\
    7  & [$10^{13.33}:10^{13.66}$)             &   363&  $4.30\times 10^{13}$   \\
    8  & [$10^{13.66}:10^{14\phantom{.22}}$)    &   155&  $8.79\times 10^{13}$  \\    
    9  & [$10^{14\phantom{.22}}   :10^{14.33}$) &    73&  $1.89\times 10^{14}$  \\
    \hline     \hline 
  \end{tabular}
\end{table}

We use the Bayesian evidence to rank the different models. The evidence was
computed with a nested sampling algorithm \citep{2004AIPC..735..395S}. As
shown by the results in Table~\ref{tab:eviden}, a TNSI is not a suitable model
for the subhalos in the MS. Our simulation does not contain baryons, which are
expected to change the mass profiles at the innermost regions. Nevertheless,
our results raises questions about the choice of the PIEMD model for lensing
observations of cluster galaxies. The best model (except for one mass bin,
which we consider to be due to statistical fluctuations) is a simple NFW
profile. The values for the truncation radii obtained by the fits were in
both cases (TNSI and truncated NFW) larger or comparable to the virial radius
of the average cluster, and therefore consistent with a lack of a detection.

The errors presented are the standard error from 10 realizations of the
measurement except for the leftmost model where we use a raw estimate coming
from the algorithm. The analysis in this case did not require a better
accuracy. We write in bold the maximum in each case, which belongs to the best
model.

From our simulations, it is not possible to detect a truncation radius. A
truncation, if present at all, happens at $\xi$ larger than
$0.2\,\Mpc$, where our measurements become unreliable due to bias. This is up
to four times larger than the truncation radii inferred by
\cite{2007A&A...461..881L}, \cite{2007MNRAS.376..180N} and
\cite{2007ApJ...656..739H} from lensing data.

\begin{table}\vspace{-10pt}
  \caption{\footnotesize Comparison of the logarithm of the evidences for the
    three models fitted to the $\Delta\Sigma$ signals. In bold we emphasize the
    best model in each mass bin.} \label{tab:eviden}
  \centering \small
\begin{tabular}{l c c c c}
  \hline
  \bf Bin &\bf NFW &\bf trunc. NFW& \bf TNSI \\
  \hline
  1  &    $-$89.71$\pm$0.05&\bf$-$89.57$\pm$0.08 & $-$157$\pm$1\\  
  2  &\bf $-$88.31$\pm$0.09&   $-$88.54$\pm$0.09 & $-$285$\pm$1\\  
  3  &\bf $-$88.38$\pm$0.09&   $-$89.38$\pm$0.12 & $-$472$\pm$1\\
  4  &\bf $-$85.90$\pm$0.07&   $-$87.72$\pm$0.13 & $-$710$\pm$3\\  
  5  &\bf $-$98.32$\pm$0.07&   $-$101.68$\pm$0.12\phantom{1}&$-$876$\pm$3\\  
  6  &\bf $-$93.38$\pm$0.05&   $-$97.04$\pm$0.17 &  $-$576$\pm$1\\  
  7  &\bf $-$99.20$\pm$0.06&   $-$102.63$\pm$0.21\phantom{1}& $-$809$\pm$2\\ 
  8  &\bf $-$88.09$\pm$0.06&   $-$92.80$\pm$0.18 & $-$528$\pm$1\\   
  9  &\bf $-$85.62$\pm$0.04&   $-$89.42$\pm$0.14 & $-$445$\pm$1\\
  \hline \hline
\end{tabular}
\end{table}                                   

\begin{figure}
  \centering
  \includegraphics[width=0.48\textwidth]{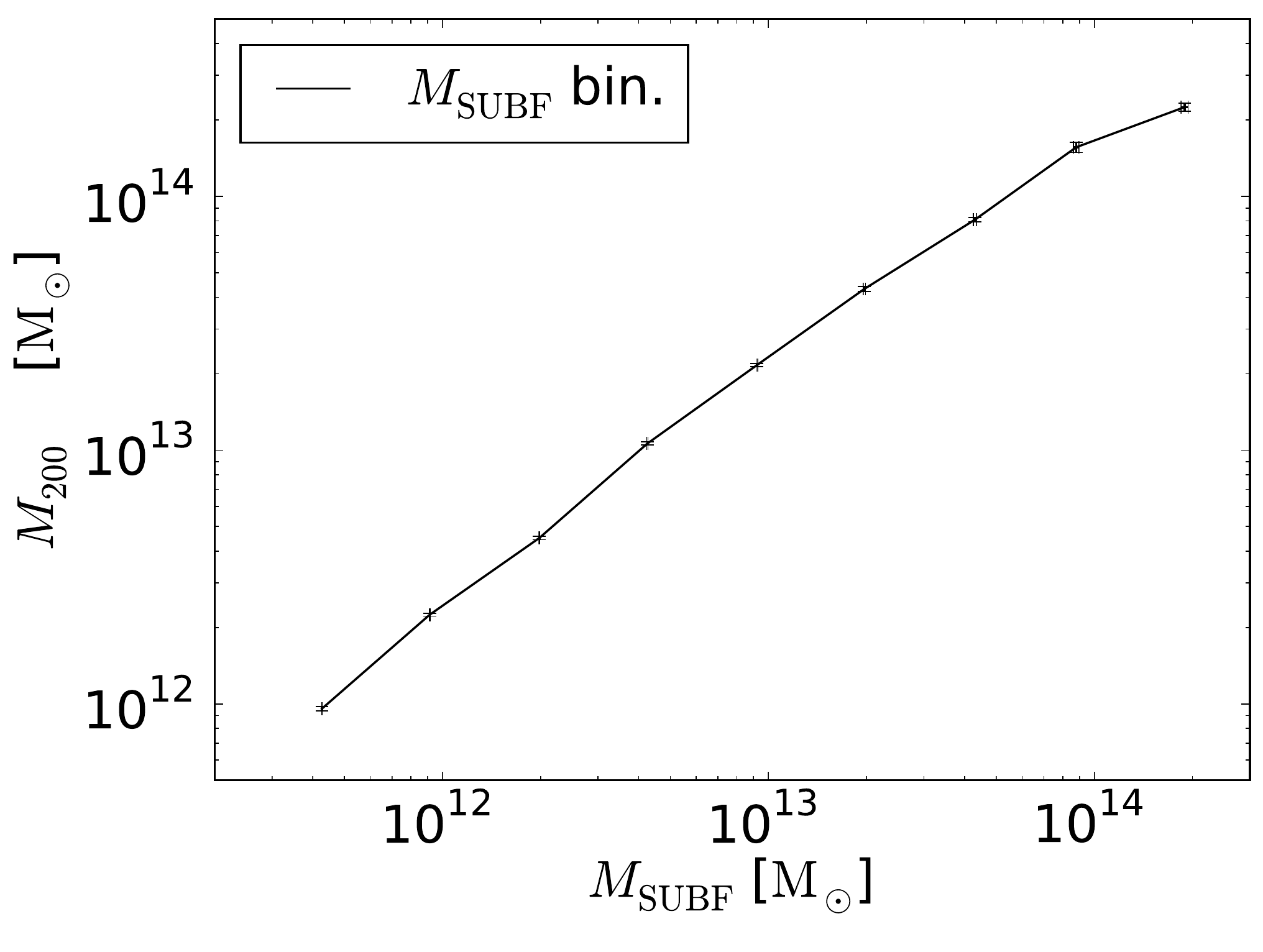}
  \includegraphics[width=0.48\textwidth]{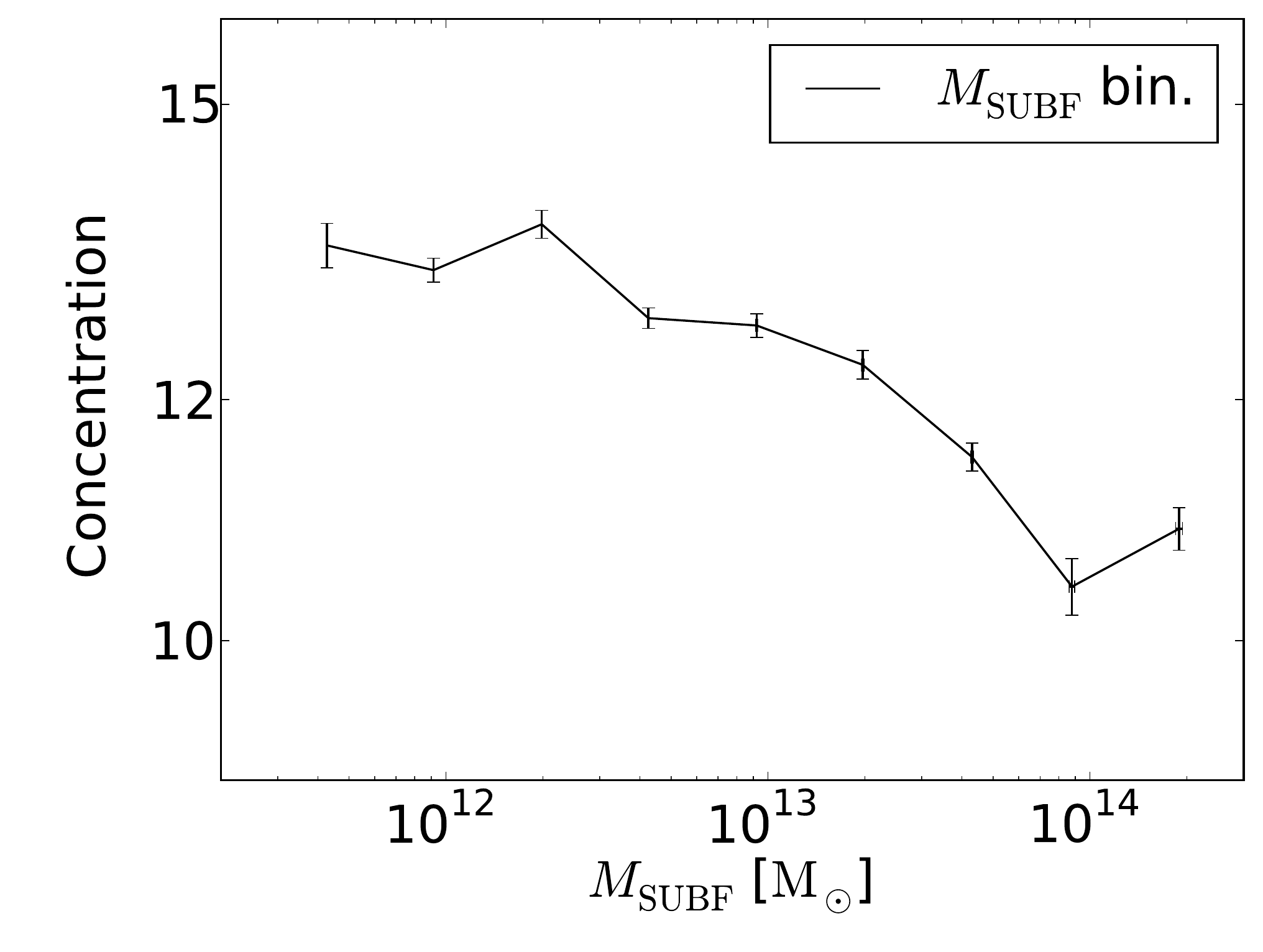}
  \caption{\footnotesize Parameterization of the subhalo profiles using NFW
    profiles. Bottom panel: concentration concentration as a function of
    $M_{\rm SUBF}$. Top panel $M_{200}$ as a function of $M_{\rm SUBF}$.}
   \vspace{-10pt}
  \label{fig:best_NFW}
\end{figure}

We take the mean of the posterior from the nested sampling algorithm as the
best-fitting values for the model parameters, and the posterior standard
deviation as parameter errors. The best-fitting values for the concentration
$c$ and  $M_{200}$ assuming an NFW profile are presented in
Fig.~\ref{fig:best_NFW}. The concentration $c$ slightly decreases with
increasing subhalo mass $\Msubf$. The mass $M_{200}$ shows a tight and
essentially linear relation with the subhalo mass $\Msubf$. Thus, $M_{200}$ is
a good {\it observable} to infer the subhalo mass $M_{\rm SUBF}$. Note, however,
that the mass $M_{200}$ is always a few times larger than $M_{\rm SUBF}$. This
indicates that the subhalos as defined by SUBFIND do not fully extend to
the radius $r_{200}$.
        
\begin{table}
  \caption{\footnotesize Binning of all subhalos according to $M_{\rm SUBF}$ in clusters for
    $z<0.9$. All subhalos with $d_{\rm M-S}>0.5$ Mpc. 
    The mass binning can be found in Tab.~\ref{tab:chara_massb_minsep}.}
  \label{tab:mass_range}
  \centering
  \begin{tabular}{c c }
    \hline 
    \bf $M_{\rm SUBF}$ bin& \bf Fitted $M_{200}$ [$\mathrm{M_\odot}$] \\
    \hline 
    1  & $(9.56\pm 0.16 )\times 10^{11}$\\
    2  & $(2.25\pm 0.03 )\times 10^{12}$\\    
    3  & $(4.51\pm 0.09 )\times 10^{12}$\\
    4  & $(1.058\pm 0.016)\times 10^{13}$\\
    5  & $(2.16\pm 0.04 )\times 10^{13}$\\
    6  & $(4.31\pm 0.11 )\times 10^{13}$\\
    7  & $(8.10\pm 0.16 )\times 10^{13}$\\
    8  & $(1.56\pm 0.07 )\times 10^{14}$\\    
    9  & $(2.24\pm 0.09 )\times 10^{14}$\\
    \hline     \hline 
  \end{tabular}\vspace{-10pt}
\end{table}

The use of $M_{200}$ as a direct mass estimator on subhalos should be
avoided. Otherwise a large fraction of the mass not bound to the subhalo
(e.g.~nearby subhalos) is attribute to it. We use $M_{200}$ only as an
\textit{observable} from which we can derive the gravitationally bound mass
$\Msubf$.

In observations, one does not know the radial range at which the \GGLC{}
signal can be reliably measured beforehand, since the range depends on the
subhalo masses to be inferred from the lensing signal. Here we use the
following method to resolve this problem. We infer $M_{200}$ from the measured
\GGLC{} signal in a preliminary radial range, i.e.~up to $1\,\Mpc$. Based on
the results in Tab.~\ref{tab:mass_range}, we use a smaller more adequate range
in $\xi$ to measure $M_{200}$ again and iterate until obtaining consistent
results.

\subsubsection{Subhalo mass estimation for magnitude-selected subhalo samples}
\label{sec:opt_chara}

So far, we have only studied mass-selected subhalo samples. We now explore
how well one can estimate subhalo masses $M_{\rm SUBF}$ from stacked subhalo
samples that are selected according to the magnitudes of the hosted galaxies,
which are more amenable to observation.

\begin{figure}
  \centering
  \includegraphics[width=0.495\textwidth]{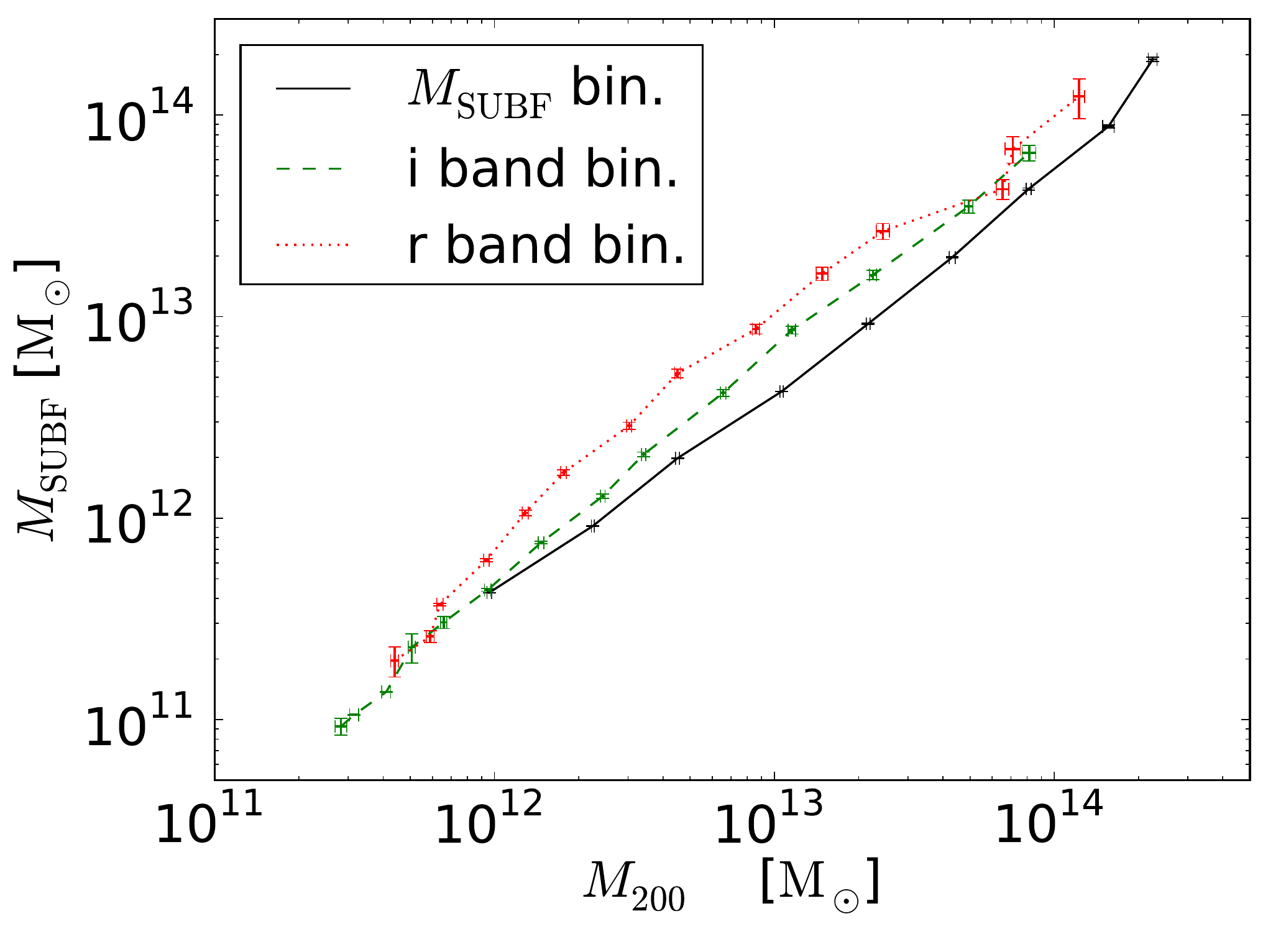}
   \vspace{-10pt}
   \caption{\footnotesize Fitted $M_{200}$ as a function of average $M_{\rm
       SUBF}$ for different binnings using the projected mass maps
     $\Sigma(\boldsymbol{\beta})$. Black solid line: binning according to
     $M_{\rm SUBF}$. Dotted red line: binning according to $r$ band absolute
     observer frame luminosity. Dashed green line: binning according to $i$
     band absolute observer frame luminosity.}
   \vspace{-5pt}
   \label{fig:bestpar}
\end{figure}

Figure.~\ref{fig:bestpar} shows the mass $M_{200}$ obtained from the \GGLC{}
versus the true mean subhalo mass $M_{\rm SUBF}$ for sub-halos binned
according to $M_{\rm SUBF}$, observer-frame absolute $r$-band magnitude, and
observer-frame absolute $i$-band magnitude. We define 14 bins in luminosity,
with a bin width of 0.33 magnitudes, between the values $-$19 and $-$23.66 for
the luminosity or both bands. The relation between the mean inferred mass
$M_{200}$ and mean subhalo mass $M_{\rm SUBF}$ is roughly linear, but
noticeably depends on the quantity used for binning.  We characterize the
relation between $M_{200}$ and $M_{\rm SUBF}$ by a power law,

\begin{equation}
\label{eq:power_law_M_SUBF_of_M_200}
  \frac{M_{\rm SUBF}}{\Msolar} = A \left(\frac{M_{200}}{\Msolar}\right)^n,
\end{equation}
with best-fitting amplitudes $A$ and exponents $n$ listed in Table~\ref{tab:bestparmod}.

The mass-binned and magnitude-binned subhalo samples differ, in particular in
the distribution of subhalo masses and mass profiles. As a consequence, even
for the same mean subhalo mass $M_{\rm SUBF}$, the mean profiles may differ
substantially. While the averaging to obtain the mean $M_{\rm SUBF}$ and the
mean \GGLC{} for each bin is a linear process, the computation of the best
fitting parameters and $M_{200}$ is not. Thus, we may not be able to estimate
$M_{\rm SUBF}$ with high precision through $M_{200}$, but we can derive useful
constraints on it.

\begin{table}
  \centering
  \caption{\footnotesize Best fitting parameters for the relation $M_{\rm
      SUBF}(M_{200})/\mathrm{M}_\odot=A\,(M_{200}/\mathrm{Mpc})^n$ coming from the data displayed in
    Fig.~\ref{fig:bestpar} in the bottom right panel. We present the mean
    and the standard error of the posterior of the parameters.}\label{tab:bestparmod}
  \begin{tabular}{c | c c c c}
    \hline
    \bf Data           & \bf n &  $\rm \log_{10}(\mathbf{ A})$ \\\hline
    Mass bin.      & 1.059$\pm$0.013   &   $-1.13\pm0.17$     \\
    $r$ band bin.  & 1.20$\pm$0.04     &   $-2.5\pm0.4$       \\
    $i$ band bin.  & 1.18$\pm$0.03     &   $-2.5\pm0.4$       \\ 
    \hline\hline
  \end{tabular}
\end{table}

\subsubsection{Subhalo evolution}\label{sec:evsubh}
\begin{table}
  \caption{\footnotesize Properties of subhalo samples used to study the time
    evolution of subhalos: symbol and line type used in plots, subhalo numbers, average $d_{\rm M-S}$ in Mpc, 
    mean subhalo mass $M_{\rm SUBF}$, and mean inferred $M_{200}$. Only subhalos in clusters with
    redshift $z\leq 0.9$ and with distance to the main halo center $d<
    0.5\,\Mpc$.}
 \label{tab:ev_Sub_mass_range}
 \centering
  \begin{tabular}{l r c c c}
    \hline 
    \bf Bin\phantom{\Big(}& \bf \# Subh.&  \bf ${d_{\rm M-S}}$ 
    &\bf ${M_{\rm SUBF}}$ [$\mathrm{M_\odot}$]& \bf ${M_{200}}$ [$\mathrm{M_\odot}$]  \\
    \hline 
    {\color{red}$\times$---}      &31408& 1.149 &$2.77\times 10^{11}$ & $(6.41\pm0.13)\times 10^{11}$\\
    {\color{red}$+$---}           &14927& 1.132 &$1.68\times 10^{11}$ &  $(4.95\pm0.17)\times 10^{11}$\\
    {\color{red}$\bullet$---}     &10482& 1.175 &$(1.236\times 10^{11})^*$ & $(3.27\pm0.17)\times 10^{11}$ \\\hline
    {\color{blue}$\times$- -}     &11487& 1.169 &$7.38 \times 10^{11}$ &  $(1.85\pm0.03)\times 10^{12}$\\
    {\color{blue}$+$- -}          & 6391& 1.136 &$4.29 \times 10^{11}$ & $(1.23\pm0.04)\times 10^{12}$  \\
    {\color{blue}$\bullet$- -}    & 4410& 1.143 &$2.9\times 10^{11}$ & $(7.5\pm0.3)\times 10^{11}$ \\\hline
    {\color{black}$\times\cdots$} &10792& 1.260 &$8.1\times 10^{12}$ & $(1.28\pm0.03)\times 10^{13}$  \\
    {\color{black}$+ \cdots$}     & 4782& 1.151 &$2.1\times 10^{12}$ & $(6.09\pm0.17)\times 10^{12}$  \\
    {\color{black}$\bullet\cdots$}& 2331& 1.150 &$1.04\times 10^{12}$ & $(2.98\pm0.11)\times 10^{12}$   \\
    \hline     \hline 
    \multicolumn{5}{p{8cm}}{\footnotesize $^{*}$The value after eliminating two misidentified subhalos.}\\
  \end{tabular}\vspace{-5pt}
\end{table}
\begin{figure}
   \includegraphics[width=0.5\textwidth]{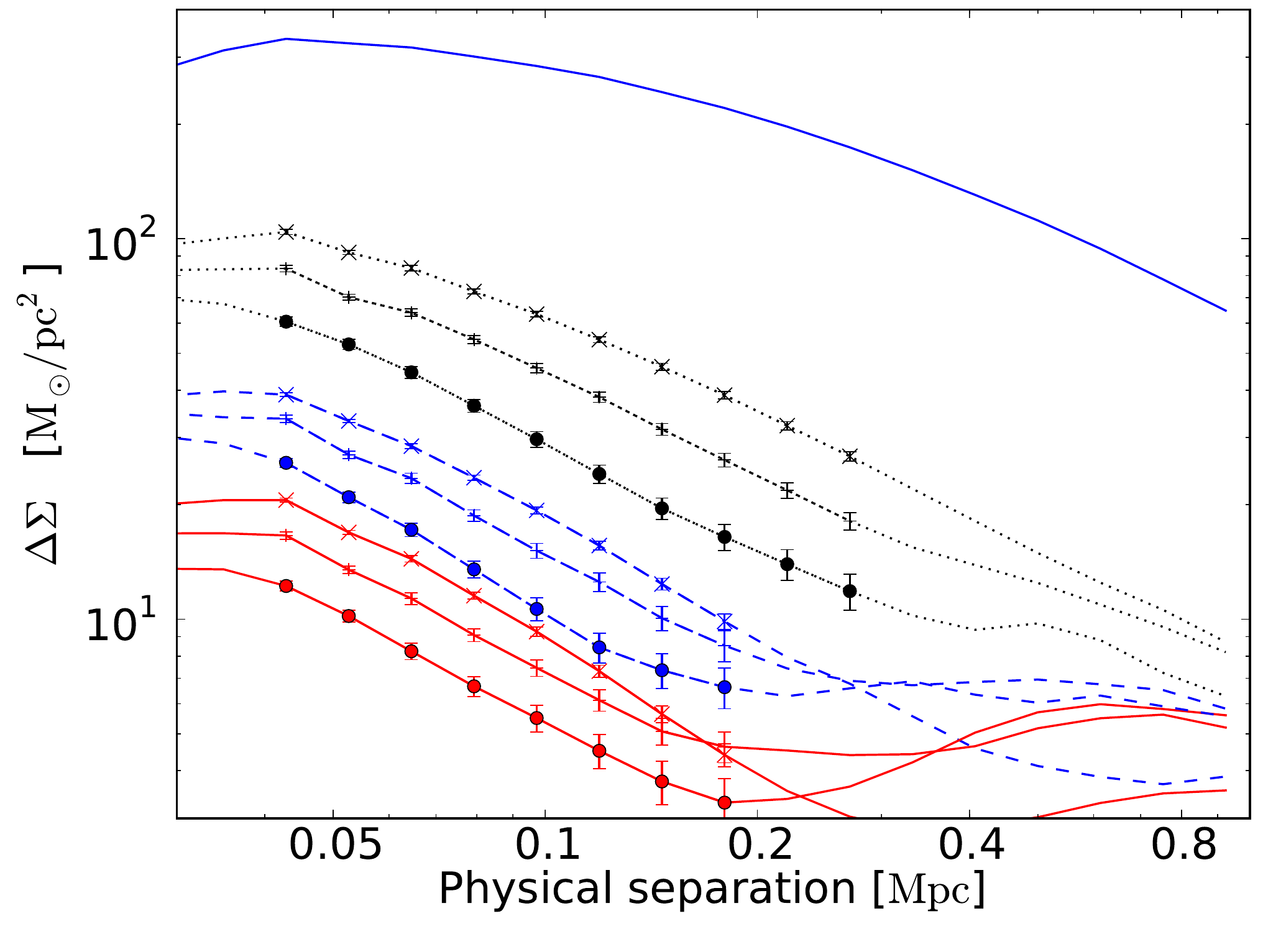}\vspace{-5pt}
  \begin{tabular}{l c c c c}
    \hline 
    \bf Bin\phantom{\Big(}& \bf Age [$\mathrm{Gys}$]&\bf  $M_{\rm inf}$    [$10^{10}\,\mathrm{M_\odot}/h$] 
    & \bf ${M_{\rm inf}}$  [$\mathrm{M_\odot}$] &  \bf  ${\rm age}$   \\
    \hline 
    {\color{red}$\times$---}      &$t\le 3$     & \multirow{3}{*}{$20\le M_{\rm inf}\le 60$}   & $4.621\times 10^{11}$ &  1.797\\ 
    {\color{red}$+$---}           &$3<t\le 4.5$ &                                             & $4.711\times 10^{11}$ & 3.723  \\
    {\color{red}$\bullet$---}     &$t>4.5$      &                                             & $4.741\times 10^{11}$ &  5.608 \\
    {\color{blue}$\times$- -}     &$t\le 3$     & \multirow{3}{*}{$60 < M_{\rm inf} \le 150$}  & $1.278\times 10^{12}$ &1.852  \\
    {\color{blue}$+$- -}          &$3<t\le 4.5$ &                                             & $1.279\times 10^{12}$ &  3.720 \\ 
    {\color{blue}$\bullet$- -}    &$t>4.5$      &                                             & $1.269\times 10^{12}$ &  5.629 \\
    {\color{black}$\times\cdots$} &$t\le 3$     &  \multirow{3}{*}{$ M_{\rm inf} > 150$}       & $1.28 \times 10^{13}$ & 1.798 \\
    {\color{black}$+ \cdots$}     &$3<t\le 4.5$ &                                             & $6.82 \times 10^{12}$ &  3.688 \\
    {\color{black}$\bullet\cdots$}&$t>4.5$      &                                             & $4.88 \times 10^{12}$ &  5.534 \\ 
    \hline     \hline 
  \end{tabular}
 \centering
 \caption{\footnotesize Profiles for different infall mass and time spend
   inside the cluster: with error bars the regions where the bias is below
   10\%. The color and the line type are the same for samples selected from
   the same infall mass range. The symbols are the same for samples defined
   from the same age bin. In solid blue we plot the average signal for the
   host cluster as a visual reference. In the table the ranges used for the
   binning and the average value for each sub-sample.}
  \label{fig:ev_SubH}\vspace{-5pt}
\end{figure}

In order to observe how subhalos evolve inside the clusters we considered two
new parameters, the infall mass $M_{\rm inf}$ and infall time (both computed
by \citealt{2006MNRAS.371..537W}). The first is the mass of the main halo that
evolved into the current subhalo at the last epoch when it was a central
dominant object, i.e.~the mass just before it was accreted by a larger
structure. The infall time is the moment at which this accretion was detected
and allows us to compute the subhalo age, i.e.~the time spent inside the
cluster at the time of its observation.

We divide our sample into three classes according to $M_{\rm inf}$, and each
of these subhalo classes into three sub-classes according to their age. In
order to define the classification, we maximized the signal-to-noise defined
as the average ratio (over $\xi$), between the measured $\Delta\Sigma(\xi)$
and its sample variance (we also avoided too small subhalo samples). The
resulting ranges for classes and sub-classes are listed in
Tab.~\ref{tab:ev_Sub_mass_range}. The resulting \GGLC{} signals are shown in
Fig.~\ref{fig:ev_SubH}.

As shown in Table~\ref{tab:ev_Sub_mass_range}, the mean subhalo masses
decrease with increasing subhalo age independent of the considered infall mass
range. This is clear evidence for subhalos losing mass inside
clusters.\footnote{Only for the subhalo samples with the largest infall mass
  range, the decrease in subhalo mass can be partially attributed to
  differences in the mean infall mass with time of the samples.} The mass loss
is also seen in the subhalo \GGLC{} signals shown in Fig.~\ref{fig:ev_SubH},
which also decrease in amplitude at smaller radii with increasing subhalo
age. The mean masses $M_{200}$ inferred from the \GGLC{} signals also reflect
the mass loss.

The \GGLC{} profiles also indicate that the mass loss is not due to a simple
truncation but rather due to a mass loss at all radii (in the range where the
subhalo profiles can be measured reliably), which corroborates the findings of
\citet{2003ApJ...584..541H}. At very large radii, the subhalo profiles tend to
increase with increasing subhalo age. This result indicates that the
surrounding subhalo's contribution and main halo asymmetries become more
important with increasing subhalo age.

\subsection{Forecasts for future surveys} \label{sec:forec}

\subsubsection{Galaxy observables} \label{sec:galobs}

\begin{figure*}
  \begin{center}
    \includegraphics[width=.49\textwidth,angle=0]{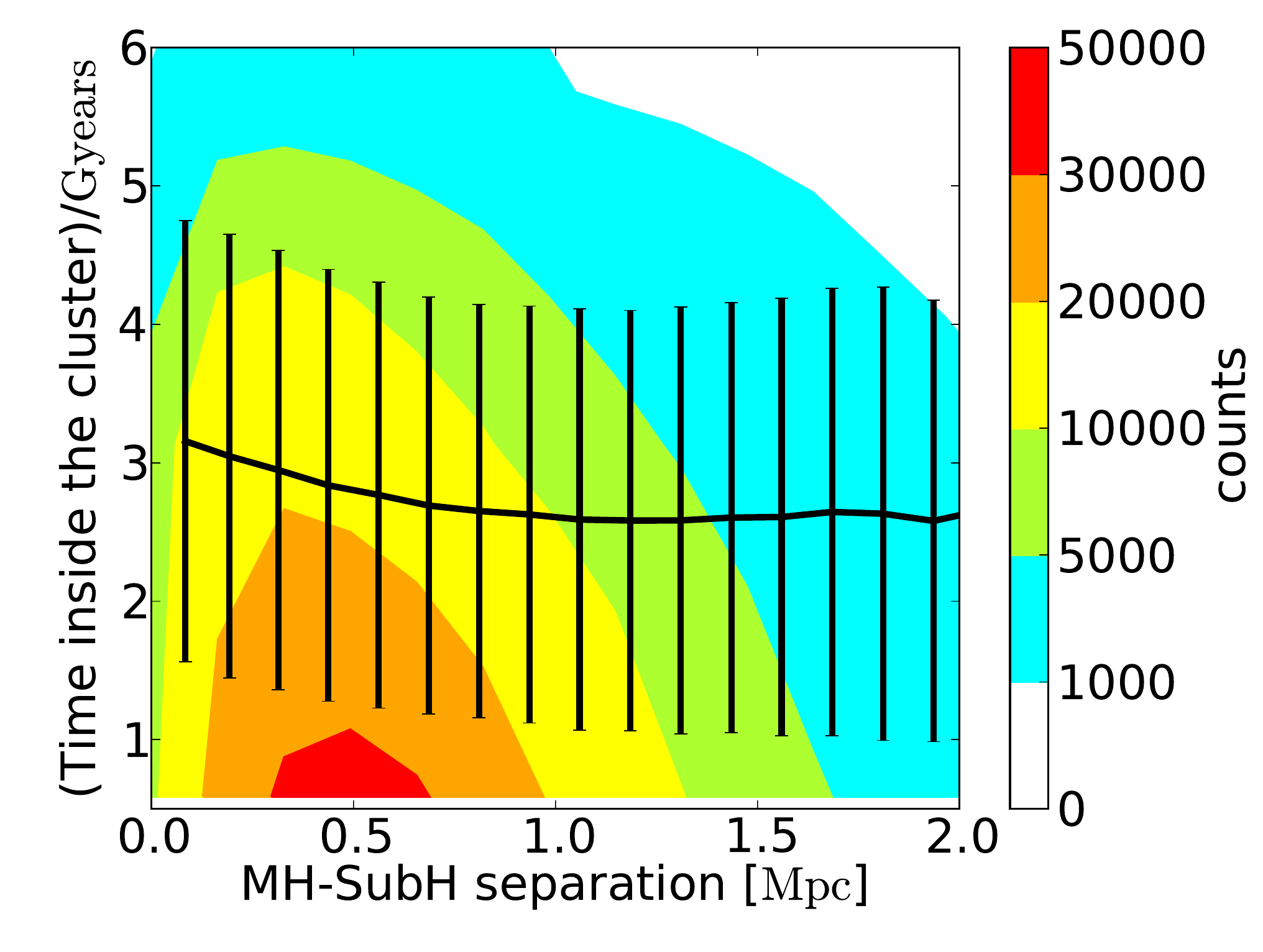}
    \includegraphics[width=.49\textwidth,angle=0]{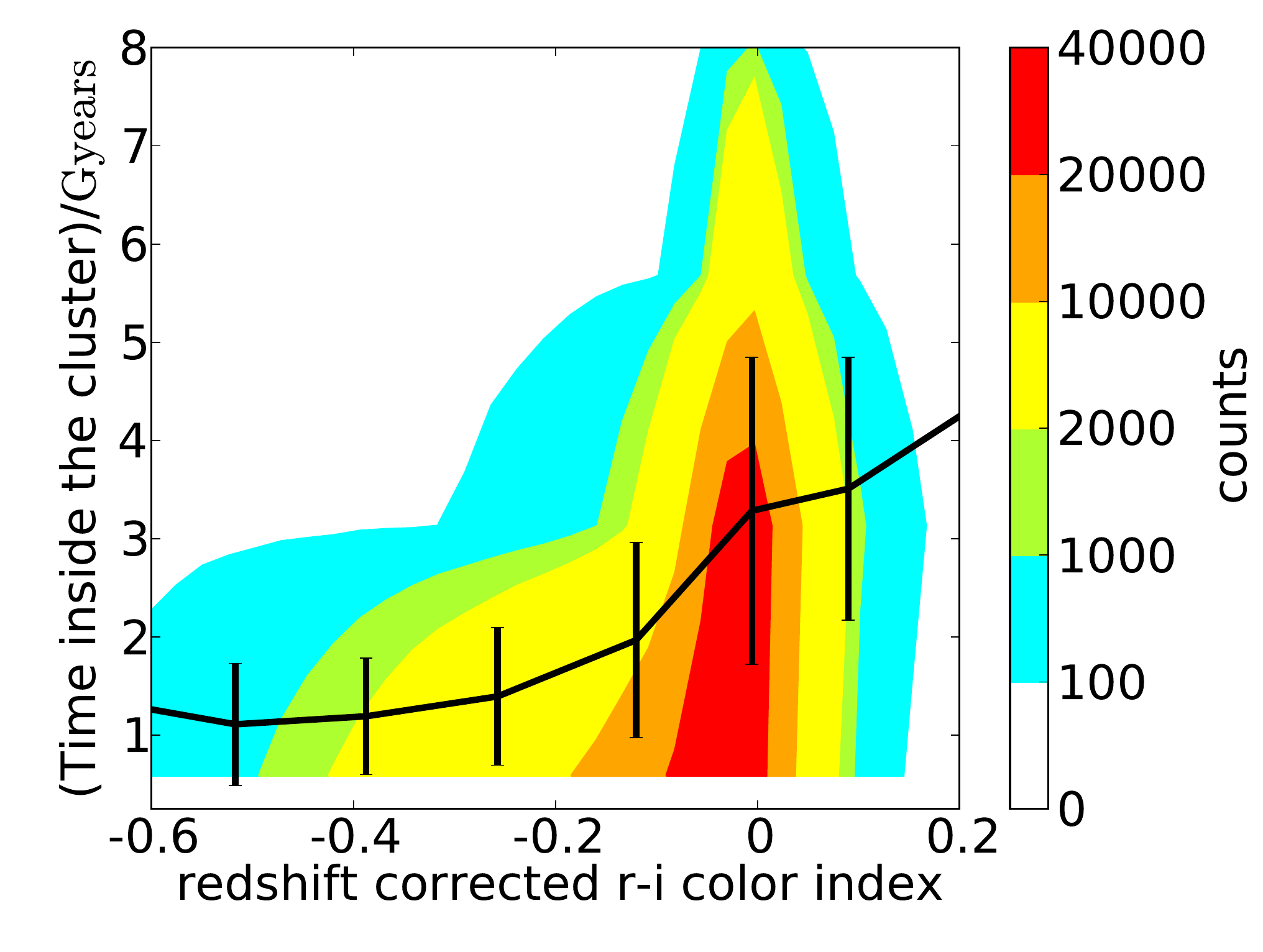}
    \includegraphics[width=.49\textwidth,angle=0]{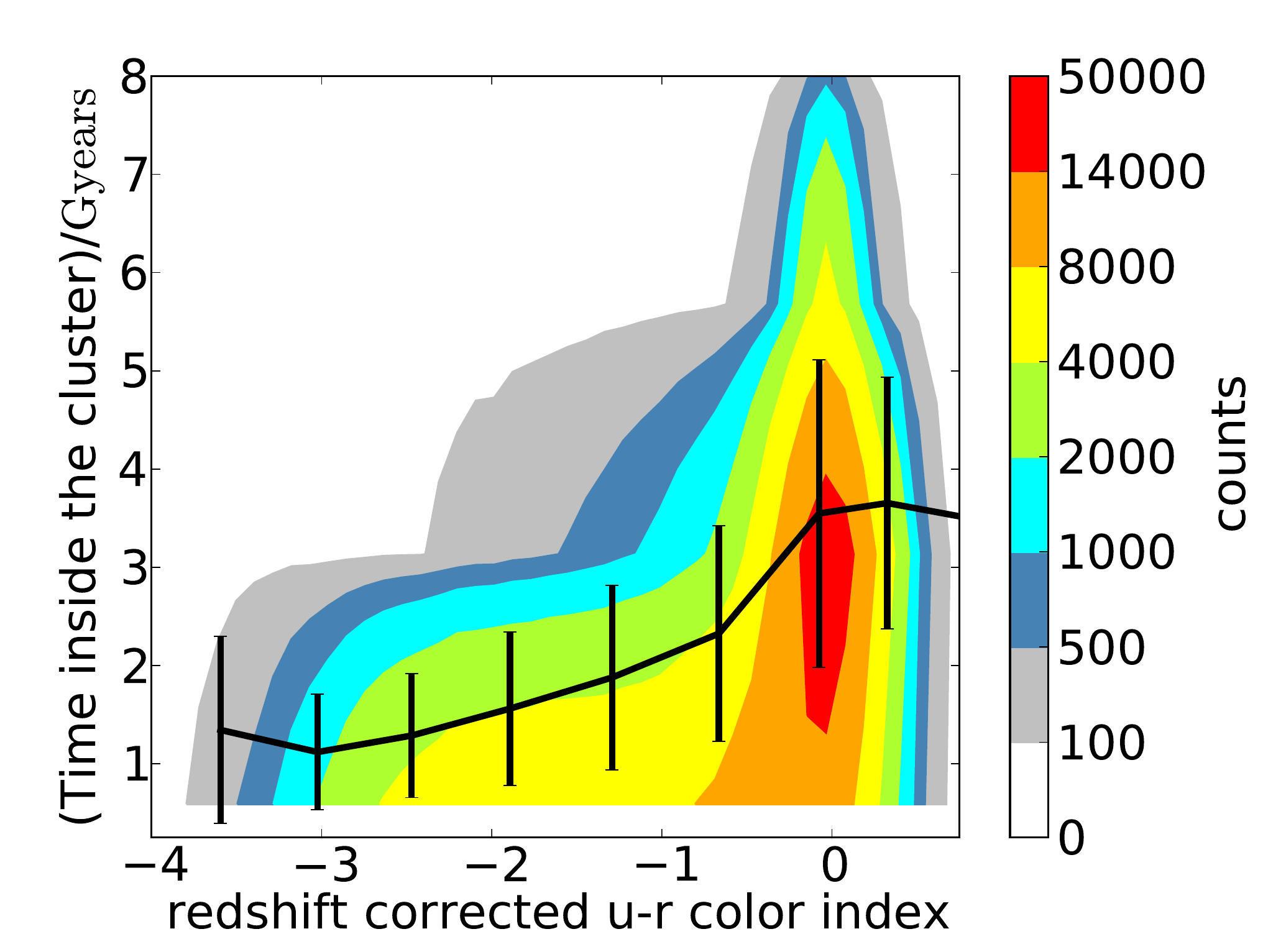}
    \fcolorbox{black}{black}{\includegraphics[width=.48\textwidth,angle=0]{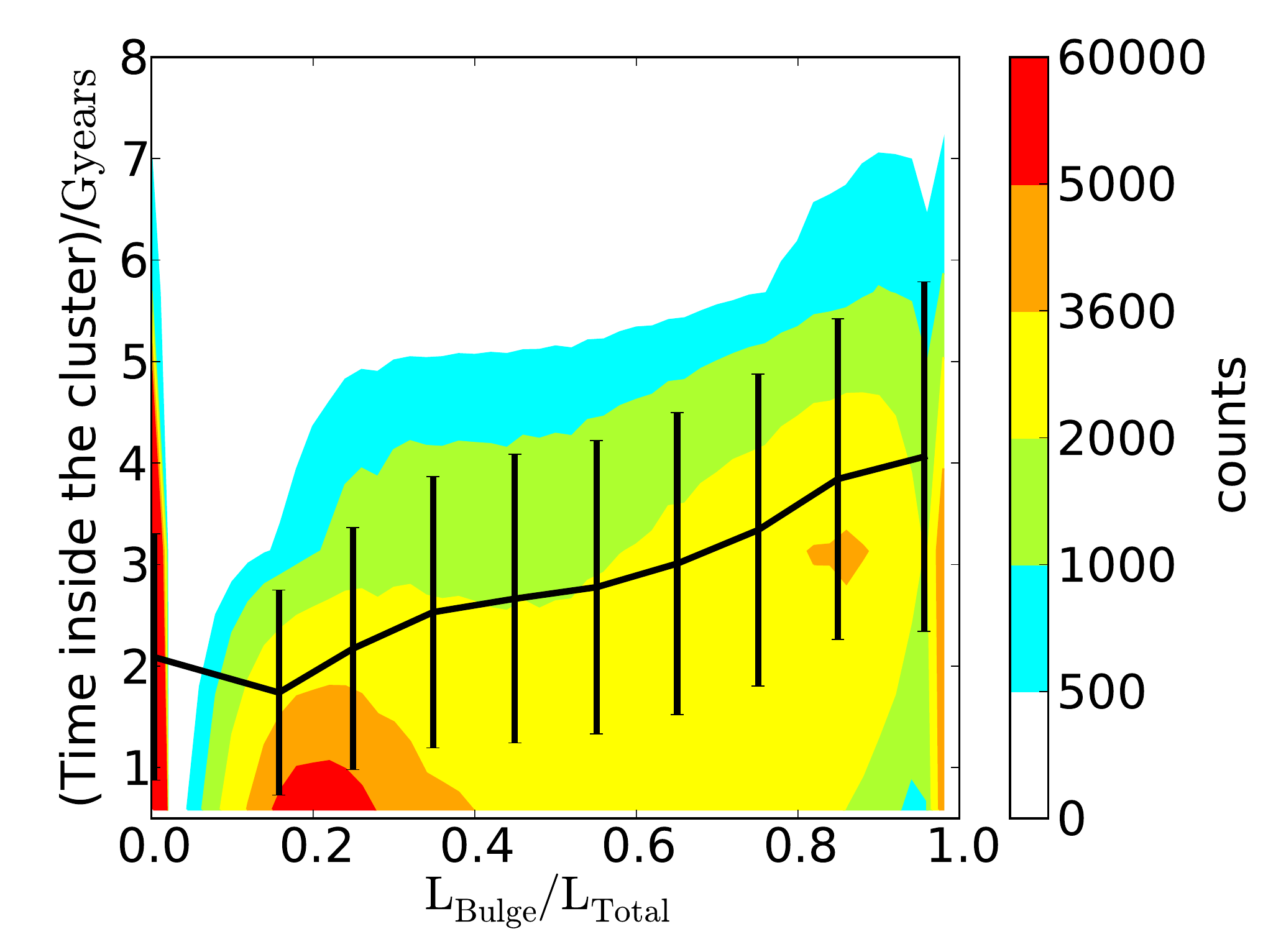}}
    \fcolorbox{black}{black}{\includegraphics[width=.48\textwidth,angle=0]{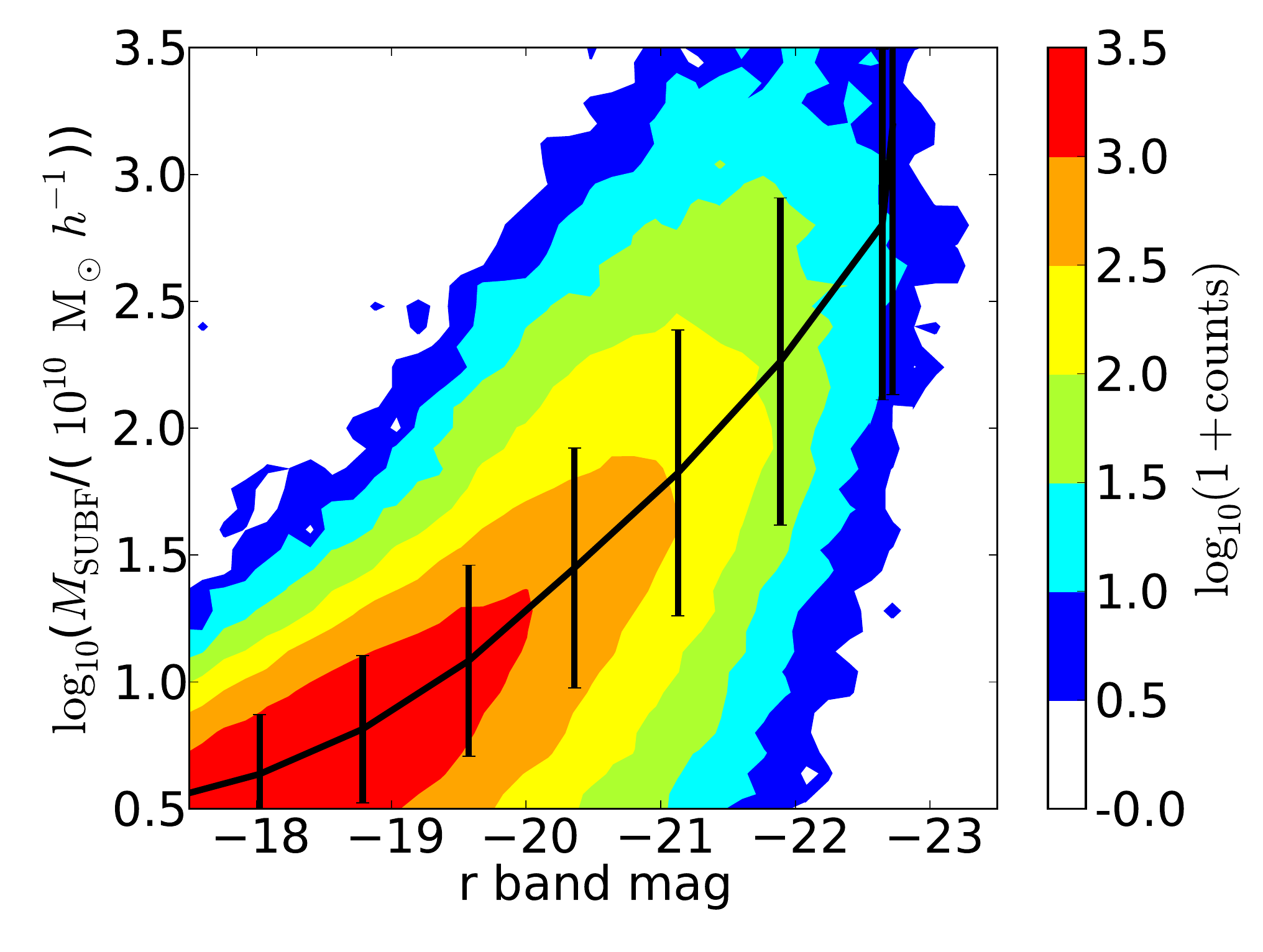}} 
    \fcolorbox{black}{black}{\includegraphics[width=.48\textwidth,angle=0]{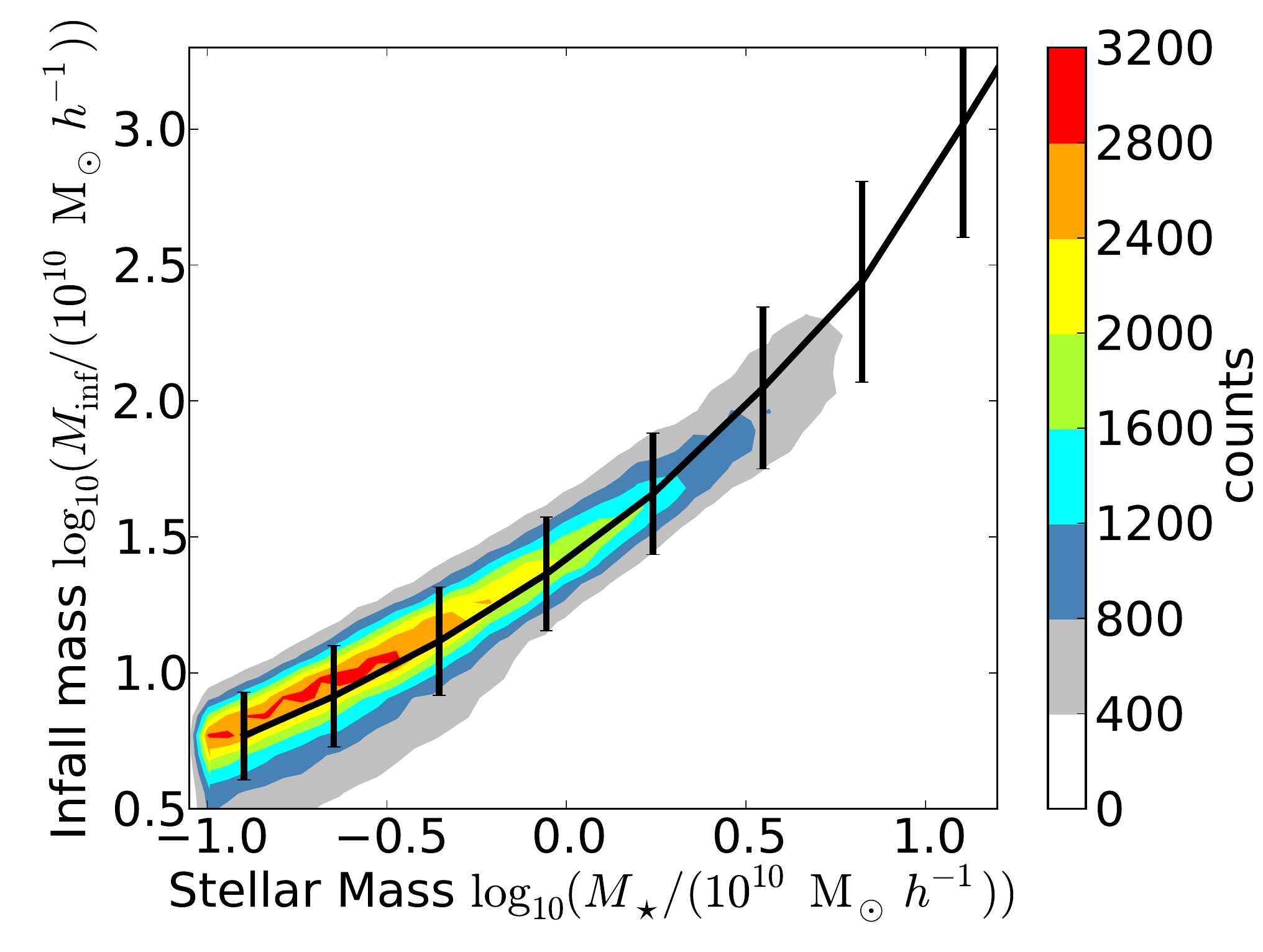}}
    \caption{\footnotesize Relations between physical subhalo properties
      (vertical axis) and observable proxies (horizontal axis). With black
      frames, we outline those finally used. Contours give the full
      distribution; we overplot the mean value of each subhalo property and
      its standard deviation for each bin in the observable (curves with error
      bars). Top left: age of the subhalo versus the projected radial
      separation of the subhalo from the main halo center. Top right: age of
      the subhalo versus redshift-corrected $r-i$. Middle left: age of the
      subhalo versus redshift-corrected $u-r$. See text for a description of
      the redshift correction. Middle right: age of the subhalo versus our
      morphology estimator. Bottom left: $M_{\rm SUBF}$ versus absolute $r$
      band magnitude. Bottom right: $M_{\rm inf}$ versus stellar mass (using
      results from \citealt{2006MNRAS.371..537W}). Note that the contour
      levels differ between the panels.}\label{fig:proxies}
  \end{center}
\end{figure*}

Understanding how different observables depend on the underlying subhalo
characteristics is essential for obtaining optimal results with our method. We
study these relations using the semi-analytic models of galaxy formation in
the MS. A discussion of how well the model predicts such relations is beyond
the scope of this work. Our aim is to show the feasibility of extracting
information with the method we propose. A good review about semi-analytic
models is discussed in \cite{2006RPPh...69.3101B} and
\cite{2010PhR...495...33B}. Galaxy clustering \citep{2006MNRAS.369.1009B},
galaxy colors and metallicity \citep{2001MNRAS.328..726S} and morphology
evolution \citep{2006MNRAS.366..499D}, just to outline a few examples, have
been successfully studied with semi-analytic models.

We only include type-1 galaxies (galaxies with a host subhalo) in our
study. We justify our choice in Sect.~\ref{sec:type2}.

For our investigation we have studied proxies for the mass as determined by
SUBFIND, the infall mass and the age of the subhalos (i.e.~the time spent
inside the cluster). The results are shown in Fig.~\ref{fig:proxies}, where we
have marked with a black frame the proxies that show the strongest
correlations and are therefore used for our final analysis, for which we only
used galaxies hosted by subhalos and imposed an apparent magnitude limit of
$r<22$.

The age of the subhalos is the most difficult to obtain; none of the
observable that we considered displays a strong correlation with this
quantity. In top left panel of Fig.~\ref{fig:proxies}, we plot the projected
radial separation of the subhalo versus its age. Subhalos fall gradually
into the main halo through dynamical friction, and so $d_{\rm M-S}$ should be
an indicator of the age. However, since the projected separation corresponds
to a wide range of three-dimensional separations, we only find a very weak
correlation.

In the top right and the middle left panels, we plot the $r-i$ and $u-r$ color
indices as a function of subhalo age. The evolution of the stellar population
and the depletion of gas to form new stars, make the color of galaxies evolve
with time. We account for the redshift evolution of the galaxy colors as
follows: we identify for each redshift the red cluster sequence
\textit{averaged over all clusters}, and subtract its mean color from all
galaxy colors at the same redshift. By doing this, we can compare galaxies at
different redshifts. Both observables are only weakly correlated with the
subhalo age. Note that the contour levels are not equally spaced.

The most sensitive proxy for the age of the subhalos is their morphology,
which we quantify using the ratio between the luminosities of the bulge and
the disk (middle right panel). There are many galaxies which only have a disk
component, or only have a bulge (25\% of the subhalos in total). In both cases
the age of these subhalos cannot be determined. We exclude these galaxies from
our analysis, as they should be easy to avoid in real surveys.

The commonly used proxy for mass is luminosity. In the bottom left panel, we
plot the $r$ band absolute magnitude as a function of $M_{\rm
  SUBF}$. Luminosity in the $i$ bands give qualitatively similar results.

Finally, using results from \cite{2006MNRAS.371..537W}, we study the stellar
mass as a proxy for the infall mass $M_{\rm inf}$. The result is displayed in
the bottom right panel, showing a strong correlation. In the semi-analytic
models, galaxies loose their gas rapidly once they become part of a cluster
and the star formation effectively stops. Therefore galaxies in clusters
maintain on average the stellar mass which they had prior to fall into the
cluster, which is then strongly correlated with the halo mass at that time.

\subsubsection{Simulated Surveys}\label{sec:simsur}

We now investigate the prospects of measuring the weak lensing signal of
subhalos in upcoming surveys. We focus in particular on DES and LSST, but we
explore also variations of the main survey characteristics. The area simulated
in our ray-tracing is not large enough to represent the total size of the
surveys. We are not able, therefore, to reproduce the expected cosmic
variance. However, our major concern is the detectability of the signal due to
the statistical noise, since we expect that our subhalo samples to be
representative. Therefore the influence in the measurements stemming from
cosmic variance is assumed to be subdominant.

We create mock galaxy catalogues similar to those expected from these
surveys. As described in Sect.~\ref{sec:rt}, the redshift distribution of our
sources follows that of \citealt{1996ApJ...466..623B}. The modulus of the
intrinsic ellipticity is drawn from a Gaussian distribution with zero mean and
standard deviation $\sigma_{\vert \epsilon_{\rm i} \vert}$. This intrinsic
ellipticity is the dominant source of noise in our measurement, resulting in
an error in the measurement $\propto\sigma_{\vert \epsilon_{\rm
    i}\vert}/\sqrt{n}$, where $n$ is the total number of foreground-background
galaxy pairs used to measure the final $\Delta\Sigma(\xi)$. For a single
galaxy, we assume $\sigma_{\vert \epsilon_{\rm i} \vert}=0.3$.  However, the
survey areas of DES and LSST are considerably larger than the area covered by
the ray-tracing. In order to simulate these surveys with their larger number
of background galaxies, we therefore use a effective intrinsic ellipticity
variance, which is reduced by the ratio of the survey area to the area of the
simulations $\left(\sigma_{\vert \epsilon_{\rm i}\vert}^{(\rm eff.)}\right)^2$
$=\sigma^2_{\vert \epsilon_{\rm i} \vert}\times R$.

For the lens sample, we impose a maximum redshift of $z=0.9$. Subhalos at
higher redshift are difficult to probe with lensing since there are fewer
background sources available. Henceforth, their inclusion offers no
improvement in the detectability of the signal. For the LSST-like survey, we
only considered lens galaxies with $r<26$ in apparent magnitude, for the other
three surveys is $r<22$. We divide these samples further into $14$ bins in
absolute $r$-band magnitude ranging from $r=-23.66$ to $r=-19$ with a bin
width of $0.33$ magnitudes.

The details of these surveys are summarized in Tab.~\ref{tab:survs}. The LSST
and the DES surveys are expected to be around 10\% larger in size than the
ones that we effectively consider here. In this way, we account for the
reduction of the usable area due to masking. Furthermore, with the ficticious
DES-WIDE survey we analyze the impact of having a survey like DES but with the
solid angle of LSST, and with the INT survey the effect of having deeper
observations on the same area as DES.

\begin{table}
  \centering\vspace{-5pt}
  \caption{\footnotesize Parameters for the different surveys considered in
    our detectability study. We present the median redshift, the background
    galaxy number density, the standard deviation of the intrinsic ellipticity
    and its equivalent survey solid angle.}\label{tab:survs}
  \begin{tabular}{l | r c r r}
    \hline
    \bf survey & \bf Median $z$ & \bf $\dfrac{\mathrm{gal}^{\phantom{2}}}{{\mathrm{arcmin}}^2_{\phantom{2}}}$ & \bf $\sigma^{(\rm eff.)}_{\vert \epsilon_{\rm i} \vert}$ & \bf  eff. $\mathrm{\bf degrees^2}$ \\\hline
    LSST       & 1.2    &  40  & 0.05  & 18432  \\
    INT        & 0.9    &  25  & 0.1   & 4608   \\
    DES-WIDE   & 0.68   &  12  & 0.05  & 18432  \\
    DES        & 0.68   &  12  & 0.1   & 4608   \\
    \hline\hline
  \end{tabular}\vspace{-15pt}
\end{table}

For simplicity, in the following the errors plotted are the combined standard
errors obtained from $\Sigma_{\rm Subh}$ and $\Sigma_{\rm cal}$. We assessed
that the correlation between $\Sigma_{\rm Subh}$ and $\Sigma_{\rm cal}$ is
subdominant. However, for the parameter fitting we use the full covariance matrix
computed with bootstrapping in each case.

\subsubsection{Signal-to-noise ratio} \label{subsec:simsur}

For each survey and luminosity bin, we compute a mean signal-to-noise ratio
$(SNR)$ by averaging over eight logarithmic bins in projected radius $\xi$ in
the range from $0.043$ to $0.12$ Mpc:
\begin{equation}
  SNR=
  \left\langle
    \frac{\epsilon_{\rm t}(\xi)}{\sigma\left[\epsilon_{\rm t}(\xi)\right]}
  \right\rangle_{\xi}.
\end{equation}
\noindent We use for $\sigma\left[\epsilon_{\rm t}(\xi)\right]$ the combined
standard error of the mean $\Sigma_{\rm Subh}$ and $\Sigma_{\rm cal}$. These
$SNR$ values neither take into account correlated noise nor express how well a
particular feature can be detected. Nonetheless, they are compact estimators
and offer a rule of thumb for the expected signal quality.

\begin{figure}
\centering
\includegraphics[width=.5\textwidth,angle=0]{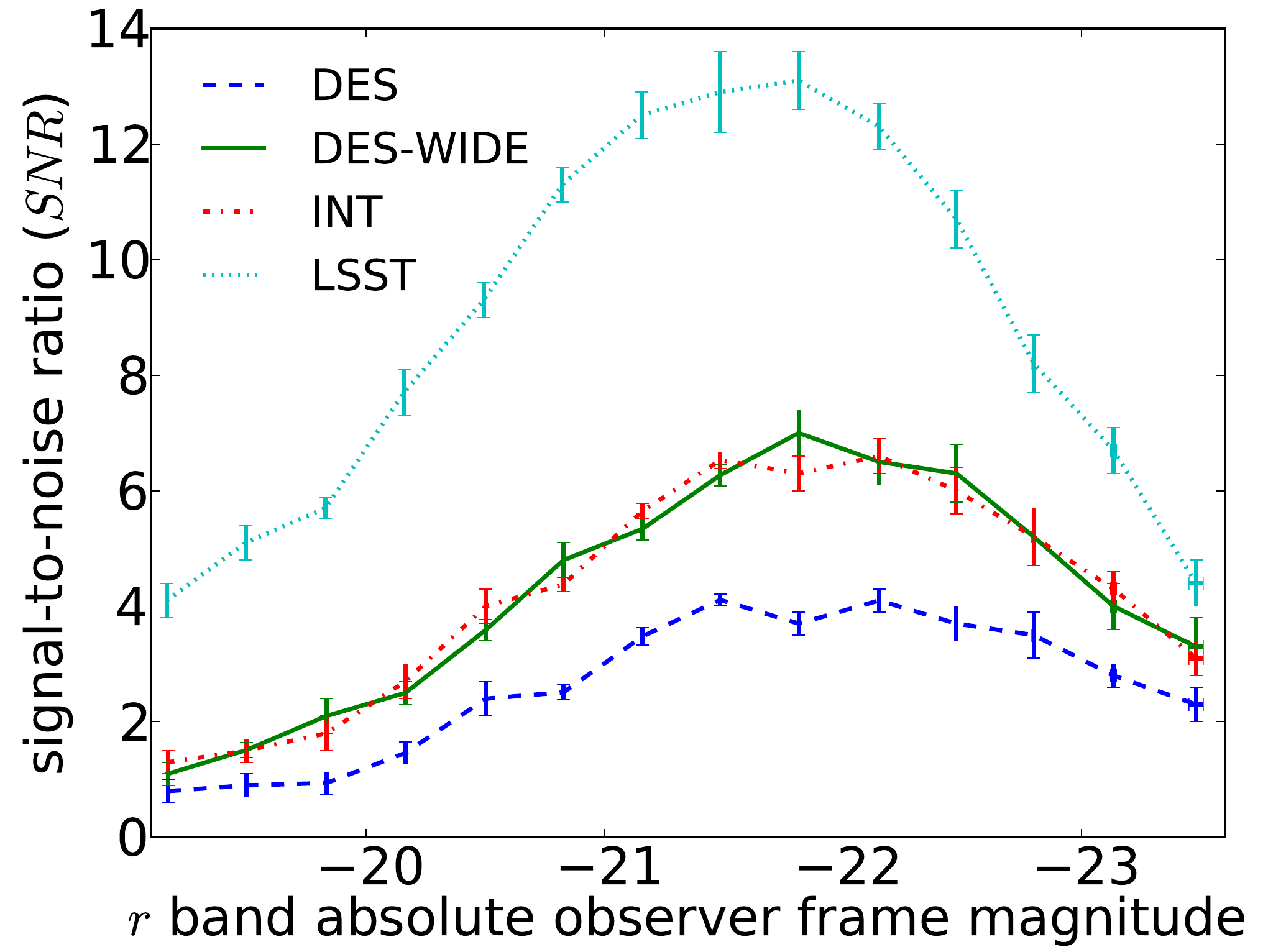}
\vspace{-20pt}
\caption{\footnotesize Our compact SNR estimator, for the different luminosity
  bins for each mock survey considered. For each luminosity bin we plot the
  mean magnitude. The errors plotted are standard errors obtained from the
  sample.}\label{fig:SNR}\vspace{-15pt}
\end{figure}

In Fig.~\ref{fig:SNR}, we present the resulting $SNR$ values for the four
surveys considered.  Luminous galaxies usually have a larger halo with a
stronger signal, whereas faint galaxies are more numerous, corresponding to a
smaller statistical error. The $SNR$ depends on these two competing effects,
resulting in a maximum of the $SNR$ signal-to-noise ratios are around $-21.5$
magnitude.

The least luminous bins are below the two-sigma level in all surveys except
for LSST; in these cases, the signal-to-noise ratio can be improved by using
broader bins.

Finally, in the left panel of Fig.~\ref{fig:lumDES}, we present the predicted
galaxy-galaxy lensing signals for a LSST-like survey, and in the right panel
for a DES-like survey. The signal-to-noise is lower in a DES-like survey,
requiring the use of fewer luminosity bins (from $r=-24.25$ to $r=19$ with a
width of $0.75$ magnitudes) than for LSST.  This figure presents a global
characterization of the expected signals, taking into account an adequate
treatment of the galaxy samples. We can see the quality of the signals and how
detailed our models can be.

\subsubsection{Mass-luminosity relation}

\begin{figure*}[h]\centering
  \vspace{-10pt}
  \includegraphics[width=.495\textwidth]{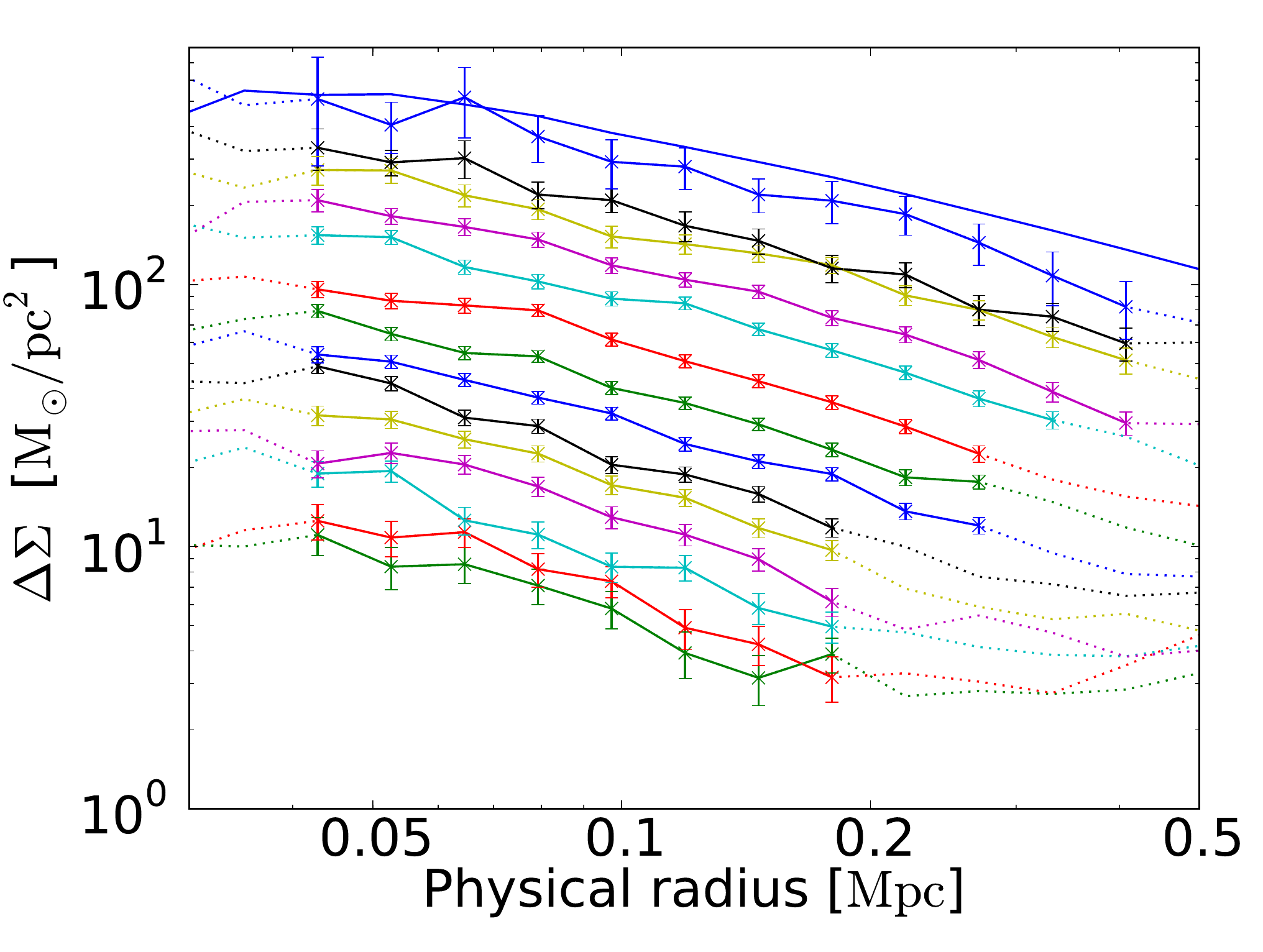}
  \includegraphics[width=.495\textwidth]{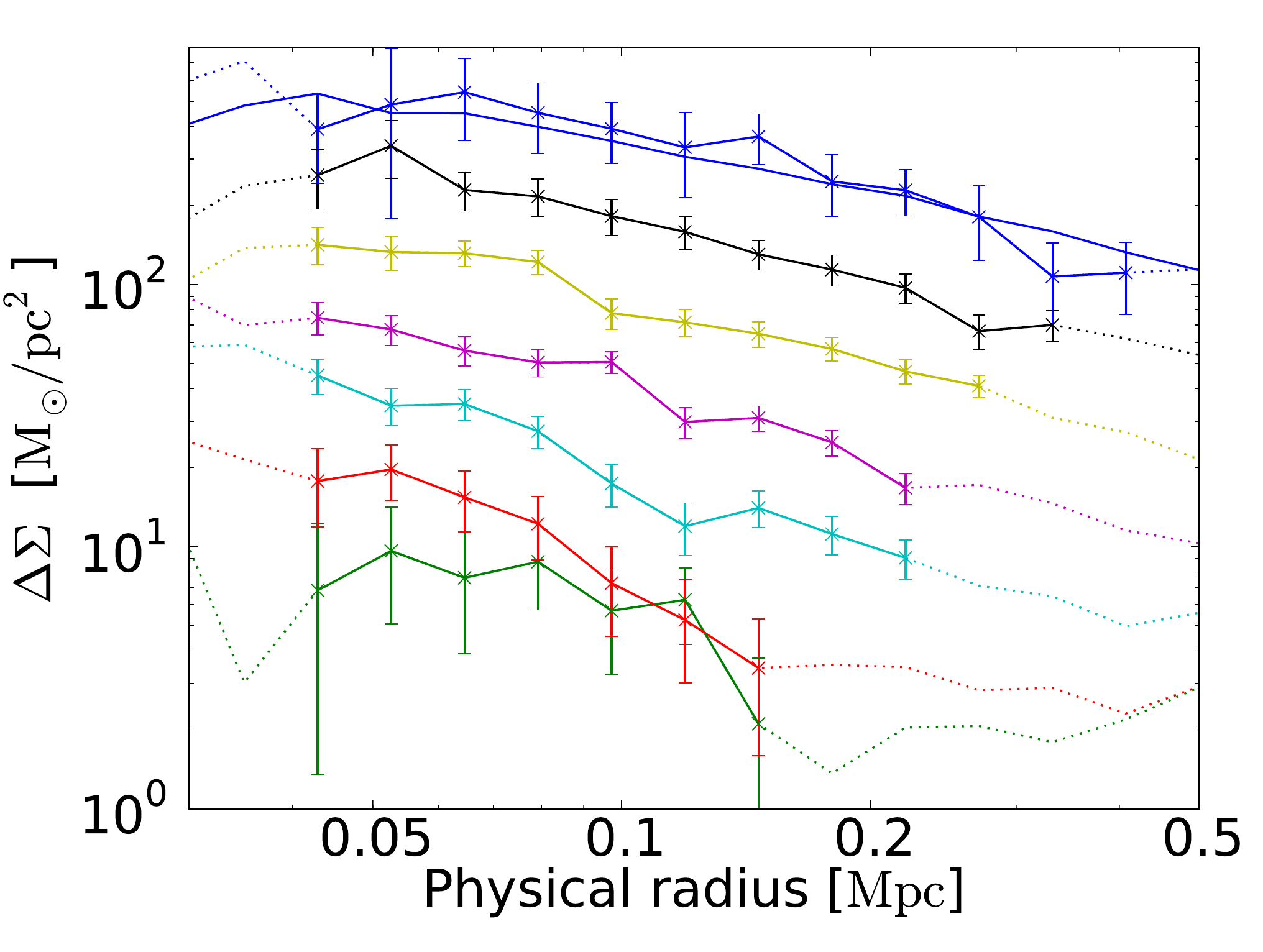}  \vspace{-10pt}
  \caption{\footnotesize Measured $\Delta\Sigma(\xi)$ profiles of galaxies
    binned by $r$-band absolute magnitude. Left panel: LSST-like survey, with
    $r<26$ in apparent magnitude. Right panel: DES-like survey, with $r<22$ in
    apparent magnitude. Error bars show the data used for the analysis later
    on, the range was derive from our results in
    Tab.~\ref{tab:mass_range}. The solid blue line without error bars is the
    measurement for the host cluster shown as a visual reference.}
  \label{fig:lumDES}
\end{figure*}

We now address the question how well a mass-luminosity relation for subhalos
can be estimated from DES and LSST. The measurement of such a relation can
help to distinguish between models of sub-structure formation. As we probe
larger scales than stellar dynamics, and we consider large samples of galaxies
compared to strong lensing analyses, our method is largely complementary to
other methods. In Fig.~\ref{fig:lumDES} we present the results for DES and
LSST. Note that we only plot error bars for the range where we consider our
measure has a bias below 10\%.

\begin{figure}[h]\centering
  \includegraphics[width=.495\textwidth]{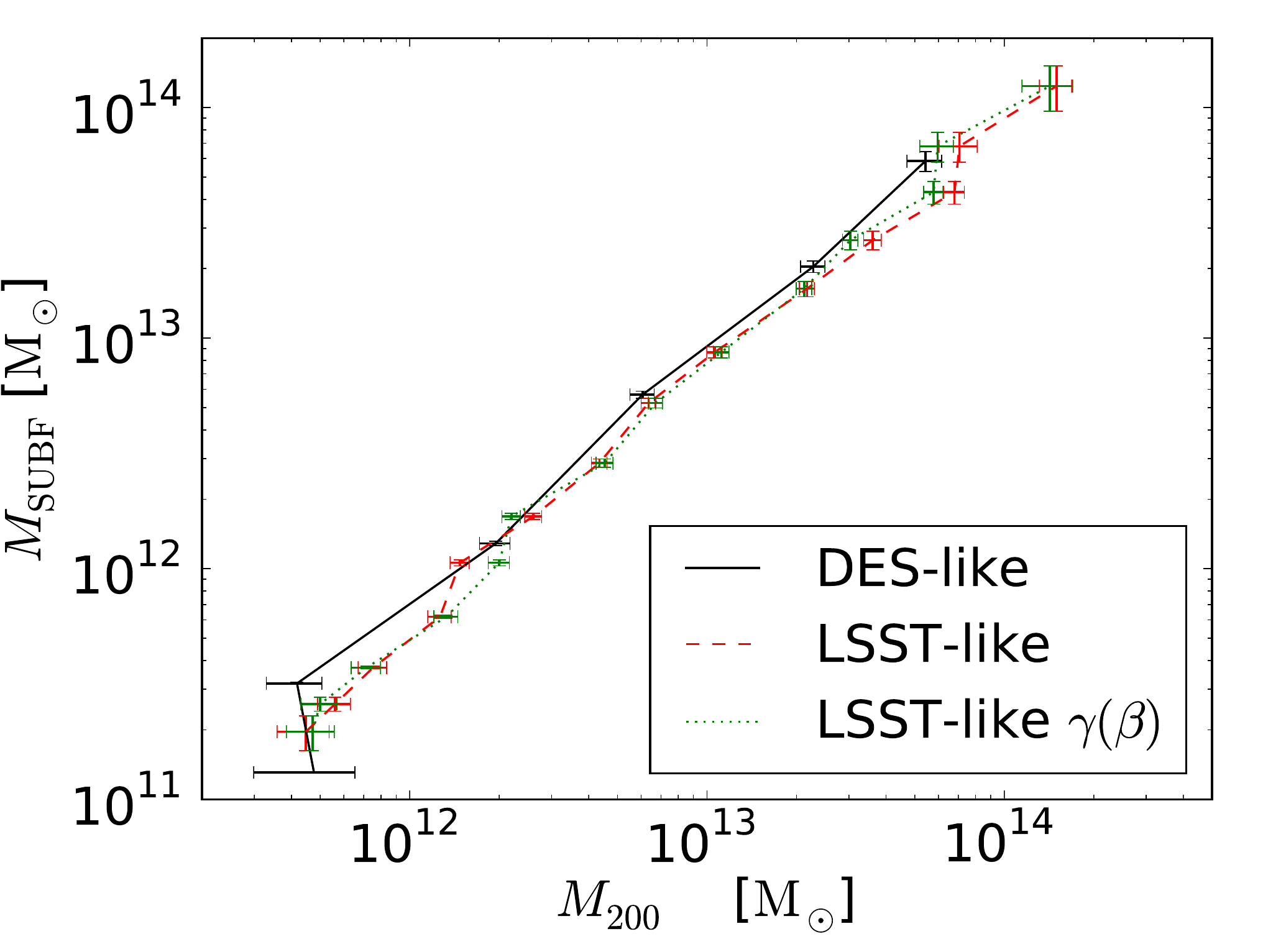}\\
  \includegraphics[width=.495\textwidth]{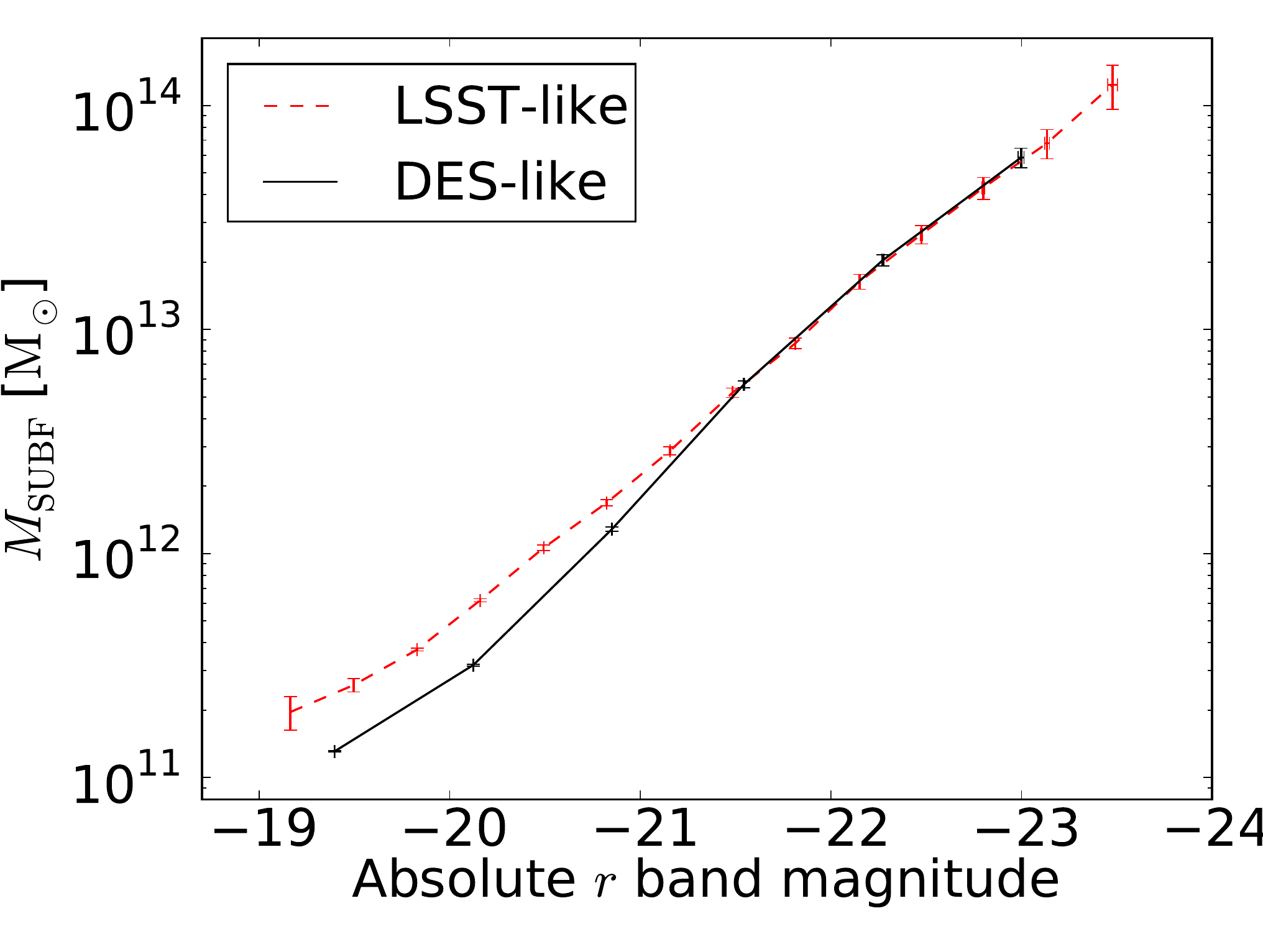}
  \caption{\footnotesize Top: relation between the fitted parameter $M_{200}$
    and $M_{\rm SUBF}$. Bottom: the mass-luminosity relation obtained from the
    catalogues. We plot in solid black the results for a DES-like survey, and
    in dashed red for a LSST-like survey.} \label{fig:lumDES2a}
\end{figure}

\begin{figure}[h]\centering
  \includegraphics[width=.495\textwidth]{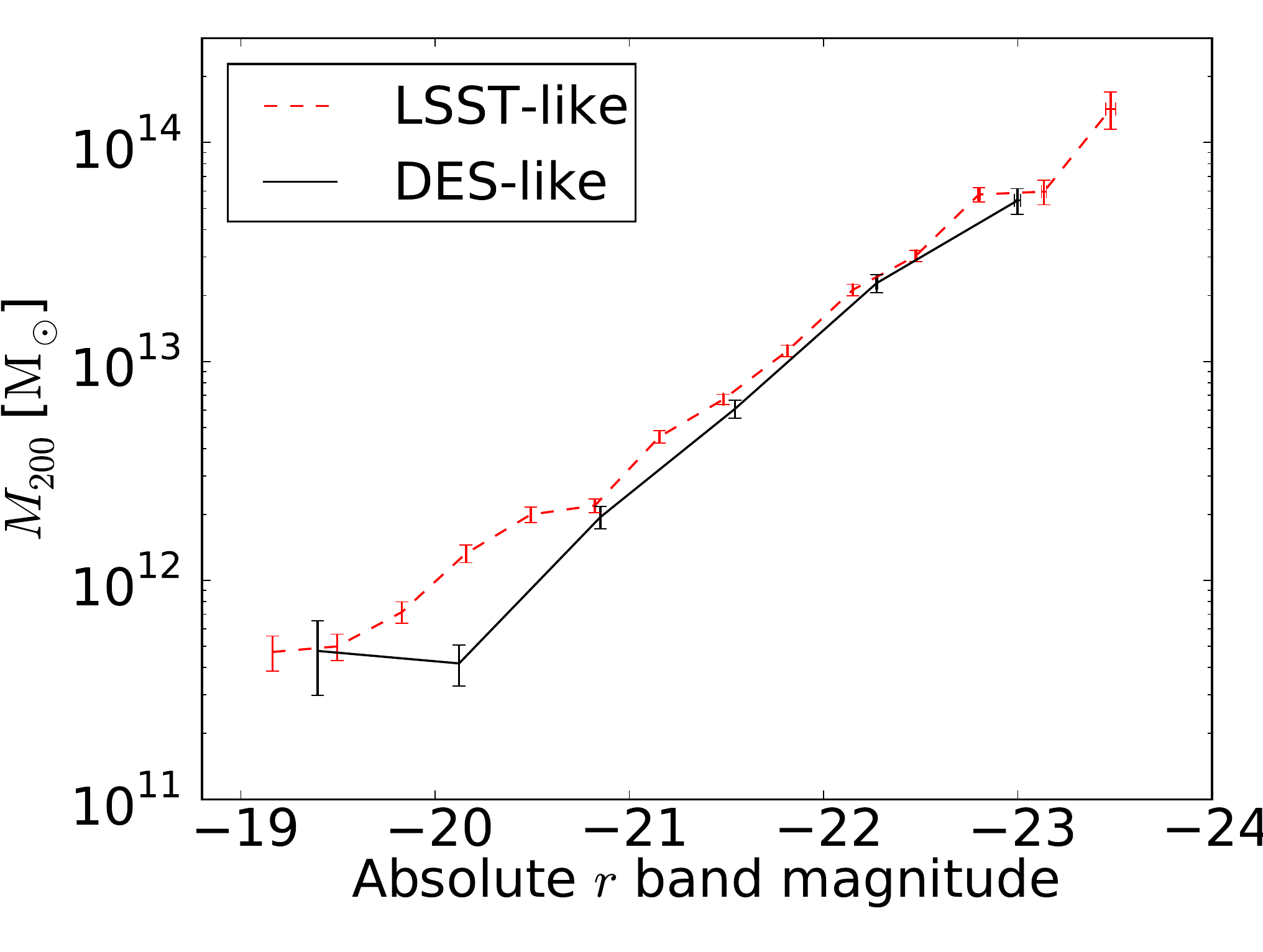}\vspace{-10pt}
  \caption{\footnotesize Bottom: the relation between the fitted parameter
    $M_{200}$ and the luminosity in absolute observer frame $r$ band
    magnitude. We plot in solid black the results for a DES-like survey, and
    in dashed red for a LSST-like survey.}
  \label{fig:lumDES2b}
\end{figure}

In order to obtain the subhalo mass $M_{\rm SUBF}$, one needs to derive
$M_{200}$ from the measured $\Delta\Sigma$ profiles. The subhalo mass can then
be computed from a scaling relation between $M_{\rm SUBF}$ and $M_{200}$,
which is shown in the top panel of Fig.~\ref{fig:lumDES2a}. The final
mass-luminosity relations, for both a LSST and a DES-like survey, are shown in
Fig.~\ref{fig:lumDES2b}. Using DES or LSST, one can constrain the
mass-luminosity relation over two decades in mass. The systematic differences
between the relation for the two surveys can be explained by the different
lens samples used. The brightest bins are populated with the same objects for
both surveys and therefore the scaling relations are similar. At the fainter
end, the different magnitude cuts in combination with the broader luminosity
bins in the DES change the mass-luminosity relation (Fig.~\ref{fig:lumDES2a},
bottom panel).


Nonetheless, the relation between $M_{200}$ and the mass is the same (within
the errors) in both cases. The smaller number density of faint galaxies in DES
compared to the LSST survey reduces in this case the range over which we
can measure the mass-luminosity relation. Note that the faintest luminosity
bin for the DES-like survey may not be usable as the measurement error becomes
too large.

\subsubsection{Evolution of subhalo profiles}

In Sect.~\ref{sec:evsubh}, we showed how the subhalo mass profiles change with
time and found that it is not possible to detect a truncation in the original
sense, and that the mass density decreased at all scales consistent with a
heating process. We now study whether this behavior can be detected with DES
or LSST using observable quantities only.

\begin{figure*}[h]
  \centering
  \includegraphics[width=.495\textwidth]{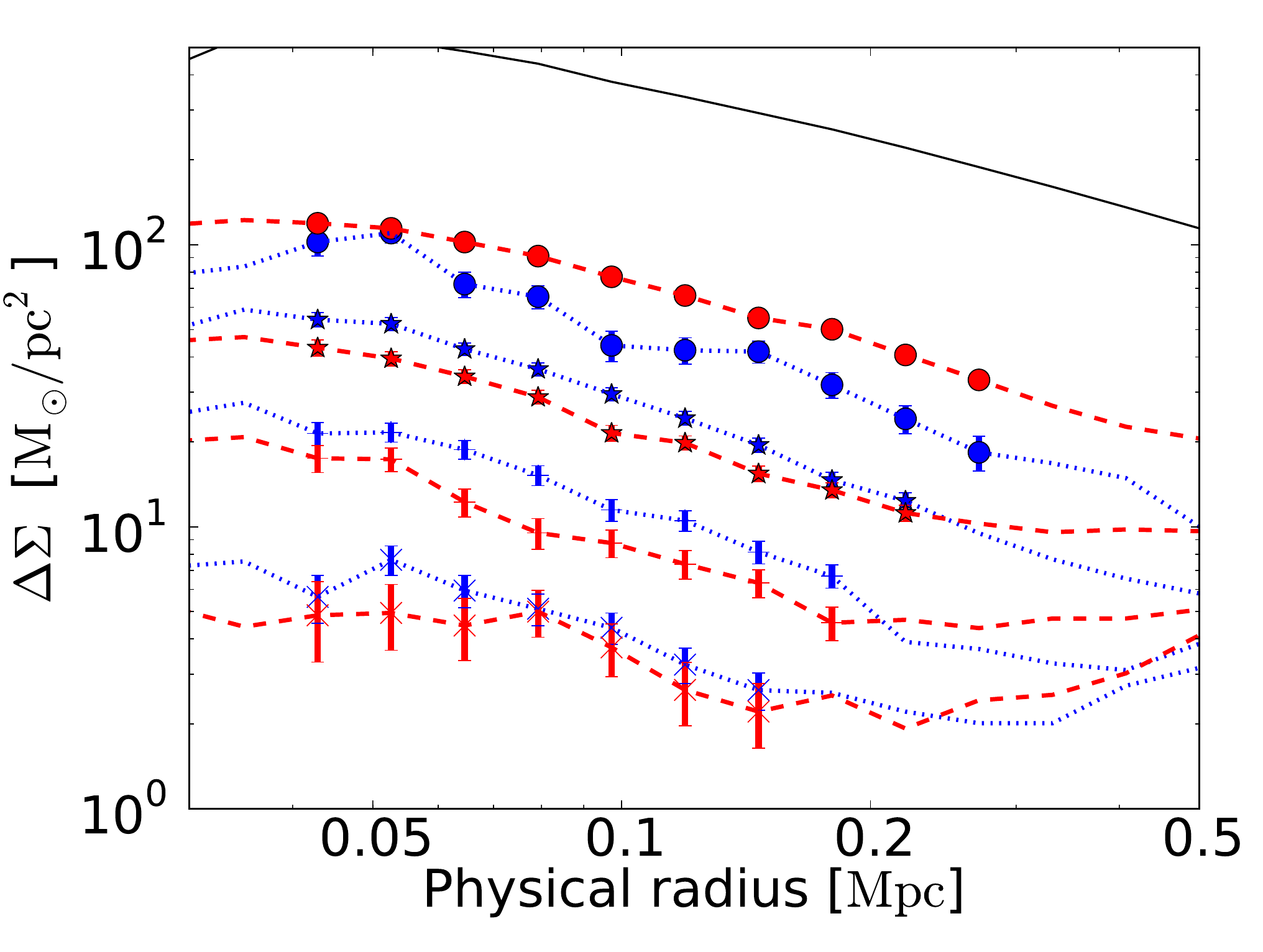}
  \includegraphics[width=.495\textwidth]{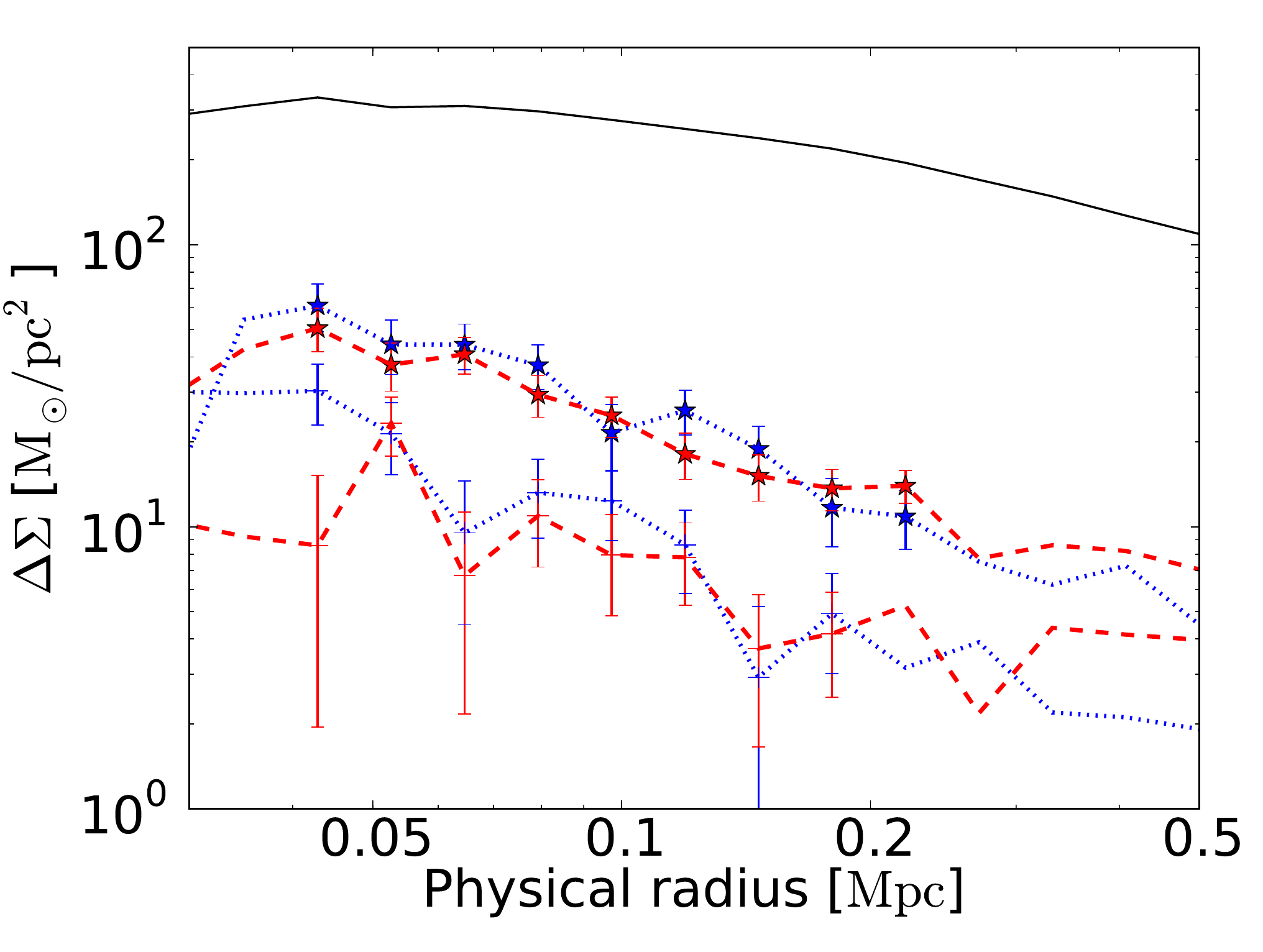}
  \caption*{\footnotesize Data for the LSST-like survey (left
    panel). Galaxies with $z<0.9$ and $r<26$. Masses in units of $10^{10}\,\mathrm{M_\odot}$.}
  \label{tab:ev_LSST_stellmass_range}
  \centering
  \begin{tabular}{l r c c c c c c c c}
    \hline 
    \bf Bin& \bf \# Subhalos  & \bf  Morph.   
    &\bf $\left\langle\frac{L^{\phantom{2}}_{\rm Bulge}}{L_{\rm Total}}\right\rangle$ & \bf  ${\mathrm{age}}$ 
    & \bf $M_{\rm \star}$ bin & \bf ${M_\star}$ & \bf ${M_{\rm inf}}$
    & \bf ${M_{\rm SUBF}}$  & ${M_{200}}$  \\
    \hline 
    {\color{blue}$\times\cdot\cdot$} & 66703 & Disk  & 0.3212 &2.362 &I  & 0.5839 & 19.94 & 11.23   &  32$\pm$4    \\
    {\color{red}$\times$- -}         & 31587 & Bulge & 0.7526 &3.329 &I  & 0.7203 & 19.91 & 8.2     &  29$\pm$10      \\\hline
    {\color{blue}$+\cdot\cdot$}      & 20733 & Disk  & 0.3562 &2.229 &II & 2.326  & 761.3 & 43.2    &  122$\pm$10    \\
    {\color{red}$+$- -}              & 21952 & Bulge & 0.7956 &3.651 &II & 2.390  & 657   & 23.1    &  71$\pm$8    \\\hline
    {\color{blue}$\star\cdot\cdot$}  & 9101  & Disk  & 0.4157 &2.160 &III& 6.299  & 2900  & 162     &  333$\pm$16    \\
    {\color{red}$\star$- -}          & 13003 & Bulge & 0.8008 &3.522 &III& 6.491  & 2840  & 119     &  294$\pm$17    \\\hline
    {\color{blue}$\bullet\cdot\cdot$}& 762   & Disk  & 0.468  &1.91  &IV & 13.45  & 860   & 485     &  763$\pm$80     \\
    {\color{red}$\bullet$- -}        & 2063  & Bulge & 0.799  &2.78  &IV & 15.60  & 2070  & 1280    &  1909$\pm$90    \\
    \hline     \hline \\
  \end{tabular}
  \vspace{-10pt}
  \label{tab:ev_DES_stellmass_range}
  \caption*{\footnotesize Data for the DES-like survey (right panel). Galaxies
    with $z<0.9$ and $r<22$. Masses in units of $10^{10}\,\mathrm{M_\odot}$.}
  \begin{tabular}{l r c c c c c c c c}
    \hline 
    \bf Bin& \bf \# Subhalos  & \bf  Morph.   
    &\bf $\left\langle\frac{L^{\phantom{2}}_{\rm Bulge}}{L_{\rm Total}}\right\rangle$ & \bf  ${\mathrm{age}}$ 
    & \bf $M_{\rm \star}$ bin& \bf ${M_\star}$ & \bf ${M_{\rm inf}}$
    & \bf ${M_{\rm SUBF}}$  & ${M_{200}}$  \\
    \hline 
    {\color{blue}$+\cdot\cdot$}       &8378 &Disk & 0.3494 & 2.386 &I & 2.233 & 69  & 42.7 &  47$\pm$7   \\
    {\color{red}$+$- -}               &8809 &Bulge& 0.8029 & 4.30  &I & 2.308 & 60.9& 20.8 &  73$\pm$11  \\\hline
    {\color{blue}$\star\cdot\cdot$}   &4577 &Disk & 0.4186 & 2.32  &II& 6.92  & 302 & 178  &  267$\pm$19 \\
    {\color{red}$\star$- -}           &7531 &Bulge& 0.8066 & 3.90  &II& 7.23  & 244 & 153  &  307$\pm$21 \\
    \hline \hline 
  \end{tabular}
  \caption{\footnotesize Morphology-stellar mass classification for a
    LSST-like survey (left panel) and for a DES-like survey (right panel).
    The black solid line is the measurement for the host cluster shown for
    visual reference. We only plot symbols and error bars for the range where
    we consider that the bias is below 10\%. The subhalos are at $d_{\rm
      M-S}<0.5$ Mpc. Blue dotted lines correspond to galaxies with a
    relatively small bulge: $\mathrm{Disk}\rightarrow 0<L_{\rm Bulge}/L_{\rm
      Total}\le 0.6$. Red dashed lines correspond to galaxies with a large
    bulge which spent more time inside a cluster: $\mathrm{Bulge}\rightarrow
    0.6<L_{\rm Bulge}/L_{\rm Total}< 0.98$. The different symbols distinguish
    different stellar mass bins (in units of $10^{10}\,\mathrm{M_\odot}$). For
    the left panel the ranges are: $\mathrm{1}\rightarrow[0.14:1.4]$,
    $\mathrm{II}\rightarrow[1.4:4.11]$, $\mathrm{III}\rightarrow[4.11:10.96]$,
    $\mathrm{4}\rightarrow[10.96:\infty]$. For the right panel the ranges are
    $\mathrm{I}\rightarrow[1.1:4.1]$,
    $\mathrm{II}\rightarrow[4.1:13.7]$.}  \label{fig:stell_morphDES}
\end{figure*}

In Fig.~\ref{fig:stell_morphDES}, the predicted galaxy-galaxy lensing signals
for both LSST (left panel) and DES (right panel) are shown with different
symbols for the different stellar mass bins. Additionally, we have divided
each of the stellar mass bins into two subsets according to morphology. The
details of the binning in morphology and stellar mass are given in tables
inside the figure. 

The dashed red lines belong to galaxies with a relatively large bulge, which
should belong to old subhalos. With blue dotted lines we plotted the signals
from galaxies with a large disk, which should belong to younger subhalos. The
number of stellar mass bins we can analyze is limited, since we need to split
them further into early-type and late-type galaxies. The stellar mass
ranges were obtained by experimentation; in the case of the DES-like survey we
focused on a particular range where the effect was strongest. Note that we
only plot error bars for the range where our measurement has a bias below 10\%.

The sub-division according to morphology does not always produce the desired
results. One can see from the last two rows in the top table, that for the
most massive satellite galaxies, the infall mass (eighth column) of the two
morphology bins are very different. Older subhalos have a significantly larger
mean infall mass, which indicates that they evolved in a different
manner. This difference masks any possible detection of mass loss due to tidal
stripping by the cluster.

With the exception of this particular bin, however,
Fig.~\ref{fig:stell_morphDES} shows that the profile amplitude of
bulge-dominated galaxies is smaller than the one of disk-dominated objects, in
accordance with our predictions from Sect.~\ref{sec:evsubh}.  This can be
detected with high significance in the LSST; for the DES-like survey the
detection is hampered by the low signal-to-noise ratio.

\section{Systematic effects}\label{sec:system}

\subsection{Validity of the weak lensing approximation}\label{sec:wap}

So far, we have assumed that the weak lensing approximation is valid in our
measurements, i.e.~that $\kappa \ll 1$ and thus that the reduced shear
effectively equals the shear. To check the validity of this approximation, we
compare the galaxy-galaxy lensing signal of subhalos measured using shear with
the signal obtained using reduced shear.

In Fig.~\ref{fig:lumDES2a} we have estimated $M_{200}$ by fitting NFW profiles
to the $\Delta\Sigma$ profiles measured from galaxy catalogues obtained from
the full ray-tracing using the shear $\gamma(\boldsymbol{\theta})$ and from
similar catalogues using the reduced shear $g(\boldsymbol{\theta})$. No
significant difference between the two methods is found. For the average
subhalo, the impact of the environment is not important. Within internal checks,
we have observed a significant discrepancy only for the most massive subhalos
for the innermost radii. This is produced by the large convergence of the
subhalo itself and not related to the host main halo.

\subsection{Type-2 galaxies}\label{sec:type2}
\begin{figure}
  \centering
  \includegraphics[width=.5\textwidth]{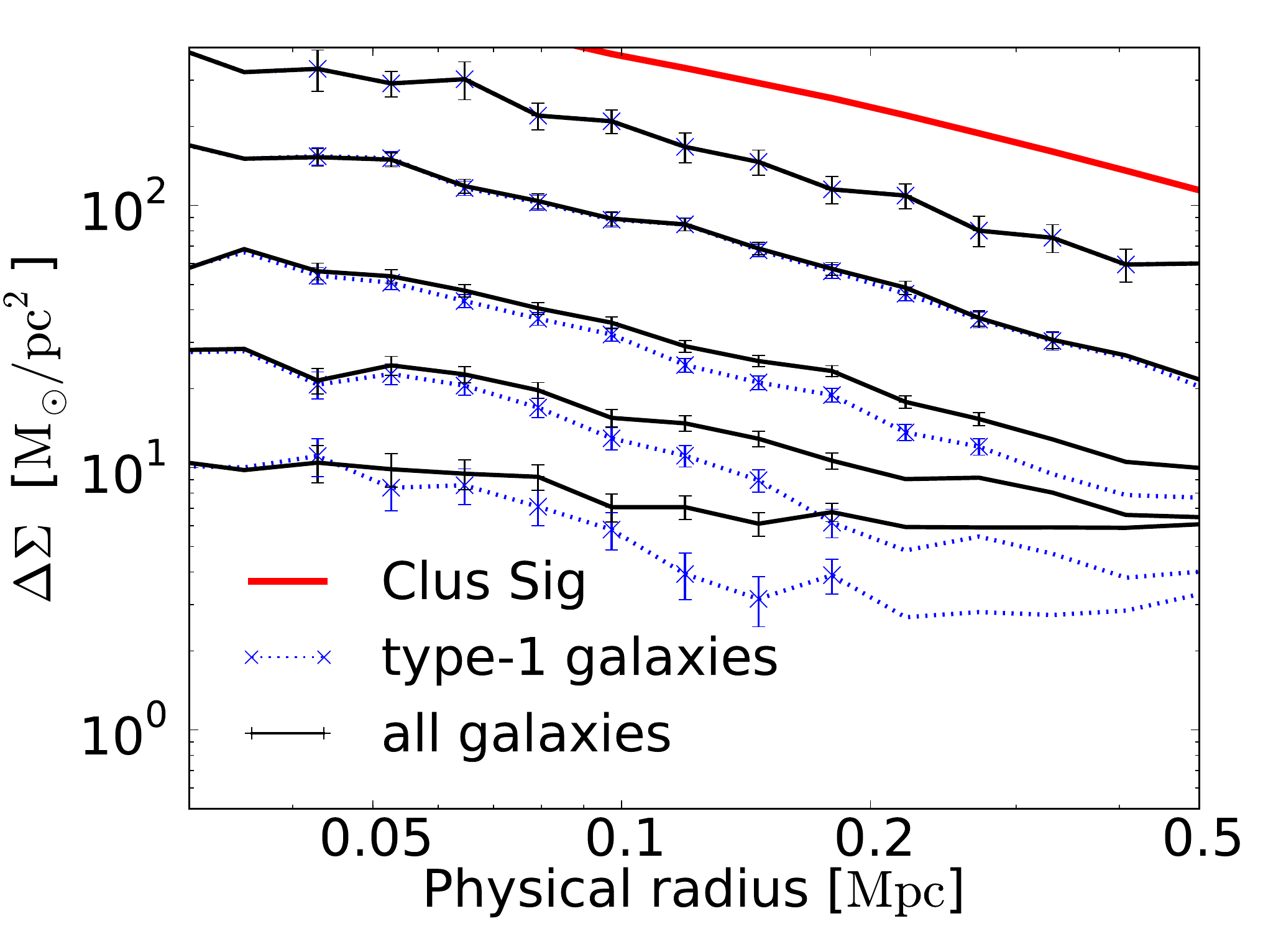}\vspace{-10pt}
  \caption{\footnotesize \footnotesize Excess surface mass density
    $\Delta\Sigma(\xi)$ for a few luminosity bins for an LSST-like survey. The
    blue dashed lines correspond to the measurements only for type-1
    galaxies. The solid black lines correspond to the combination of both
    type-1 and type-2 galaxies. The range where we assume the measurement has
    a bias below 10\% is highlighted, (derived using our results for type-1
    only profiles). The solid thick red line is the signal for the average
    cluster, shown for visual reference.}
  \label{fig:type2lum}
\end{figure}
\begin{figure}
 \centering
  \includegraphics[width=.5\textwidth]{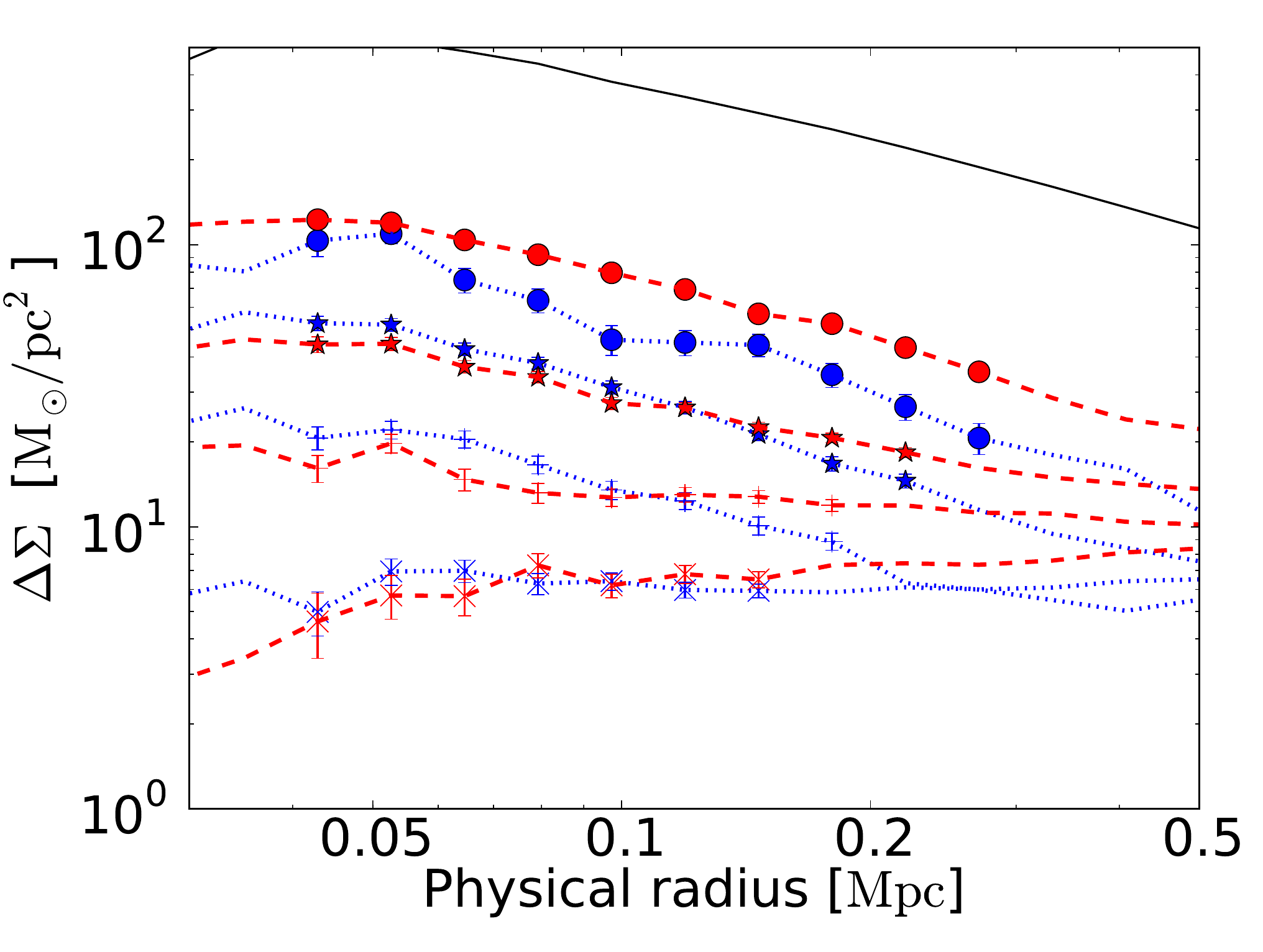}\vspace{-10pt}
  \caption{\footnotesize Morphology-stellar mass classification for a
    LSST-like survey including both type-1 and type-2 galaxies. Blue dotted
    lines correspond to galaxies with a relatively small bulge, red dashed to
    those with a large bulge. The different symbols distinguish different
    stellar mass bins. The range where we assume the measurement has a bias
    below 10\% is highlighted, (derived using our results from the type-1 only
    profiles). The classification ranges are described in the top table in
    Fig.~\ref{fig:stell_morphDES}. The black solid line is the measurement for
    the host cluster shown for visual reference.}
  \label{fig:type2stellm}
\end{figure}
\begin{figure}
   \centering
  \includegraphics[width=.5\textwidth]{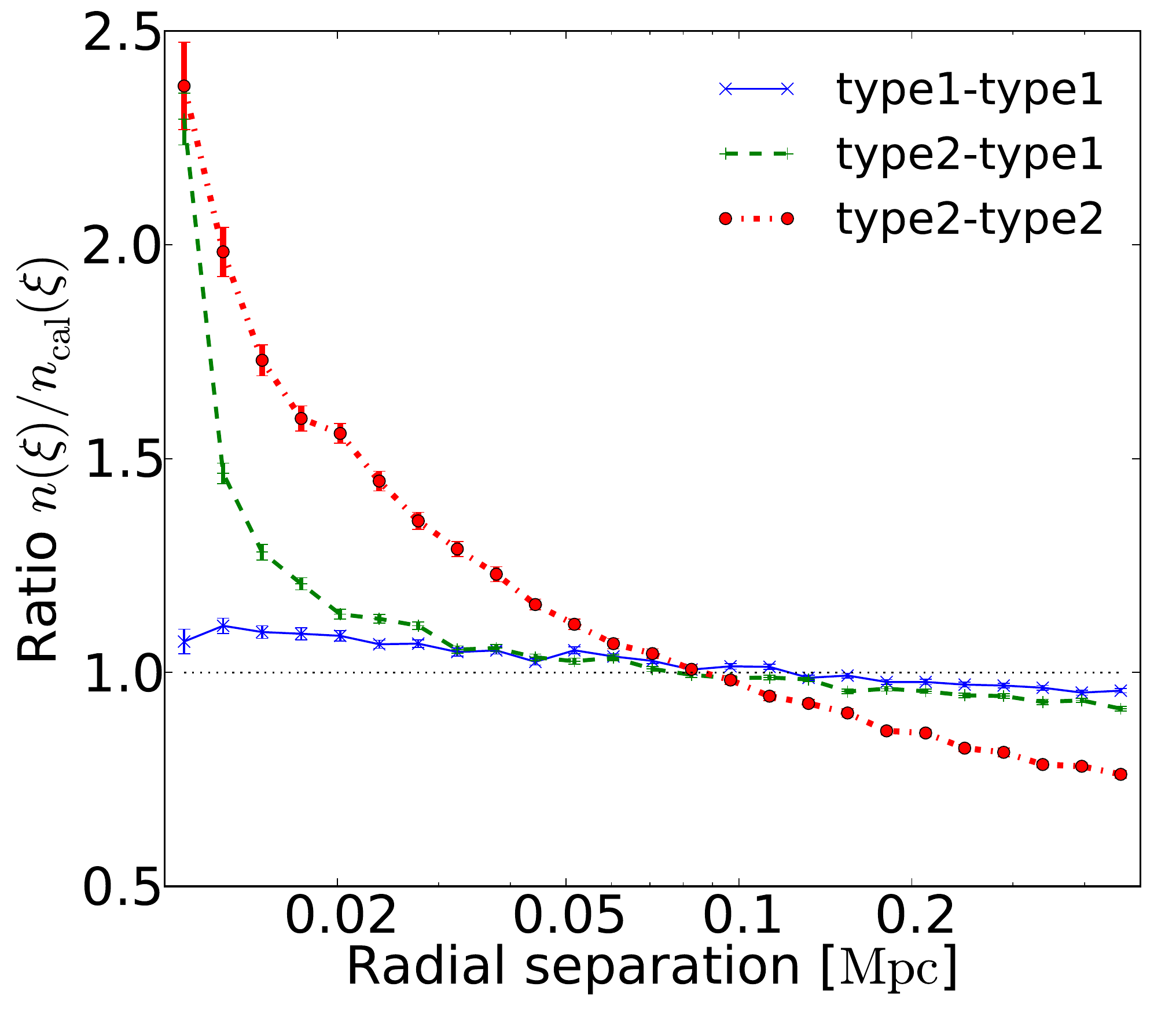}\vspace{-10pt}
  \caption{\footnotesize Ratio between the galaxy number density around a
    galaxy and around its corresponding calibration point, as a function of
    distance. We also plot as visual reference the line where the ratio is
    $1$. Dotted-dashed red: excess type-2 number density around type-2. Solid
    blue: type-1 number density around type-1. Green dashed: excess type-1
    number density around type-2. $d_{\rm M-S}<0.5$ Mpc.}
  \label{fig:type2excesscomp}
\end{figure}

The semi-analytic model of galaxy formation by \cite{2007MNRAS.375....2D}
contains galaxies that have ``lost'' their subhalos. Whenever a subhalo is
no longer detected (e.g. because it has lost most of its mass due to tidal
stripping), the hosted galaxy is still maintained and placed at the position
of the most bound particle of the former subhalo. After the dynamical
friction timescale, these galaxies eventually disappear as they merge into the
main halo galaxy.

The results presented so far were computed excluding type-2 galaxies for the
following reasons:\vspace{-5pt}
\begin{itemize}
\item they are hosted by subhalos below or around the mass resolution limit,
\item they exist under the assumption that the stripping of a large part of
  the subhalo does not destroy the hosted galaxy,
\item the survival time is heavily influenced by the model.\vspace{-5pt}
\end{itemize}

Nevertheless, to understand the impact of such galaxies on our measurements,
we analyze how our results are altered when including them. We present how our
measurements would be affected assuming that type-2 are real and that we
cannot distinguish between them and type-1 galaxies.

In Fig.~\ref{fig:type2lum}, we present $\Delta\Sigma(\xi)$ for a selection of
luminosity bins for type-1 galaxies (with dashed blue lines) and the same
luminosity bins but for the combination of type-2 and type-1 galaxies (with
solid black lines). The inclusion of type-2 galaxies change the average profile,
especially at large radii. The change is larger for fainter galaxies. The new
profiles are flatter, and there is a systematic error in the determination of
the mass profile. This effect can also interfere with our ability to assign
any kind of mass estimate.

In Fig.~\ref{fig:type2stellm}, we present the same analysis as in the top
panel in Fig.~\ref{fig:stell_morphDES}, but with both type-2 and type-1
galaxies. Recall that the line and color type encodes the morphology (as a
proxy for the subhalos' age), and the symbol encodes the stellar mass (as a
proxy for the subhalo infall mass). The inclusion of type-2 galaxies is more
important if we aim to estimate the mass loss on subhalos induced by tidal
stripping. Our former ability to measure this effect was small, and the change
in the profiles that type-2 galaxies produce makes it impossible.

We are able to trace the mass responsible for the lensing signal in type-2
galaxies. In Fig.~\ref{fig:type2excesscomp}, we quantify how likely is to find
a companion near a type-2 galaxy. We compare the number density as function of
radius, around a galaxy and its calibration point.  Type-2 galaxies are on
average close to a subhalo which accounts for the lensing signal measured. We
were able as well to detect type-2 galaxies associated with projected
overdensities which were not detected as subhalos by SUBFIND.

The presence of galaxies with no halos makes our measurements less
precise. However, if such galaxies exists, we could use our analysis to
determine the fraction of galaxies of such type, and to distinguish them from
galaxies with a host halo.

\section{Summary and conclusions}
\label{sec:sumcon}

In order to understand the evolution of galaxy clusters, it is essential to
study the matter profiles of satellite galaxies. Analyzing how these
profiles evolve with time can improve our knowledge on galaxy
evolution and on structure formation.

In this work we analyzed the use of weak gravitational lensing on satellite
galaxies inside clusters, in particular the use of galaxy-galaxy lensing. With
this probe, we can measure in a non-parametric way the average projected mass
profiles of the host matter halos of galaxies. This cosmological probe
correlates the image distortion (shear) of a background galaxy with the mass
of a foreground galaxy.

Galaxy-galaxy lensing needs large galaxy samples. For this reason it has not
yet been fully exploited on cluster satellite galaxies. In our work, we
forecast results for future surveys using the Millennium Simulation
(\citealt{2005Natur.435..629S}), ray-tracing simulations
(\citealt{2009A&A...499...31H}) and the galaxy semi-analytic catalogues by
\cite{2007MNRAS.375....2D}.

We presented the details of galaxy-galaxy lensing measurements on satellite
galaxies and its contamination from the host cluster. In order to overcome the
contamination from the host cluster we proposed a calibration method, which
can solve the problem up to a certain range in radius. For each subhalo, we
define a point at the same distance from the main halo as the subhalo, but in
the opposite direction as seen from the cluster center. We can estimate the
contribution of the main halo by measuring the tangential shear around this
new point, under the assumption that the cluster is point-symmetric.

We created mass maps of our clusters using theoretical profiles with
known parameters. With this mock cluster sample we were able to test our
calibration method taking into account realistic characteristics for our
cluster samples such as halo spatial distribution or mass function. We
estimated the performance of our measurements at different radii and we
defined a minimal separation between the subhalo and the main halo center to
optimize the signals.

With the previous tests, we could characterize the subhalos in the Millennium
Simulation using projected mass maps. The weak lensing signal of subhalos is
well described by a simple NFW profile, and it was not possible to detect a
truncation radius. Our results are consistent with an abrupt truncation of the
mass profile at radii larger than 0.2 Mpc. There are certain discrepancies
between our work and the previously published works from
\cite{2007A&A...461..881L}, \cite{2007MNRAS.376..180N},
\cite{2007ApJ...656..739H} and \cite{2010arXiv1007.4815S}. These authors
measured a much smaller extent of the subhalos using gravitational lensing on
a few observed clusters. The results were derived using parametric models for
the mass profiles of the subhalos and the main halo. The models used by these
authors do not fit subhalos in the Millennium Simulation. Since the
Millennium Simulation only contains dark matter, there is not absolute
certainty that NFW profiles should be the best description for halos of real
galaxies. On the other hand our method did not include strong lensing
constraints in comparison to the previous works. Nevertheless, our results
challenge the choices made in the afore-mentioned works. The truncation radii
that they measured can be also interpreted as a consequence of the parametric
method used. Therefore, further analysis and better data are needed in order
to solve the problem.

We also were able to characterize the evolution of the mass profiles. We show
that the lensing profiles decrease in amplitude with time. This is consistent
with a mass loss at all scales. Due to the tidal forces exerted by the host halo,
subhalos are stripped of the mass at the outermost radii, and at the same
time the mass at the inner regions is redistributed. This is consistent with
the work by \cite{2003ApJ...584..541H} and also supports the idea of
large truncation radii.

After describing the subhalo profiles we used simulated galaxy catalogues to
forecast signals for future surveys, focusing on DES
\citep{2005astro.ph.10346T} and LSST \citep{2008arXiv0805.2366I}. We analyzed
the semi-analytic catalogues in order to classify the galaxy samples to
optimize the measurements. With the result from this analysis we derived
the following results.

We predict the detectability of the signals using a compact estimator of the
signal-to-noise ratio. The cluster sample required is very large, but we can
already expect signals from a DES-like survey roughly above the three sigma
level. The data from a LSST-like survey is well-suited for the studies that we
proposed.

There is not a unique way of separating between subhalo and host halo mass. In
this work, we considered that the mass of the subhalo is well estimated in our
simulations by the SUBFIND algorithm \citep{2001MNRAS.328..726S} ($M_{\rm
  SUBF}$). We modeled the relation between the measured NFW profiles and the
gravitationally bound mass of the subhalo mass $M_{\rm SUBF}$. We also checked
that the weak lensing approximation is valid.

According to the semi-analytic catalogues, luminosity in the SDSS $r$ band is
a good proxy for mass. We predict that it is possible even for a DES-like
survey to constrain the mass-luminosity relations of subhalos over two
decades in mass, from around $5\times10^{11}\:\mathrm{M_\odot}$ to
$10^{14}\:\mathrm{M_\odot}$.

In order to study the time evolution of the profiles using realistic data, we
binned the galaxies according the mass of the subhalo prior to falling into
the cluster and the time spent inside the cluster (subhalo age). Within the
semi-analytic catalogues, it is possible to infer their initial mass from
their observed stellar mass. On the other hand we could not find any galaxy
observable that is strongly correlated with the subhalo age. With these
observables, we could only put weak constraints to the evolution of subhalo
matter profiles.

In our analyses we neglected galaxies without a host subhalo which one finds
in the semi-analytic catalogues of \cite{2007MNRAS.375....2D}. A priori, these
galaxies are not fully reliable as they populate mass overdensities below the
resolution limit of the Millennium Simulation. Nevertheless, for completeness
we considered their influence. These galaxies show a lensing signal with a
high amplitude, which can be explained as being produced by correlated
halos. We also were able to quantify the correlation between the position of
type-2 galaxies and other subhalos. Finally, we presented how our previous
analysis is affected by assuming that these galaxies are realistic and that we
are not able to distinguish them from galaxies with a host subhalo. However,
further simulations are needed in order to investigate galaxies of such
characteristics.

\begin{acknowledgements}
  We thank Lan Wang for providing the infall mass and infall snapshot
  catalogues for this project. We also thank Tim Eifler for moral support and
  comments, Simon White for his comments. EPM acknowledges support by the
  Argelander-Institut f\"ur Astronomie and by the EU Project DUEL,
  Projektnr. 36133. JH and SH acknowledge support by the Deutsche
  Forschungsgemeinschaft within the Priority Programme 1177 under the project
  SCHN 342/6 and the Transregional Collaborative Research Centre TRR 33 ``The
  Dark Universe''. The Millennium Simulation databases used in this paper and
  the web application providing online access to them were constructed as part
  of the German Astrophysical Virtual Observatory.

\end{acknowledgements}

\bibliographystyle{aa}
\bibliography{references}

\begin{thebibliography}{52}
\expandafter\ifx\csname natexlab\endcsname\relax\def\natexlab#1{#1}\fi

\bibitem[{{Baltz} {et~al.}(2009){Baltz}, {Marshall}, \&
  {Oguri}}]{2009JCAP...01..015B}
{Baltz}, E.~A., {Marshall}, P., \& {Oguri}, M. 2009, Journal of Cosmology and
  Astro-Particle Physics, 1, 15

\bibitem[{{Bartelmann}(1996)}]{1996A&A...313..697B}
{Bartelmann}, M. 1996, \aap, 313, 697

\bibitem[{{Bartelmann} \& {Schneider}(2001)}]{2001PhR...340..291B}
{Bartelmann}, M. \& {Schneider}, P. 2001, \physrep, 340, 291

\bibitem[{{Baugh}(2006)}]{2006RPPh...69.3101B}
{Baugh}, C.~M. 2006, Reports on Progress in Physics, 69, 3101

\bibitem[{{Benjamin} {et~al.}(2007){Benjamin}, {Heymans}, {Semboloni}, {van
  Waerbeke}, {Hoekstra}, {Erben}, {Gladders}, {Hetterscheidt}, {Mellier}, \&
  {Yee}}]{2007MNRAS.381..702B}
{Benjamin}, J., {Heymans}, C., {Semboloni}, E., {et~al.} 2007, \mnras, 381, 702

\bibitem[{{Benson}(2010)}]{2010PhR...495...33B}
{Benson}, A.~J. 2010, \physrep, 495, 33

\bibitem[{{Blaizot} {et~al.}(2006){Blaizot}, {Szapudi}, {Colombi},
  {Budav{\`a}ri}, {Bouchet}, {Devriendt}, {Guiderdoni}, {Pan}, \&
  {Szalay}}]{2006MNRAS.369.1009B}
{Blaizot}, J., {Szapudi}, I., {Colombi}, S., {et~al.} 2006, \mnras, 369, 1009

\bibitem[{{Brada{\v c}} {et~al.}(2006){Brada{\v c}}, {Clowe}, {Gonzalez},
  {Marshall}, {Forman}, {Jones}, {Markevitch}, {Randall}, {Schrabback}, \&
  {Zaritsky}}]{2006ApJ...652..937B}
{Brada{\v c}}, M., {Clowe}, D., {Gonzalez}, A.~H., {et~al.} 2006, \apj, 652,
  937

\bibitem[{{Brainerd} {et~al.}(1996){Brainerd}, {Blandford}, \&
  {Smail}}]{1996ApJ...466..623B}
{Brainerd}, T.~G., {Blandford}, R.~D., \& {Smail}, I. 1996, \apj, 466, 623

\bibitem[{{Clowe} {et~al.}(2006){Clowe}, {Brada{\v c}}, {Gonzalez},
  {Markevitch}, {Randall}, {Jones}, \& {Zaritsky}}]{2006ApJ...648L.109C}
{Clowe}, D., {Brada{\v c}}, M., {Gonzalez}, A.~H., {et~al.} 2006, \apjl, 648,
  L109

\bibitem[{{De Lucia} \& {Blaizot}(2007)}]{2007MNRAS.375....2D}
{De Lucia}, G. \& {Blaizot}, J. 2007, \mnras, 375, 2

\bibitem[{{De Lucia} {et~al.}(2004){De Lucia}, {Kauffmann}, \&
  {White}}]{2004MNRAS.349.1101D}
{De Lucia}, G., {Kauffmann}, G., \& {White}, S.~D.~M. 2004, \mnras, 349, 1101

\bibitem[{{De Lucia} {et~al.}(2006){De Lucia}, {Springel}, {White}, {Croton},
  \& {Kauffmann}}]{2006MNRAS.366..499D}
{De Lucia}, G., {Springel}, V., {White}, S.~D.~M., {Croton}, D., \&
  {Kauffmann}, G. 2006, \mnras, 366, 499

\bibitem[{{Fu} {et~al.}(2008){Fu}, {Semboloni}, {Hoekstra}, {Kilbinger}, {van
  Waerbeke}, {Tereno}, {Mellier}, {Heymans}, {Coupon}, {Benabed}, {Benjamin},
  {Bertin}, {Dor{\'e}}, {Hudson}, {Ilbert}, {Maoli}, {Marmo}, {McCracken}, \&
  {M{\'e}nard}}]{2008A&A...479....9F}
{Fu}, L., {Semboloni}, E., {Hoekstra}, H., {et~al.} 2008, \aap, 479, 9

\bibitem[{{Ghigna} {et~al.}(1998){Ghigna}, {Moore}, {Governato}, {Lake},
  {Quinn}, \& {Stadel}}]{1998MNRAS.300..146G}
{Ghigna}, S., {Moore}, B., {Governato}, F., {et~al.} 1998, \mnras, 300, 146

\bibitem[{{Ghigna} {et~al.}(2000){Ghigna}, {Moore}, {Governato}, {Lake},
  {Quinn}, \& {Stadel}}]{2000ApJ...544..616G}
{Ghigna}, S., {Moore}, B., {Governato}, F., {et~al.} 2000, \apj, 544, 616

\bibitem[{{Halkola} {et~al.}(2008){Halkola}, {Hildebrandt}, {Schrabback},
  {Lombardi}, {Brada{\v c}}, {Erben}, {Schneider}, \&
  {Wuttke}}]{2008A&A...481...65H}
{Halkola}, A., {Hildebrandt}, H., {Schrabback}, T., {et~al.} 2008, \aap, 481,
  65

\bibitem[{{Halkola} {et~al.}(2006){Halkola}, {Seitz}, \&
  {Pannella}}]{2006MNRAS.372.1425H}
{Halkola}, A., {Seitz}, S., \& {Pannella}, M. 2006, \mnras, 372, 1425

\bibitem[{{Halkola} {et~al.}(2007){Halkola}, {Seitz}, \&
  {Pannella}}]{2007ApJ...656..739H}
{Halkola}, A., {Seitz}, S., \& {Pannella}, M. 2007, \apj, 656, 739

\bibitem[{{Hayashi} {et~al.}(2003){Hayashi}, {Navarro}, {Taylor}, {Stadel}, \&
  {Quinn}}]{2003ApJ...584..541H}
{Hayashi}, E., {Navarro}, J.~F., {Taylor}, J.~E., {Stadel}, J., \& {Quinn}, T.
  2003, \apj, 584, 541

\bibitem[{{Hilbert} {et~al.}(2009){Hilbert}, {Hartlap}, {White}, \&
  {Schneider}}]{2009A&A...499...31H}
{Hilbert}, S., {Hartlap}, J., {White}, S.~D.~M., \& {Schneider}, P. 2009, \aap,
  499, 31

\bibitem[{{Hilbert} \& {White}(2010)}]{2010MNRAS.404..486H}
{Hilbert}, S. \& {White}, S.~D.~M. 2010, \mnras, 404, 486

\bibitem[{{Hoekstra} {et~al.}(2004){Hoekstra}, {Yee}, \&
  {Gladders}}]{2004ApJ...606...67H}
{Hoekstra}, H., {Yee}, H.~K.~C., \& {Gladders}, M.~D. 2004, \apj, 606, 67

\bibitem[{{Israel} {et~al.}(2010){Israel}, {Erben}, {Reiprich}, {Vikhlinin},
  {Hildebrandt}, {Hudson}, {McLeod}, {Sarazin}, {Schneider}, \&
  {Zhang}}]{2010A&A...520A..58I}
{Israel}, H., {Erben}, T., {Reiprich}, T.~H., {et~al.} 2010, \aap, 520, A58

\bibitem[{{Ivezic} {et~al.}(2008){Ivezic}, {Tyson}, {Allsman}, {Andrew},
  {Angel}, \& {for the LSST Collaboration}}]{2008arXiv0805.2366I}
{Ivezic}, Z., {Tyson}, J.~A., {Allsman}, R., {et~al.} 2008, astro-ph/0805.2366

\bibitem[{{Jee} {et~al.}(2009){Jee}, {Rosati}, {Ford}, {Dawson}, {Lidman},
  {Perlmutter}, {Demarco}, {Strazzullo}, {Mullis}, {B{\"o}hringer}, \&
  {Fassbender}}]{2009ApJ...704..672J}
{Jee}, M.~J., {Rosati}, P., {Ford}, H.~C., {et~al.} 2009, \apj, 704, 672

\bibitem[{{Kassiola} \& {Kovner}(1993)}]{1993ApJ...417..450K}
{Kassiola}, A. \& {Kovner}, I. 1993, \apj, 417, 450

\bibitem[{{Komatsu} {et~al.}(2010){Komatsu}, {Smith}, {Dunkley}, {Bennett},
  {Gold}, {Hinshaw}, {Jarosik}, {Larson}, {Nolta}, {Page}, {Spergel},
  {Halpern}, {Hill}, {Kogut}, {Limon}, {Meyer}, {Odegard}, {Tucker}, {Weiland},
  {Wollack}, \& {Wright}}]{2010arXiv1001.4538K}
{Komatsu}, E., {Smith}, K.~M., {Dunkley}, J., {et~al.} 2010, astro-ph/1001.4538

\bibitem[{{Limousin} {et~al.}(2007){Limousin}, {Kneib}, {Bardeau}, {Natarajan},
  {Czoske}, {Smail}, {Ebeling}, \& {Smith}}]{2007A&A...461..881L}
{Limousin}, M., {Kneib}, J.~P., {Bardeau}, S., {et~al.} 2007, \aap, 461, 881

\bibitem[{{Mandelbaum} {et~al.}(2008){Mandelbaum}, {Seljak}, \&
  {Hirata}}]{2008JCAP...08..006M}
{Mandelbaum}, R., {Seljak}, U., \& {Hirata}, C.~M. 2008, \jcap, 8, 6

\bibitem[{{Mandelbaum} {et~al.}(2006){Mandelbaum}, {Seljak}, {Kauffmann},
  {Hirata}, \& {Brinkmann}}]{2006MNRAS.368..715M}
{Mandelbaum}, R., {Seljak}, U., {Kauffmann}, G., {Hirata}, C.~M., \&
  {Brinkmann}, J. 2006, \mnras, 368, 715

\bibitem[{{Muldrew} {et~al.}(2010){Muldrew}, {Pearce}, \&
  {Power}}]{2010arXiv1008.2903M}
{Muldrew}, S.~I., {Pearce}, F.~R., \& {Power}, C. 2010, astro-ph/1008.2903

\bibitem[{{Natarajan} {et~al.}(2007){Natarajan}, {De Lucia}, \&
  {Springel}}]{2007MNRAS.376..180N}
{Natarajan}, P., {De Lucia}, G., \& {Springel}, V. 2007, \mnras, 376, 180

\bibitem[{{Navarro} {et~al.}(1997){Navarro}, {Frenk}, \&
  {White}}]{1997ApJ...490..493N}
{Navarro}, J.~F., {Frenk}, C.~S., \& {White}, S.~D.~M. 1997, \apj, 490, 493

\bibitem[{{Okabe} {et~al.}(2010){Okabe}, {Takada}, {Umetsu}, {Futamase}, \&
  {Smith}}]{2010PASJ...62..811O}
{Okabe}, N., {Takada}, M., {Umetsu}, K., {Futamase}, T., \& {Smith}, G.~P.
  2010, \pasj, 62, 811

\bibitem[{{Parker} {et~al.}(2007){Parker}, {Hoekstra}, {Hudson}, {van
  Waerbeke}, \& {Mellier}}]{2007ApJ...669...21P}
{Parker}, L.~C., {Hoekstra}, H., {Hudson}, M.~J., {van Waerbeke}, L., \&
  {Mellier}, Y. 2007, \apj, 669, 21

\bibitem[{{Percival} {et~al.}(2010){Percival}, {Reid}, {Eisenstein}, {Bahcall},
  {Budavari}, {Frieman}, {Fukugita}, {Gunn}, {Ivezi{\'c}}, {Knapp}, {Kron},
  {Loveday}, {Lupton}, {McKay}, {Meiksin}, {Nichol}, {Pope}, {Schlegel},
  {Schneider}, {Spergel}, {Stoughton}, {Strauss}, {Szalay}, {Tegmark},
  {Vogeley}, {Weinberg}, {York}, \& {Zehavi}}]{2010MNRAS.401.2148P}
{Percival}, W.~J., {Reid}, B.~A., {Eisenstein}, D.~J., {et~al.} 2010, \mnras,
  401, 2148

\bibitem[{{Riess} {et~al.}(2009){Riess}, {Macri}, {Casertano}, {Sosey},
  {Lampeitl}, {Ferguson}, {Filippenko}, {Jha}, {Li}, {Chornock}, \&
  {Sarkar}}]{2009ApJ...699..539R}
{Riess}, A.~G., {Macri}, L., {Casertano}, S., {et~al.} 2009, \apj, 699, 539

\bibitem[{{Schirmer} {et~al.}(2010){Schirmer}, {Suyu}, {Schrabback},
  {Hildebrandt}, {Erben}, \& {Halkola}}]{2010A&A...514A..60S}
{Schirmer}, M., {Suyu}, S., {Schrabback}, T., {et~al.} 2010, \aap, 514, A60

\bibitem[{{Schneider}(2006)}]{2006glsw.conf..269S}
{Schneider}, P. 2006, in Saas-Fee Advanced Course 33: Gravitational Lensing:
  Strong, Weak and Micro, ed. {G.~Meylan, P.~Jetzer, P.~North, P.~Schneider,
  C.~S.~Kochanek, \& J.~Wambsganss}, 269--451

\bibitem[{{Schneider} {et~al.}(2006){Schneider}, {Kochanek}, \&
  {Wambsganss}}]{SchneiderKochanekWambsganss_book}
{Schneider}, P., {Kochanek}, C., \& {Wambsganss}, J. 2006, {Gravitational
  Lensing: Strong, Weak and Micro}, Saas-Fee Advanced Course 33 (Berlin:
  Springer)

\bibitem[{{Schrabback} {et~al.}(2010){Schrabback}, {Hartlap}, {Joachimi},
  {Kilbinger}, {Simon}, {Benabed}, {Brada{\v c}}, {Eifler}, {Erben},
  {Fassnacht}, {High}, {Hilbert}, {Hildebrandt}, {Hoekstra}, {Kuijken},
  {Marshall}, {Mellier}, {Morganson}, {Schneider}, {Semboloni}, {van Waerbeke},
  \& {Velander}}]{2010A&A...516A..63S}
{Schrabback}, T., {Hartlap}, J., {Joachimi}, B., {et~al.} 2010, \aap, 516, A63+

\bibitem[{{Sheldon} {et~al.}(2009){Sheldon}, {Johnston}, {Masjedi}, {McKay},
  {Blanton}, {Scranton}, {Wechsler}, {Koester}, {Hansen}, {Frieman}, \&
  {Annis}}]{2009ApJ...703.2232S}
{Sheldon}, E.~S., {Johnston}, D.~E., {Masjedi}, M., {et~al.} 2009, \apj, 703,
  2232

\bibitem[{{Simon} {et~al.}(2008){Simon}, {Watts}, {Schneider}, {Hoekstra},
  {Gladders}, {Yee}, {Hsieh}, \& {Lin}}]{2008A&A...479..655S}
{Simon}, P., {Watts}, P., {Schneider}, P., {et~al.} 2008, \aap, 479, 655

\bibitem[{{Skilling}(2004)}]{2004AIPC..735..395S}
{Skilling}, J. 2004, in American Institute of Physics Conference Series, Vol.
  735, American Institute of Physics Conference Series, ed. {R.~Fischer,
  R.~Preuss, \& U.~V.~Toussaint}, 395--405

\bibitem[{{Springel} {et~al.}(2005){Springel}, {White}, {Jenkins}, {Frenk},
  {Yoshida}, {Gao}, {Navarro}, {Thacker}, {Croton}, {Helly}, {Peacock}, {Cole},
  {Thomas}, {Couchman}, {Evrard}, {Colberg}, \& {Pearce}}]{2005Natur.435..629S}
{Springel}, V., {White}, S.~D.~M., {Jenkins}, A., {et~al.} 2005, \nat, 435, 629

\bibitem[{{Springel} {et~al.}(2001){Springel}, {White}, {Tormen}, \&
  {Kauffmann}}]{2001MNRAS.328..726S}
{Springel}, V., {White}, S.~D.~M., {Tormen}, G., \& {Kauffmann}, G. 2001,
  \mnras, 328, 726

\bibitem[{{Suyu} \& {Halkola}(2010)}]{2010arXiv1007.4815S}
{Suyu}, S.~H. \& {Halkola}, A. 2010, astro-ph/1007.4815

\bibitem[{{The Dark Energy Survey Collaboration}(2005)}]{2005astro.ph.10346T}
{The Dark Energy Survey Collaboration}. 2005, astro-ph/0510346

\bibitem[{{Tian} {et~al.}(2009){Tian}, {Hoekstra}, \&
  {Zhao}}]{2009MNRAS.393..885T}
{Tian}, L., {Hoekstra}, H., \& {Zhao}, H. 2009, \mnras, 393, 885

\bibitem[{{Wang} {et~al.}(2006){Wang}, {Li}, {Kauffmann}, \& {De
  Lucia}}]{2006MNRAS.371..537W}
{Wang}, L., {Li}, C., {Kauffmann}, G., \& {De Lucia}, G. 2006, \mnras, 371, 537

\bibitem[{{Wright} \& {Brainerd}(2000)}]{2000ApJ...534...34W}
{Wright}, C.~O. \& {Brainerd}, T.~G. 2000, \apj, 534, 34

\end{thebibliography}

\appendix

\end{document}